\documentclass{cernyrep}
\usepackage{subcaption}
\usepackage{xcolor}
\usepackage{float}

\usepackage{texnames}
\usepackage[T1]{fontenc}
\usepackage[bookmarks, colorlinks=true, linktoc=page, pdftex, linkcolor=black, citecolor=black, urlcolor=blue]{hyperref}
\usepackage{hepnames}
\usepackage{esint}
\usepackage{physics}

\newcommand{\highlight}[1]{%
  \colorbox{green!50}{$\displaystyle#1$}}

\newcommand\independent{\protect\mathpalette{\protect\independenT}{\perp}}
\def\independenT#1#2{\mathrel{\rlap{$#1#2$}\mkern2mu{#1#2}}}

\usepackage{fancyhdr}
\fancyhfoffset{4 mm}
\fancypagestyle{ARTTITLE}{% 
\fancyhf{} % clear all header and footer fields
\lhead{\small{Proceedings of the 2018 CERN--Accelerator--School course on \it{Beam Instrumentation}, Tuusula, (Finland)}}
\lfoot{Available online at \url{https://cas.web.cern.ch/previous-schools}}
\rfoot{\thepage\hspace*{3mm}}
 
}

\begin{document}
\title{BPM Systems: A brief Introduction to Beam Position Monitoring}
\author{M.~Wendt}
\institute{CERN, Geneva, Switzerland}

\begin{abstract}
This introduction on beam position monitors (BPM) summarizes the fundamental parts of 
the tutorial presented at the CAS 2018 
on beam instrumentation.
The focus is on the signal detection and normalization, and on the principle of operation of commonly 
used broadband pickups, i.e.\ button and stripline BPMs. 
Other BPM types, such as split-plane and cavity BPMs are also discussed,
as well as the detection of low-$\beta$ beams.
Finally, a note on BPM signal processing techniques is given.  
\end{abstract}

%\keywords{beam position monitors.}

\maketitle

\thispagestyle{ARTTITLE}

\section{Introduction}

\begin{figure}[ht]
\begin{center}
\includegraphics[width=0.65\textwidth]{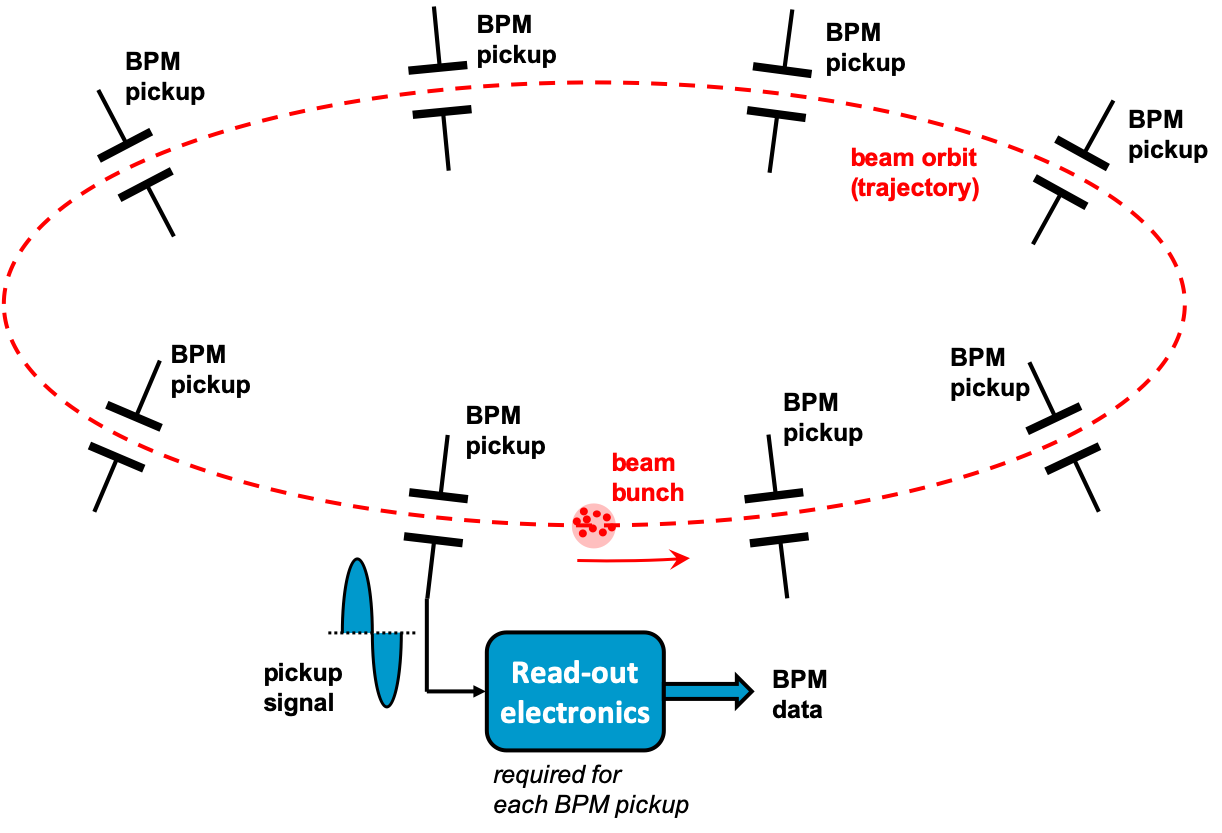}
\caption{A BPM system}
\label{fig:BPMsys}
\end{center}
\end{figure}

The beam position monitor (BPM) system is one of the most utilized and powerful beam instrumentation tools 
in a particle accelerator or accelerator beam-line.
A BPM system consists out of many beam position monitors distributed along the accelerator beam-line,
see Fig.~\ref{fig:BPMsys} for a ring accelerator, monitoring the passing beam.
Each beam position monitor consists our of:
\begin{itemize}
\item a BPM \textbf{pickup}, which is part of the vacuum chamber and consists out of two or four symmetrically arranged 
electrodes which couple to the the electromagnetic fields 
of the beam and generate electrical signals at their output ports.
Typically, a BPM pickup is located near each quadrupole magnet along the beam-line.
\item the \textbf{read-out electronics}, which is used to condition and process the signals from the pickup electrodes
 to provide the beam position information in a digital data format, such that it can be acquired and 
 further processed by the accelerator control system.
\end{itemize}

Popular BPM pickup styles are of button or stripline type, have broadband characteristics, and generate
a pulse-like output signal for each passing bunch.
This enables bunch-by-bunch BPM signal processing possibilities, but requires the synchronization of the BPM data
of all monitors in the system.
In a ring accelerator a synchronized turn-by-turn monitoring of the beam position is of interest to follow the response
of the beam to a kick or chirp excitation, e.g.\ for beam optic studies and analysis.
Averaging the BPM data over many turns (ring accelerator) or over many beam pulses (linac) at each BPM 
provides a high resolution measurement of the beam orbit or beam trajectory, used for the alignment of the beam,
for an orbit feedback, etc.

This summary of the CAS 2018 tutorial on ``BPM Systems'' gives a short introduction on the principles and
techniques of a beam position monitor.
The principle of operation of broadband BPM pickups are covered in more detail, as they are widely used,
other types of BPM pickups are mentioned briefly. 
Some information on the BPM signal processing is also provided, however, as this part is subject to the advances
in electronics technologies and certainly will change, or will be outdated soon, therefore the discussion on read-out
electronics is kept brief. 

\section{Principle of Operation of a Beam Position Monitor}

\begin{figure}[ht]
\begin{center}
\includegraphics[width=0.55\textwidth]{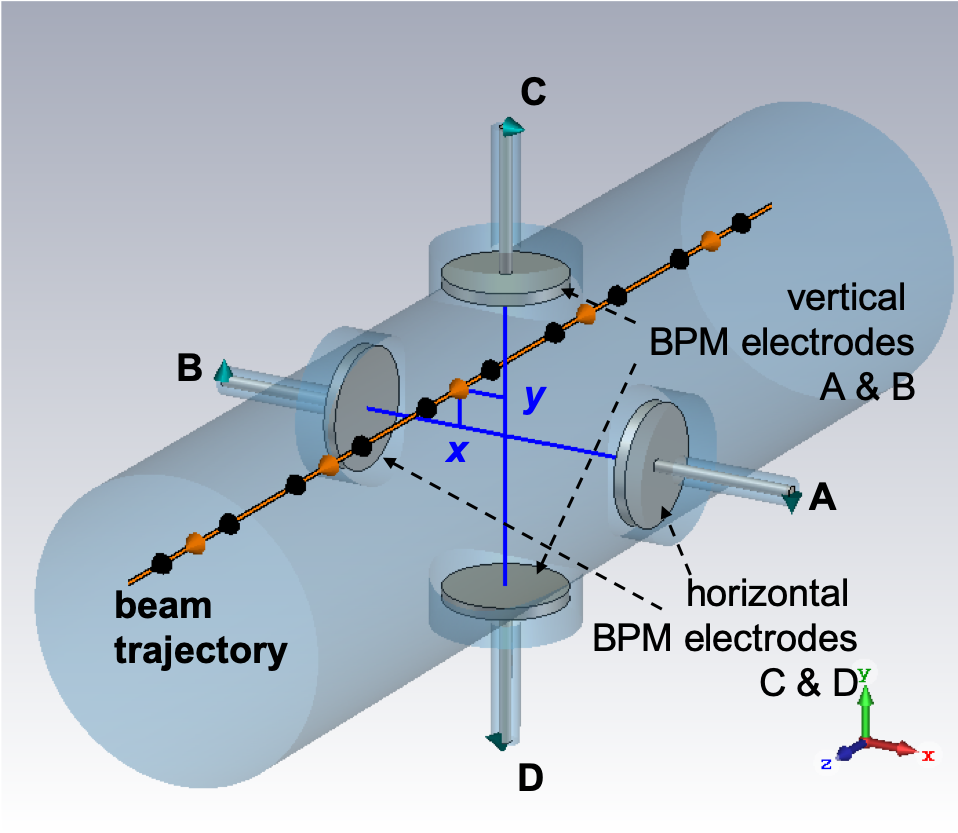}
\caption{An electromagnetic BPM pickup}
\label{fig:BPMpu}
\end{center}
\end{figure}

This introduction on beam position monitoring covers BPM systems based on non-invasive, 
electro\-magnetic-type beam pickups,
however it should be noticed, there are other methods and ways to detect the position of the beam.
Figure~\ref{fig:BPMpu} illustrates a typical pickup, in this example the so-called ``button''-style BPM, 
which consists out of four round, metallic, coin-like
electrodes (the ``buttons''), which are arranged symmetrically along the horizontal and vertical axes in a 
vacuum chamber of circular cross-section.
The horizontal ($x$) and vertical ($y$) offset of the beam trajectory wrt.\ the center of the vacuum chamber is
the \emph{beam position} or \emph{beam displacement}, and has to
be monitored with high resolution, accuracy and repeatability, this is the goal of the beam position measurement. 

\begin{figure}[ht]
\begin{subfigure}{.4\textwidth}
  \centering
  \vspace*{-1mm}
  \includegraphics[width=0.9\linewidth]{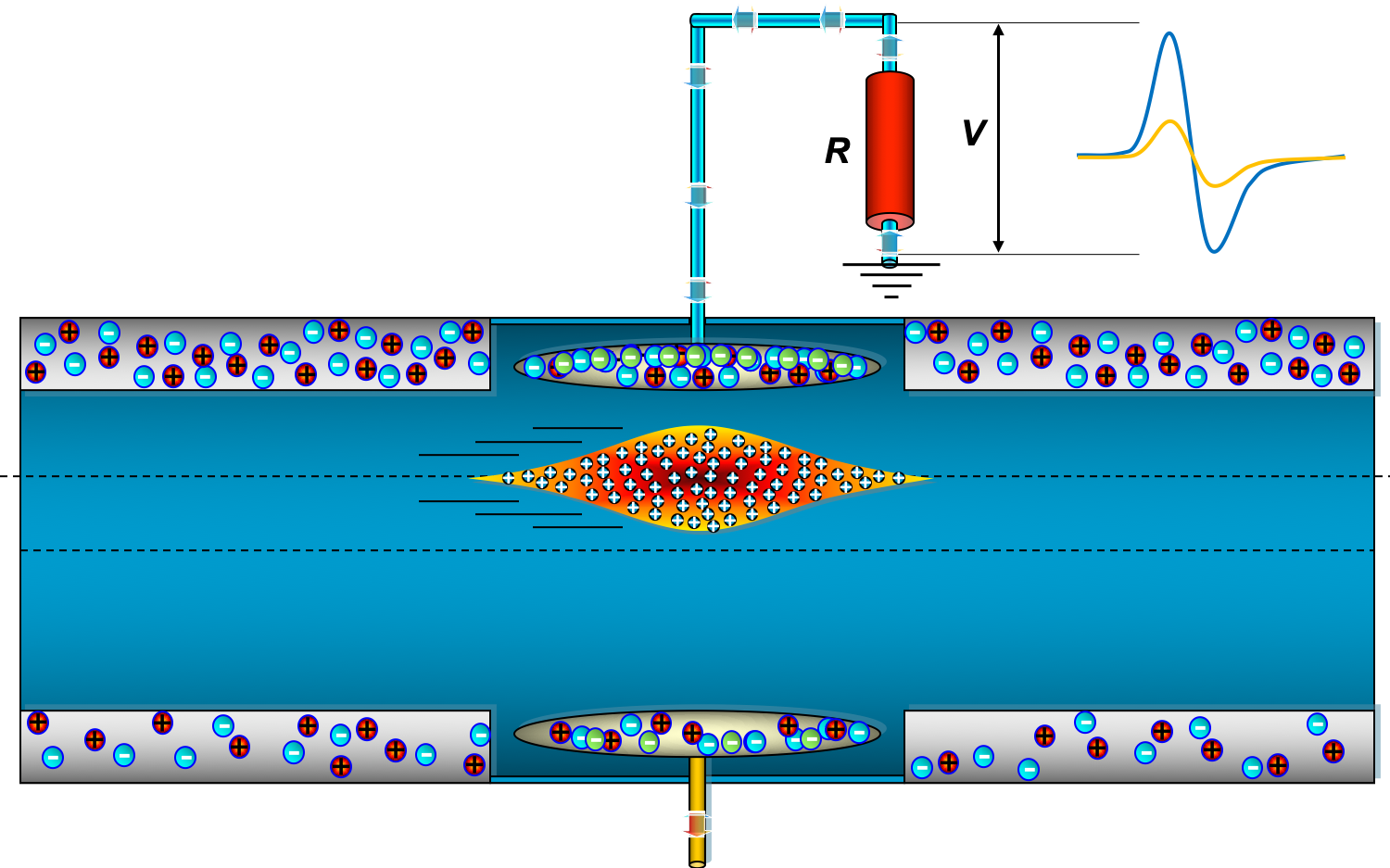}
  \caption{Image charges and \\ BPM signals for a displaced beam bunch.}
  \label{fig:sfig3a}
\end{subfigure}%
\begin{subfigure}{0.6\textwidth}
  \centering
  \includegraphics[width=0.9\linewidth]{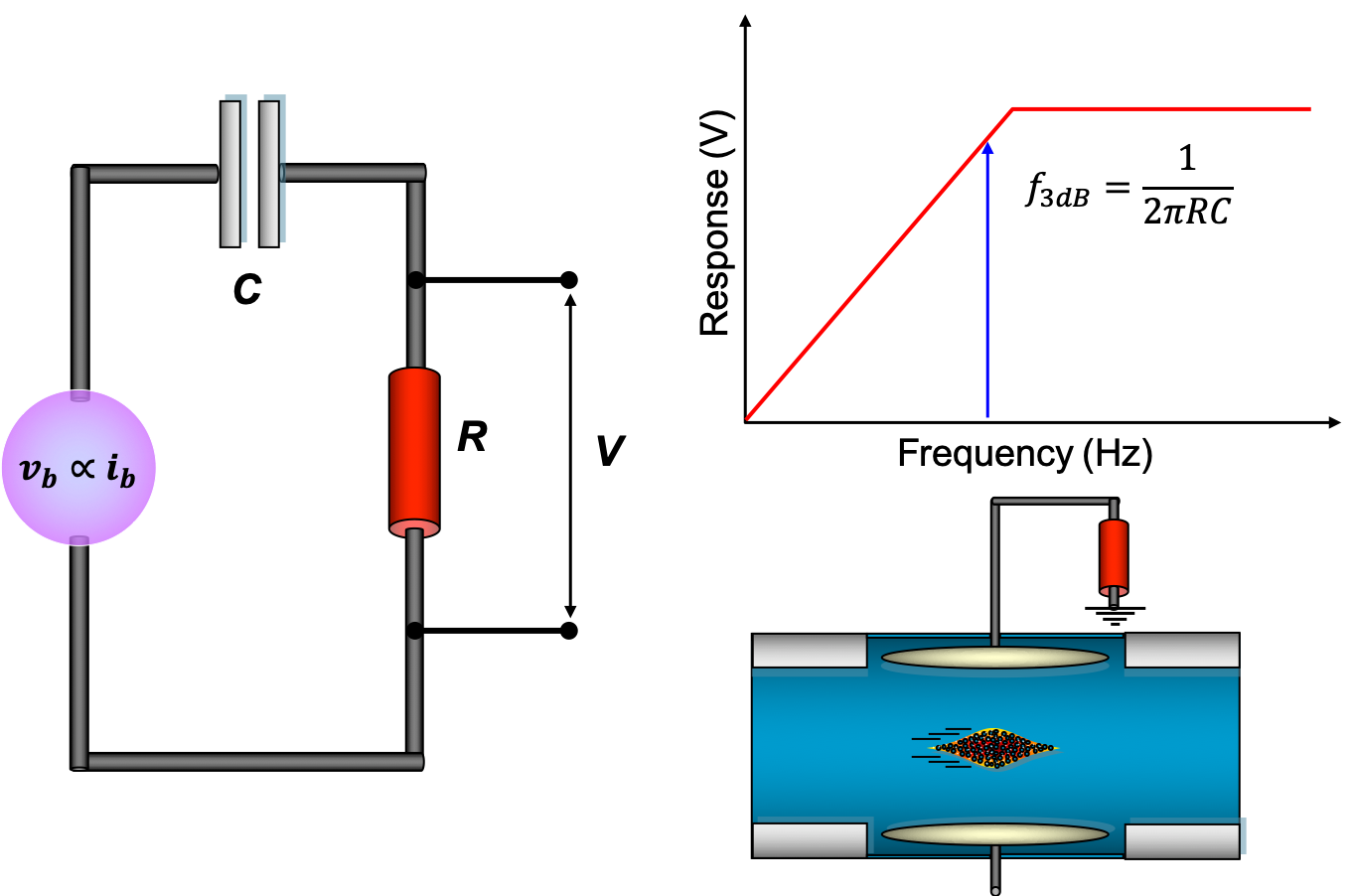}
  \caption{Equivalent circuit of an electrostatic BPM electrode.}
  \label{fig:sfig3b}
\end{subfigure}
\caption{Operation principle of a BPM pickup (courtesy \textit{O.R.\ Jones}).}
\label{fig:principle}
\end{figure}

Broadband BPM pickups operate on the principle of the image currents (or image charges), 
each charged particle of the beam is ``compensated'' by an image charge of opposite sign in the metallic vacuum chamber.
Figure~\ref{fig:sfig3a} shows the induced image charges on a pair of electrostatic BPM electrodes
and the resulting bunch response voltage signal for the upper electrode on a load resistor. 
%the pulse-waveform of both signals is identical, 
%but the amplitude depends on the distance between bunched beam and electrode, i.e.\ the transverse beam position.
The time-domain response of a BPM electrode to a bunched beam appears as a differentiated pulse 
as the image charges are induced as a displacement current through the capacitive button electrode. 
Figure~\ref{fig:sfig3b} shows the equivalent circuit
and the frequency-domain response of this capacitive coupling BPM electrode, which is equivalent to that
of a simple, 1$^{\mathrm{st}}$-order high-pass filter.
There is no ``DC-coupling'', and therefore broadband BPM pickups cannot operated with debunched,
direct-current (DC) beams.

\begin{figure}[b]
\begin{center}
\includegraphics[width=0.7\textwidth]{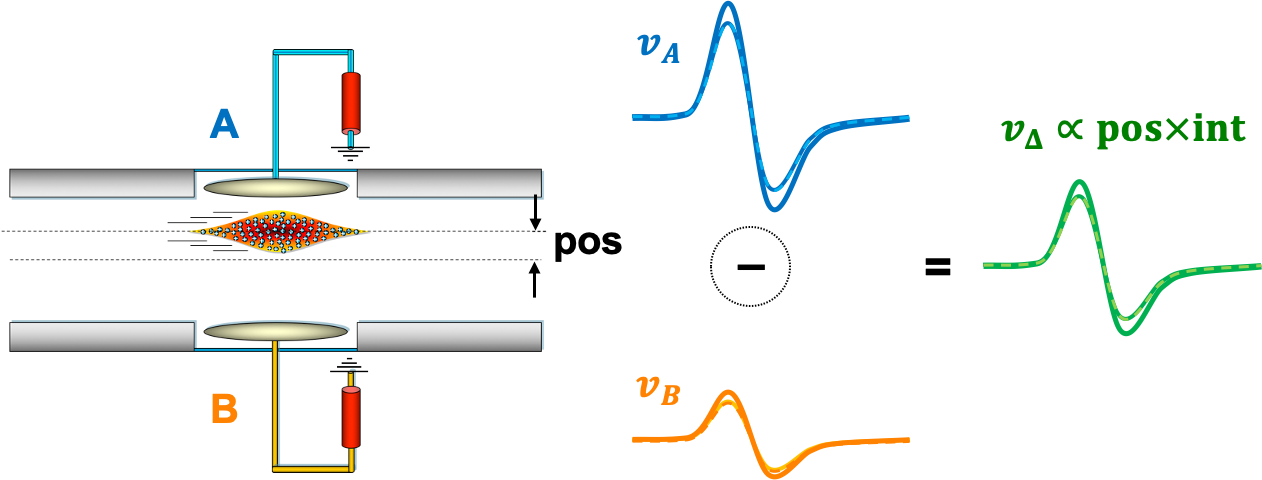}
\caption{The beam (bunch) intensity effect on the difference signal of a BPM pickup.}
\label{fig:DeltaSignal}
\end{center}
\end{figure}

As illustrated in Fig.~\ref{fig:DeltaSignal}, the pulse-like signal waveforms out of the symmetric pair of 
BPM electrodes are always identical,
but as for each electrode 
$$
v_{\mathrm{elec}}\propto\mathrm{pos}\times\mathrm{int}
$$
the signal amplitude is proportional to the distance between
BPM electrode and beam, i.e.\ the beam position ($\mathrm{pos}$) \emph{and} the intensity ($\mathrm{int}$) of the beam,
it is obvious to subtract the two voltage signals of the symmetrically arranged BPM electrodes, 
lets call them $A$ and $B$,
to get a signal proportional to the beam position ($\mathrm{pos}$):
$$
v_{\Delta} = (v_A-v_b) \propto \mathrm{pos}
$$
While this is true, the signal of each BPM electrode remains proportional to the beam intensity
\linebreak
$v_{\mathrm{elec}} \propto \mathrm{int}$,
and due to this so-called \emph{common-mode} signal %-- as Fig.~\ref{fig:DeltaSignal} illustrates -- 
the difference signal $v_{\Delta}$ of the BPM electrodes 
still contains the beam intensity:
 \begin{equation}
 v_{\Delta} = (v_A-v_B) \propto \mathrm{pos} \times \mathrm{int}
 \label{eq:delta}
 \end{equation}
 A \emph{normalization} procedure is required to extract a pure, beam intensity independent position signal from
 the BPM electrodes:
 $$
 \frac{v_{\Delta}}{v_{\mathrm{int}}} \propto \frac{\mathrm{pos} \times \mathrm{int}}{\mathrm{int}} = \mathrm{pos}
 $$
 Typically, the intensity signal required for the normalization is derived from the BPM electrodes as well, e.g.\
 $v_{\mathrm{int}}=v_{\Sigma}=v_A+v_B$, which leads to the well-known method to normalize the beam position signal:
 \begin{equation}
 \frac{v_{\Delta}}{v_{\Sigma}}=\frac{v_A-v_B}{v_A+v_B} \propto \mathrm{pos} 
 \label{eq:SumDiff}
 \end{equation} 
 however, many other methods for the beam intensity normalization also exist. 
 
 \section{The BPM Pickup}
 
A BPM pickup is a passive, linear electromagnetic coupler, each of its electrodes delivers a
signal (here defined in the frequency-domain):
\begin{equation}
V_{\mathrm{elec}}(x,y,\omega) = s(x,y)Z(\omega)I_b(\omega)
\label{eq:transfer}
\end{equation}
with $Z(\omega)$ being the transfer impedance of a BPM pickup electrode (unit: $\Omega$), 
$I_b(\omega)$ being the beam or bunch current, i.e.\ the frequency-domain
equivalent of the time-domain envelope function of the longitudinal particle distribution.
Often the bunch current $I_{\mathrm{bunch}}(\omega)$ is assumed as a \textit{Gaussian} envelope function, which is a good
approximation for electron bunches, but may have some limitations in case of bunched proton or ion beams. 
$s(x,y)$ is a sensitivity function reflecting the cross-section geometry of the BPM pickup, 
to describe the transverse position characteristic, i.e.\
the strength of the coupling between beam and BPM electrode as function of the transverse 
beam position $(x,y)$.

Hidden in Eq.~ (\ref{eq:transfer}) is an electromagnetic \emph{coverage factor}
\begin{equation}
\phi=\frac{\int J_w dA_{\mathrm{elec}}}{ \int J_w dA_{\mathrm{BPM}}}
\label{eq:cover}
\end{equation}
which is defined as ratio of the integrated wall current $J_w$ between the surface of the 
pickup electrode $A_{\mathrm{elec}}$ and the total surface area of the BPM pickup $A_{\mathrm{BPM}}$, 
expressed for a centered beam. 

\begin{figure}[ht]
\begin{subfigure}{.22\textwidth}
  \centering
  \vspace*{-1mm}
  \includegraphics[width=0.95\linewidth]{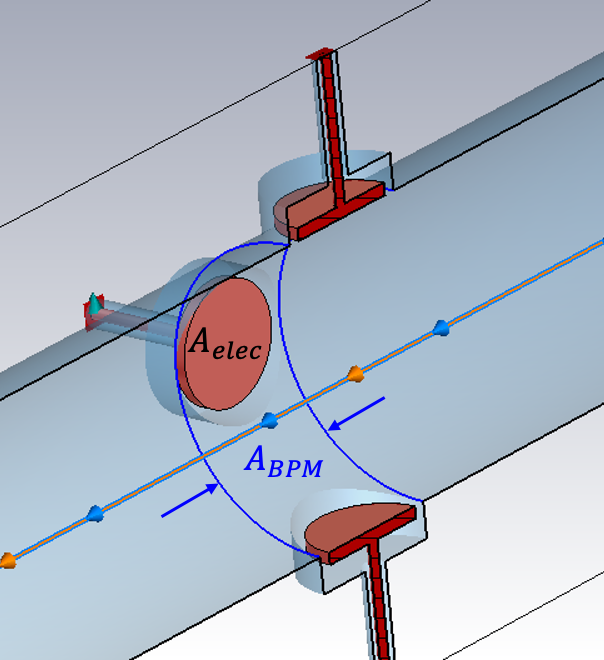}
  \caption{Coverage areas.}
  \label{fig:sfig5a}
\end{subfigure}%
\begin{subfigure}{0.55\textwidth}
  \centering
  \includegraphics[width=0.95\linewidth]{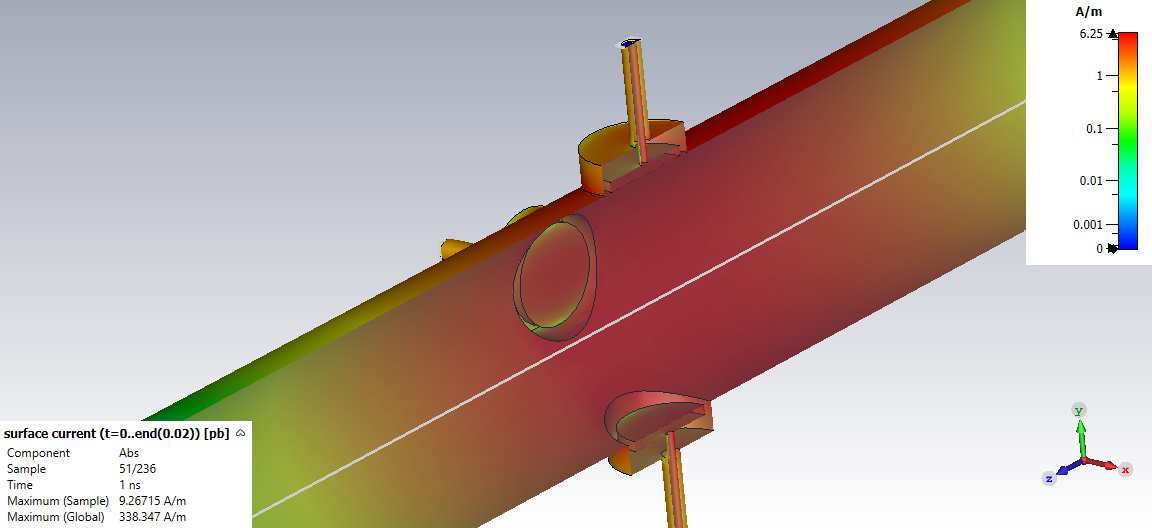}
  \caption{Image (wall) current density $J_w$.}
  \label{fig:sfig5c}
\end{subfigure}
\begin{subfigure}{0.22\textwidth}
  \centering
  \includegraphics[width=0.95\linewidth]{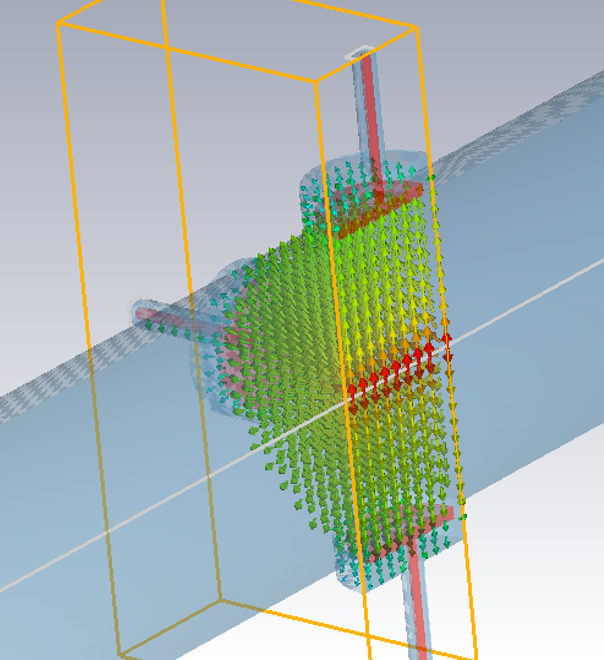}
  \caption{Local E-field.}
  \label{fig:sfig5b}
\end{subfigure}
\caption{Electromagnetic coverage of button BPM electrodes.}
\label{fig:coverage}
\end{figure}

Figure~\ref{fig:coverage} illustrates the coverage factor on the longitudinal section of a button-style BPM pickup:
\newline
Fig.~\ref{fig:coverage}a indicates the surface areas $A_{\mathrm{elec}}$ and $A_{\mathrm{BPM}}$.
\newline
Fig.~\ref{fig:coverage}b shows a ``snap-shot'' of the image (wall) current density $J_w$ for a \textit{Gaussian} bunch 
as it is passing the BPM pickup, and
\newline
Fig.~\ref{fig:coverage}c shows the E-field local in the BPM volume.

\begin{figure}[t!]
\begin{center}
\includegraphics[width=0.325\textwidth]{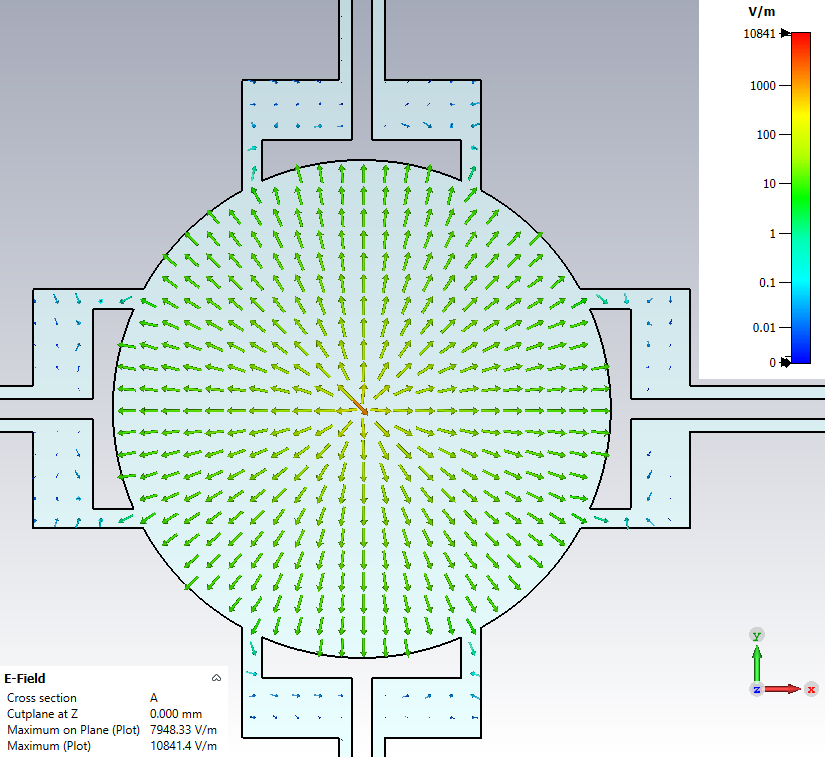} 
\hfill
\includegraphics[width=0.335\textwidth]{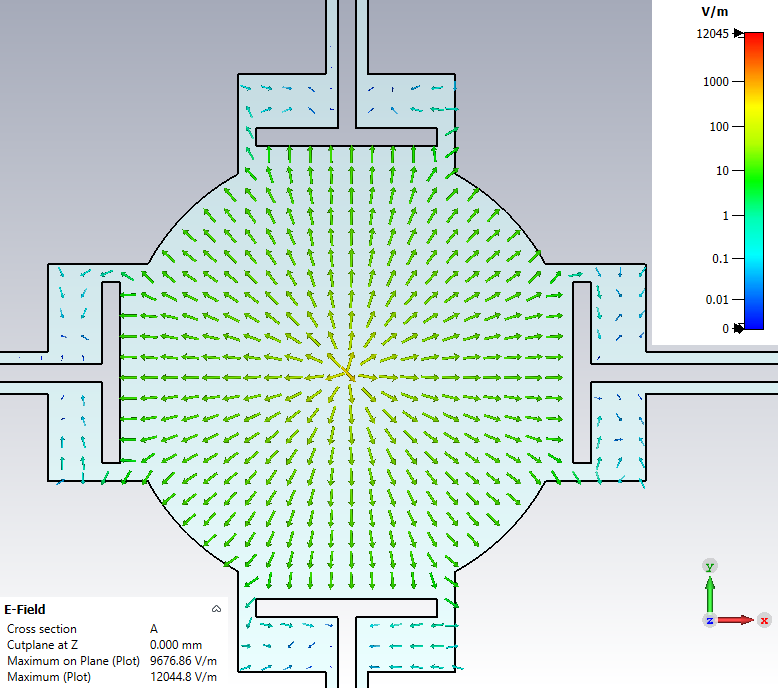} 
\hfill
\includegraphics[width=0.31\textwidth]{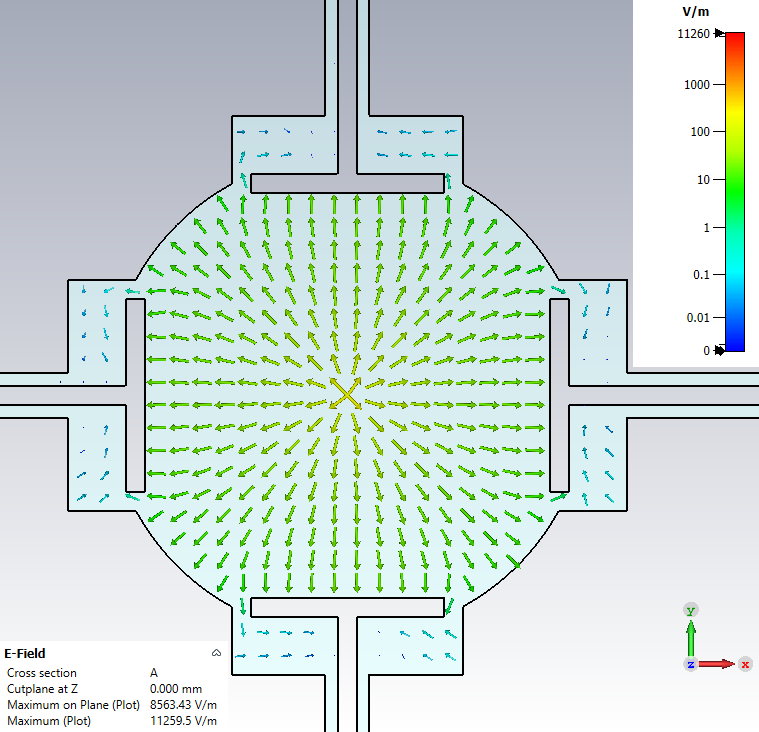}
\caption{E-field for different BPM electrode arrangements.}
\label{fig:coverage_cross}
\end{center}
\end{figure}

While similar, the \emph{geometric} and the \emph{electromagnetic} coverage are not identical,
$$
\frac{\int J_w dA_{\mathrm{elec}}}{ \int J_w dA_{\mathrm{BPM}}} \neq 
\frac{A_{\mathrm{elec}}}{ A_{\mathrm{BPM}}}
$$
which is illustrated for the cross-section of three different button BPM arrangements in Fig.~\ref{fig:coverage_cross}.
Evidently, for the the curved, flush BPM electrodes,  Fig.~\ref{fig:coverage_cross} (left), geometric and electromagnetic coverage are very similar, but this is not true for the other examples with flat BPM electrodes, 
arranged flush with the beam pipe,
Fig.~\ref{fig:coverage_cross} (center), or raised, Fig.~\ref{fig:coverage_cross} (right).

As Eq.~(\ref{eq:cover}) expresses the coupling ratio between beam and BPM electrode for a centered beam,
it is equivalent to $s(x=0,y=0)=\phi$. 
But while it is an intrinsic part of the BPM position characteristic,
there is no practical meaning linked as it vanishes through the normalization procedure. 
The coverage factor $\phi$ is based on size and shape of the BPM electrode, which also defines the frequency depending
beam coupling characteristic, therefore it also appears in the
the discussion of the transfer impedance $Z(\omega)$ for broadband BPM pickup electrodes,
relevant to compute the signal levels of the BPM electrodes. 

%However, as this coverage factor can also be expressed in geometric terms, care has to be taken:\newline 
%In the following sections $\phi$ appears
%twice, in the discussion of the sensitivity function $s(x,y)$ based on the image charges, as well as
%in the discussion of the transfer impedance $Z(\omega)$ for broadband BPM pickup electrodes. 

Furthermore, please notice that Eq.~(\ref{eq:transfer}) applies only to broadband BPM pickups, where the transverse
position sensitivity $s \independent \omega$ is frequency independent
%More restricted, Eq.~(\ref{eq:transfer}) is limited to 
for relativistic beams ($\beta \to 1$). 
For low-$\beta$ beams $s(x,y)$
is a function of the beam velocity $v=\beta c_0$, and is discussed at the end of this section.

Equation~(\ref{eq:transfer}) separates the frequency dependent, but beam position independent 
response function $Z(\omega)$ from the frequency independent beam position characteristic $s(x,y)$, which is based
on the image charge model (for broadband BPM pickups).
This methodology simplifies the understanding and analysis of a BPM pickup.
The characteristics of a resonant (cavity) BPMs does not simply match to Eq.~(\ref{eq:transfer}),
and is discussed separately.
 
 \subsection{Image charges}
 
\begin{figure}[ht]
\hfill
\begin{subfigure}{.34\textwidth}
  \centering
  \includegraphics[width=0.9\linewidth]{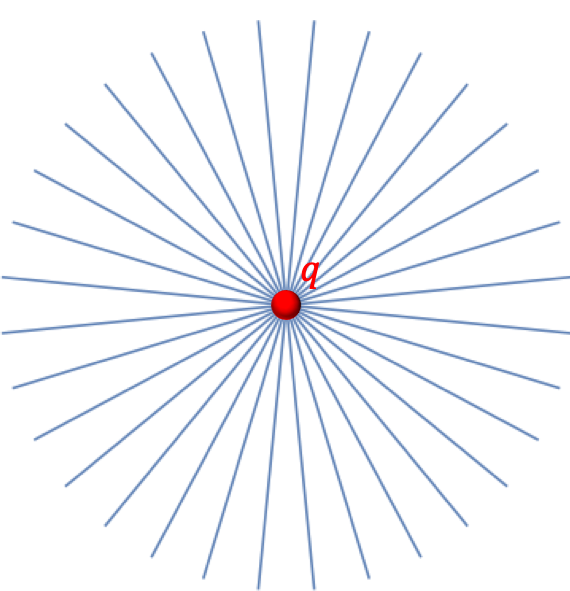}
  \caption{point charge at rest}
  \label{fig:sfig7a}
\end{subfigure}%
\hfill
\begin{subfigure}{0.3\textwidth}
  \centering
  \includegraphics[width=0.65\linewidth]{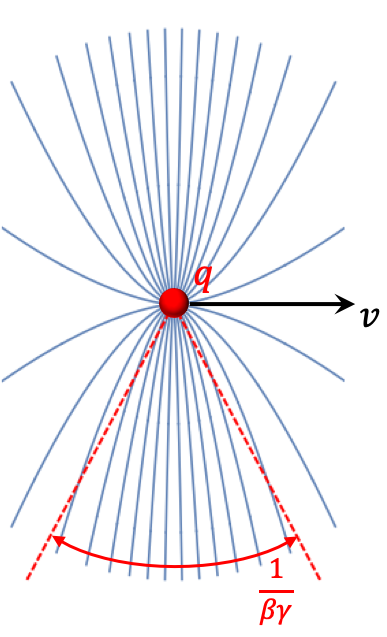}
  \caption{point charge at $\beta=0.6$.}
  \label{fig:sfig7b}
\end{subfigure}%
\hfill
\begin{subfigure}{0.3\textwidth}
  \centering
  \includegraphics[width=0.27\linewidth]{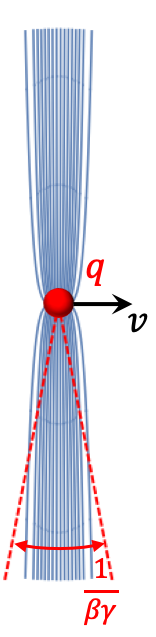}
  \caption{point charge at $\beta=0.9$.}
  \label{fig:sfig7c}
\end{subfigure}
\caption{Lines of constant electric field (in 2D) for a point charge in free space moving at constant velocity.}
\label{fig:PointCharge}
\end{figure}

From electrostatic principles (\textit{Gauss} law, \textit{Poisson} equation) we find for a point charge in the 
rest frame of the charge:
$$
\nabla \vec{E} =-\nabla^2 \phi = \frac{\rho}{\epsilon} \implies \phi'=\frac{q}{4\pi\epsilon_0} \frac{1}{r'} 
$$
the potential $\phi'$ follows a uniformly distributed static electric field, see also Fig.~\ref{fig:sfig7a}
$$
\vec{E'}=\frac{q}{4\pi\epsilon_0} \frac{\vec{r'}}{r'^3}
$$
where $|\vec{r}|=\sqrt{x^2+y^2+z^2}$ is the modulus of the vector between the charge and the observer.
From the \textit{Lorentz} transformation for the electromagnetic fields we find the $E$-field components
for a point charge moving with constant velocity $v=\beta c_0$ along the $z$-axis:
$E_x=\gamma E'_x$, $E_y=\gamma E'_y$, and $E_z=E'_z$, with $\gamma=1/\sqrt{1-\beta^2}$.
The transformation $r'=\sqrt{x^2+y^2+\gamma^2 (z-vt)^2}$ results at $t=0$ in the electric field
$$
\vec{E}=\frac{q}{4\pi\epsilon_0}\frac{\gamma}{(x^2+y^2+\gamma^2 z^2)^{3/2}}\vec{r}
$$
where $\vec{r}=(x,y,z-vt)$ is the vector from the current position of the particle to the observer.
The magnetic field is found by the relation $\vec{B}=1/c_0^2 (\vec{v}\times \vec{E})$.
Figure~\ref{fig:sfig7b} and \ref{fig:sfig7c} show how the electric field gets ``compressed'' as the velocity of the
a point charge increases.
The opening angle of the EM-field is $\propto 1/(\beta\gamma)$.

\begin{figure}[ht]
\begin{center}
\includegraphics[width=0.4\textwidth]{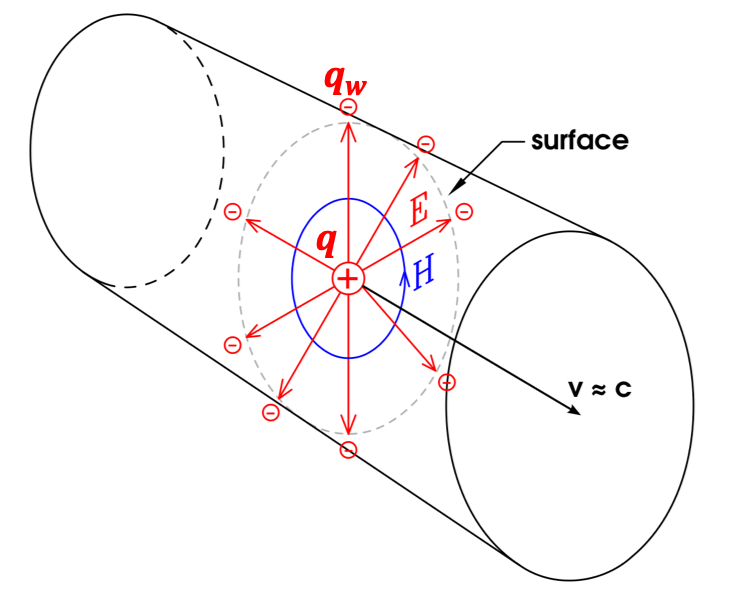}
\caption{TEM field of a point charge moving with $v\simeq c_0$ in a conductive beam pipe.}
\label{fig:WallCurrent}
\end{center}
\end{figure}
At the relativistic limit $\beta\to 1$ the electromagnetic field of the point charge is purely transverse, 
a so-called transverse electro-magnetic (TEM) field (``pancake''-like field),
which also applies if the charge moves with relativistic velocity in a conducting vacuum chamber, see Fig.~\ref{fig:WallCurrent}.
In this case the point charge $q$ is compensated by image (wall) charges $q_w$ which are distributed at
the inner surface of the beam pipe. 
$q$ and $q_w$ are always linked through the electromagnetic field.

\subsection{Position characteristic in a circular vacuum chamber}
\label{sec:PosChar}
\begin{figure}[ht]
\begin{center}
\includegraphics[width=0.45\textwidth]{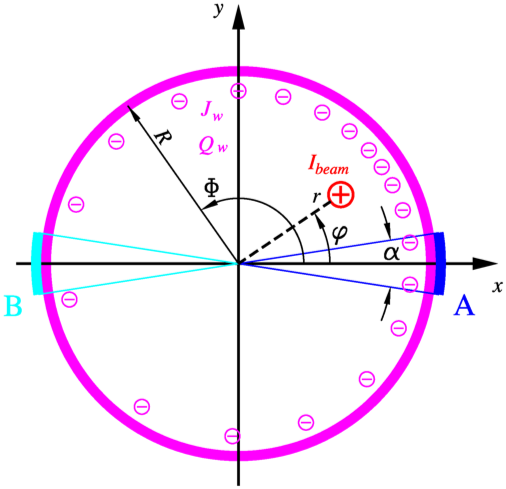}
\caption{Image charges for an off-centered beam (line-charge) in a beam pipe with circular cross-section.}
\label{fig:BPMcircular}
\end{center}
\end{figure}

A charged particle beam $I_{\mathrm{beam}}$ with infinite small transverse dimension, a so-called pencil beam 
which is equivalent to a line charge, 
traveling with relativistic velocity
($\beta \to 1$) in a beam pipe with circular cross-section of radius $R$ has only transverse electromagnetic field 
components (TEM field), and is equivalent to a line charge.
Figure~\ref{fig:BPMcircular} illustrates the equivalent 2-dimensional electrostatic image charge problem
for a beam charge at the position $(x=r\cos\varphi,y=r\cos\varphi)$.
The solution for the wall current distribution can be found in form of a series expansion:
\begin{align}
J_w(R, \Phi, r, \varphi)&=
-\frac{I_{\mathrm{beam}}}{2\pi R}\left [ 1+\sum_{n=1}^{\infty} \left (\frac{r}{R} \right )^n \cos n (\Phi-\varphi) \right ]
\label{eq:WCappr} \\
\intertext{or as a closed-form expression:}
J_w(R, \Phi, r, \varphi)&=
-\frac{I_{\mathrm{beam}}}{2\pi R} \frac{R^2-r^2}{R^2+r^2-2Rr\cos (\Phi-\varphi)}
\label{eq:WCclosed}
\end{align}
A wall current related fraction $I_{\mathrm{elec}}$ is induced on a BPM electrode covering an arc $\alpha$:
\begin{equation}
I_{\mathrm{elec}}=R\int_{-\alpha/2}^{+\alpha/2} J_w(R, \Phi, r, \varphi) \mathrm{d}\Phi
\label{eq:Ielec}
\end{equation}
For the two horizontal arranged electrodes $A$ and $B$ of Fig,~\ref{fig:BPMcircular} we find:
\begin{equation}
I_{\mathrm{elec}}=-\frac{I_{beam}}{2\pi} s_{\mathrm{elec}} \left ( r/R,\varphi, \alpha \right )
\label{eq:ElecAppr}
\end{equation}
with the sensitivity functions for the $A$ and $B$ electrodes based on (\ref{eq:WCappr}):
\begin{align}
s_A \left( r/R,\varphi, \alpha \right ) &= 
\alpha + 4 \sum_{n=1}^{\infty} \frac{1}{n} \left ( \frac{r}{R} \right )^n \cos ( n\varphi ) \sin \left ( \frac{n\alpha}{2} \right ) 
\label{eq:Aelec} \\
s_B \left( r/R,\varphi, \alpha \right ) &= 
\alpha + 4 \sum_{n=1}^{\infty} \frac{1}{n} \left ( \frac{r}{R} \right )^n \cos ( n\varphi ) 
\sin \left [n \left ( \pi+\frac{\alpha}{2} \right ) \right ]
\label{eq:Belec}
\end{align}
Following the intensity normalization concept of Eq.~(\ref{eq:SumDiff}) for our two symmetric, horizontal 
electrodes $A$ and $B$, using (\ref{eq:WCappr}), (\ref{eq:ElecAppr}), (\ref{eq:Aelec}), (\ref{eq:Belec}),
we can approximate the horizontal position characteristic 
for a circular beam pipe of radius $R$ as:
\begin{equation}
 \frac{\Delta}{\Sigma} = \frac{A-B}{A+B}  \simeq \frac{4\sin(\alpha /2)}{\alpha} \frac{x}{R}
+ \mathrm{higher\: order \: terms} = \mathrm{hor. \: position}
\label{eq:HorPosApprox}
\end{equation}
with $x/R$ being the normalized horizontal beam position (normalized to the beam pipe radius $R$), 
and $\alpha$ being the coverage angle of the electrodes, intercepting some fraction of the image current $J_w$.
In general, the normalized position characteristic is non-linear, however, for narrow electrodes 
$\sin(\alpha /2)\approx\alpha /2$ and small beam displacements $x^2+y^2 \ll R^2$ 
the normalized beam position response
follows approximatively: 
\begin{equation}
\mathrm{hor. \: position} \approx \frac{2}{R}x
\label{eq:HorPosLin} 
\end{equation}
with $2/R=k_{\mathrm{PU}}$ sometimes called the \emph{monitor constant}.
From (\ref{eq:WCclosed}) and {\ref{eq:Ielec}) we can also find a closed form expression for the
normalized position characteristic in \textit{Cartesian} coordinates:
\begin{equation}
 \frac{\Delta}{\Sigma} = \frac{A-B}{A+B}  =
\frac{s(x,y,R,\alpha)-s(-x,y,R,\alpha)}{s(x,y,R,\alpha)+s(-x,y,R,\alpha)} = \mathrm{hor. \: position}
\label{eq:HorPos}
\end{equation}
with
\begin{equation}
\begin{aligned}
s(x,y,R,\alpha) &= \pi \int_{-\alpha /2}^{+\alpha /2} J_w(R, \Phi) \mathrm{d} \Phi \\
  &= \arctan  \frac{[\left(R+x \right)^2+y^2 ] \tan \left( \frac{\alpha}{4} \right) - 2Ry}{x^2+y^2-R^2} +
   \arctan  \frac{[\left(R+x \right)^2+y^2 ] \tan \left( \frac{\alpha}{4} \right) + 2Ry}{x^2+y^2-R^2} 
\end{aligned}
\label{eq:func}
\end{equation}
being the sensitivity function.

\begin{figure}[t!]
\hfill
\begin{subfigure}{0.35\textwidth}
  \centering
  \includegraphics[width=0.9\linewidth]{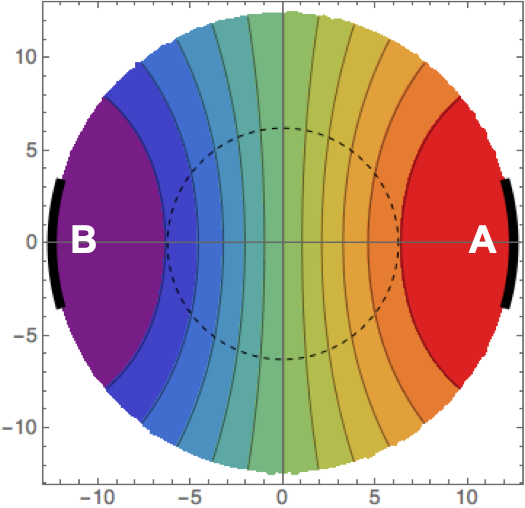}
  \caption{$f(x,y)=\mbox{const } \Delta / \Sigma$ \\ (dashed circle: half aperture)}
  \label{fig:sfig10a}
\end{subfigure}%
\hfill
\begin{subfigure}{0.58\textwidth}
  \centering
  \includegraphics[width=0.9\linewidth]{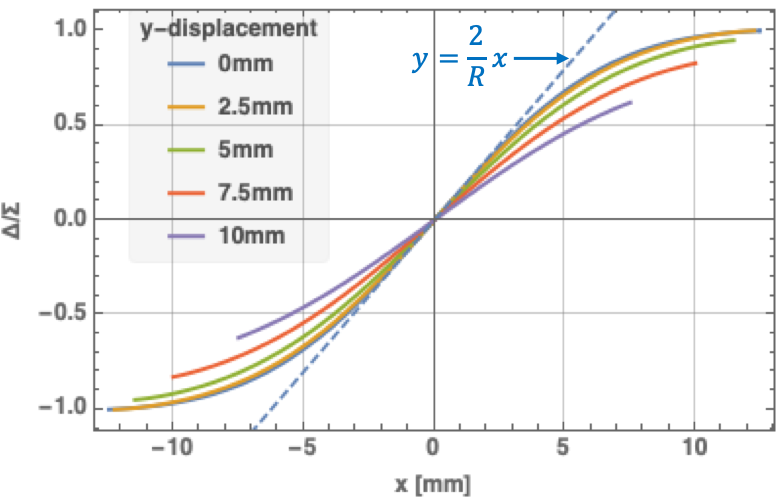}
  \caption{$\Delta / \Sigma = f(x)$ for different vertical beam displacements.}
  \label{fig:sfig10b}
\end{subfigure}%
\hspace*{1mm}
\caption{Horizontal $\Delta / \Sigma$ position characteristic in a circular beam pipe of radius $R=12.5~mm$, \\
with electrodes each covering an arc of $\alpha=30^{\circ}$.}
\label{fig:PosCharDOS}
\end{figure}

An example of the horizontal position characteristic based on (\ref{eq:HorPos}), (\ref{eq:func}) 
is illustrated in Figure~\ref{fig:PosCharDOS}, 
on the left side (Fig.~\ref{fig:sfig10a}) as parametric plot with contours of constant $\Delta /\Sigma$,
on the right side (Fig.~\ref{fig:sfig10b}) as function of the horizontal beam position.
The small covering angle ($\alpha=30^{\circ}$) of the electrodes lead to high non-linear
effects in the position characteristic, with substantial cross-coupling to the vertical plane. 
From that perspective a covering angle of $\alpha\approx 60^{\circ}$ is preferable,
however, other aspects like a minimum beam coupling impedance often have to be considered
and call for small BPM electrodes.

For accelerators covering a large range of beam intensities, a pair of logarithmic detectors in the BPM read-out electronics
can be beneficial.
In this case the intensity normalization is performed by:
\begin{align}
\mathrm{hor. \: position} &\propto 20 \log_{10} (v_A) - 20 \log_{10} (v_B) 
= 20\log_{10} \left( \frac{A}{B} \right) \label{eq:log} \\
&\simeq 20\log_{10} \left(\frac{4\sin(\alpha /2)}{\alpha} \frac{x}{R}
+ \mathrm{higher\: order \: terms}\right) \\
&= 20\log_{10} \frac{s(x,y,R,\alpha)}{s(-z,y,R,\alpha)}
\end{align} 
As Eq.~(\ref{eq:log}) indicates, the signals of the BPM electrodes $A$ and $B$ are detected with separate logarithmic
amplifiers and their detected output signals are subtracted, resulting in a normalized beam position signal.
Each logarithmic detector follows the characteristic $v_{out}=20\log_{10} (v_{in})$ over a large input signal range,
typically $40\ldots 60~dB$.
\begin{figure}[t]
\hfill
\begin{subfigure}{.35\textwidth}
  \centering
  \vspace*{-3mm}
  \includegraphics[width=0.9\linewidth]{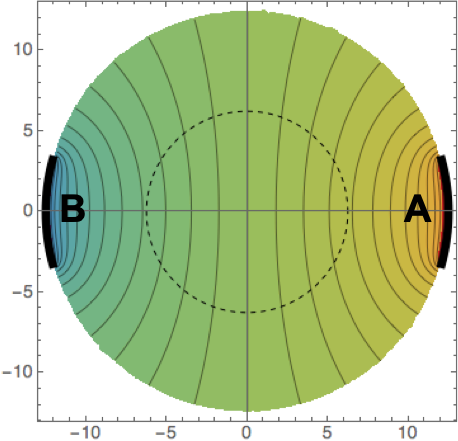}
  \vspace*{5mm}
  \caption{$f(x,y)=\mbox{const } 20\log_{10}(A/B)$}
  \label{fig:sfig11a}
\end{subfigure}%
\hfill
\begin{subfigure}{0.58\textwidth}
  \centering
  \includegraphics[width=0.9\linewidth]{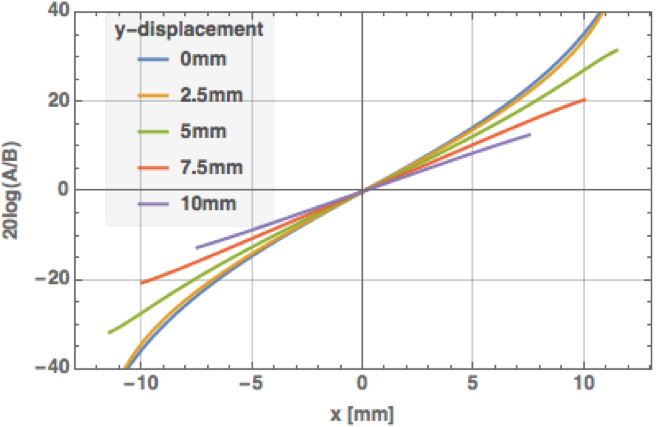}
  \caption{$20\log_{10}(A/B) = f(x)$ for different vertical beam displacements. 
                (BPM sensitivity at the origin: $2.75~dB/mm$))}
  \label{fig:sfig11b}
\end{subfigure}%
\hspace*{1mm}
\caption{Horizontal position log-ratio characteristic in a circular beam pipe of radius $R=12.5~mm$, \\
with electrodes each covering an arc of $\alpha=30^{\circ}$.}
\label{fig:PosCharLog}
\end{figure}
Figure~\ref{fig:PosCharLog} shows the position characteristic of the log-ratio normalization for the same
cross-section geometry of the BPM pickup as Fig.~\ref{fig:PosCharDOS},
which appears to be more linear compared to the $\Delta /\Sigma$ method.

\subsubsection{BPM cross-section with ``rotated'' electrodes}

\begin{figure}[b!]
\begin{subfigure}{.41\textwidth}
  \centering
  \includegraphics[width=0.8\linewidth]{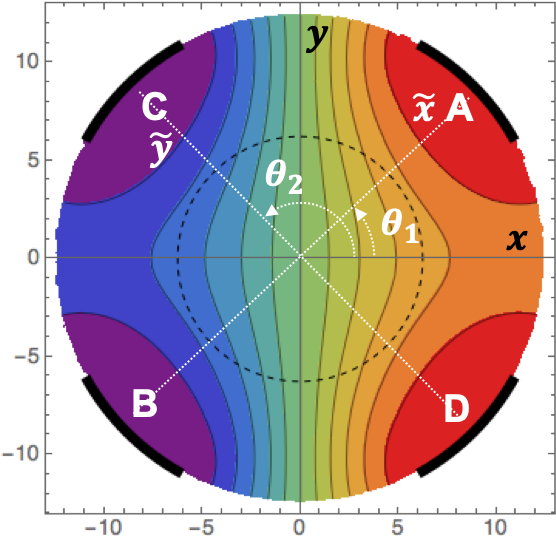}
  \caption{$f(x,y)=\mbox{const } \Delta / \Sigma$ \\ (dashed circle: half aperture)}
  \label{fig:sfig12a}
\end{subfigure}%
\begin{subfigure}{0.59\textwidth}
  \centering
  \vspace*{3mm}
  \includegraphics[width=0.8\linewidth]{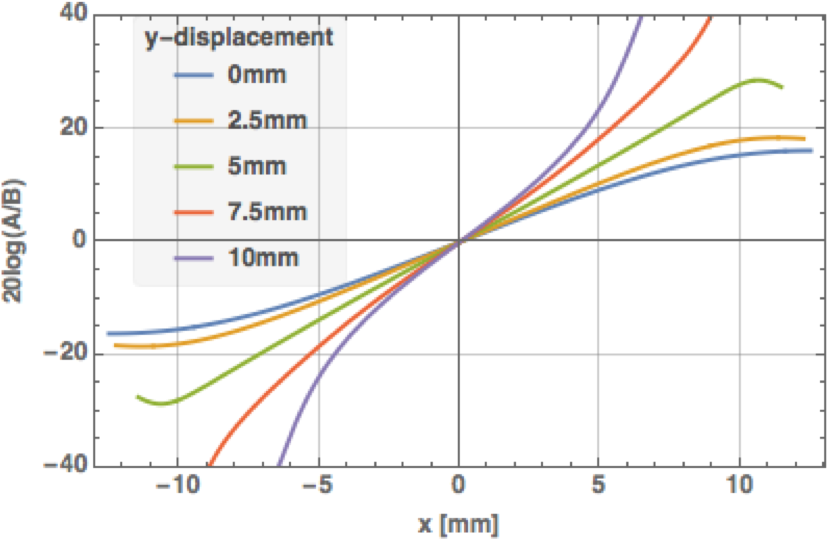}
  \caption{$20\log_{10}(A/B) = f(x)$ for different vertical beam displacements. 
                (BPM sensitivity at the origin: $1.94~dB/mm$))}
  \label{fig:sfig12b}
\end{subfigure}%
\caption{Horizontal position characteristic in a circular beam pipe of radius $R=12.5~mm$, \\
with electrodes rotated by $45^{\circ}$, each covering an arc of $\alpha=30^{\circ}$.}
\label{fig:PosChar45}
\end{figure}

In some cases the BPM electrodes cannot be located along the horizontal and vertical planes, e.g.\ 
in electron storage rings due to the synchrotron light fan which can have unwanted effects on the horizontal
electrodes, or due to collision debris at BPM pickups near the interaction point of a particle collider, 
or just due to real-estate restrictions.
Intense photons, charged primary or secondary particles on the BPM electrodes have to be strictly avoided,
they would alter the signal response in a non-predictable way.

For those situations, the BPM pickup electrodes can be arranged in a different way, but still symmetric to the
horizontal and vertical plane. 
For a simple circular cross-section we can still use Eq.~(\ref{eq:func}), applying:
\begin{align}
x &= \tilde{x}\cos \theta_1 - \tilde{x}\sin \theta_2 \nonumber \\
y &= \tilde{x}\sin \theta_1 + \tilde{x}\cos \theta_2 \nonumber
\end{align}
for the rotated, not necessarily orthogonal coordinates ($\tilde{x}$, $\tilde{y}$) of the BPM electrodes. 
Most popular is an arrangement of all electrodes rotated by $\theta_1=\theta_2=\theta=45^{\circ}$,
see Fig.~\ref{fig:PosChar45}.
We used the same cross-section dimensions ($R=12.5~mm$, $\alpha=30^{\circ}$), but by comparing
Fig.~\ref{fig:sfig10a} with Fig.~\ref{fig:sfig12a} (or Fig.~\ref{fig:sfig11b} with Fig.~\ref{fig:sfig12b})
it is evident that the position non-linearities increase substantially for a BPM with rotated electrodes.

\subsection{Numerical analysis of the position characteristic}

For the general case of an arbitrary shape of the BPM pickup and its electrodes the analytical analysis has limitations,
we just may not find an analytical expression for the position characteristic $s(x,y)$. 
In this case the BPM pickup must by
characterized in a different way, e.g.:
\begin{enumerate}
\item by measurements with a stretched wire, which requires a BPM test-stand and specific lab equipment.
\item by numerical computation, either by solving the Laplace equation in two-dimensions (2D) to find the electrostatic potential, 
or by solving the Maxwell equations in three-dimansions (3D) to find coupled electrode signals through the EM fields excited 
by a transient beam stimulus.
\end{enumerate}
While the stretched wire measurement method is useful as verification of the final prototype, the
numerical computation is already helpful during the construction phase, enabling the optimization of 
shapes towards specific goals and the study of the effect of manufacturing tolerances.  

\subsubsection{Electrostatic analysis in 2D}
 
\begin{figure}[b!]
\hfill
\begin{subfigure}{0.43\textwidth}
  \centering
  \includegraphics[width=0.8\linewidth]{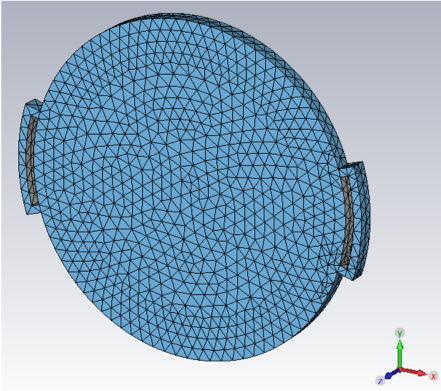}
  \caption{2D ``slice'' of the BPM pickup with a tetrahedral mesh, prepared for the numerical analysis}
  \label{fig:sfig13a}
\end{subfigure}
\hfill
\begin{subfigure}{0.46\textwidth}
  \centering
  \includegraphics[width=0.8\linewidth]{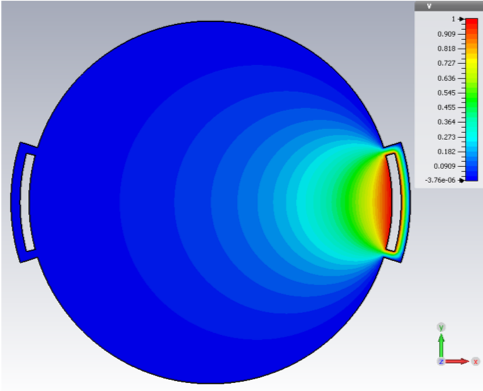}
  \caption{Result of the quasi-2D numerical analysis: equipotentials of electrode $A$.}
  \label{fig:sfig13b}
\end{subfigure}\hspace*{5mm}
\caption{Numerical analysis of a circular BPM pickup of radius $R=12.5~mm$, \\
with a pair of horizontal electrodes, each spanning $\alpha=30^{\circ}$.}
\label{fig:NumAnalysis}
\end{figure} 
For a relativistic beam with sufficiently long bunches compared to the beam pipe aperture ($\sigma_l \gg R$) 
and large values of $\gamma$ ($1/\gamma^2 \ll (\sigma_l/R)^2$), 
the analysis of a BPM electrode configuration can be reduced to a 
simple two-dimansional electrostatic problem:
\begin{equation} \label{eq:simpleesproblem}
	\nabla_{\perp}^{2}\mathbf{\Phi_{elec}(r)} = \frac{\rho}{\phi} \mathbf{\delta(r-r_0)},
\end{equation}
\noindent with the \textit{Fourier} expansion of line charge density $\rho$ and potential $\phi$ 
being a constant factor ($\rho \propto \phi$) of no influence.
Instead of solving Eq.~\ref{eq:simpleesproblem} for many positions $\mathbf{r} \in (x,y)$ 
of the beam equivalent line charge density $\rho$, we apply \textit{Green's reciprocity theorem} 
and solve the \textit{Laplace} equation numerically for one of the BPM electrodes, 
see Fig.~\ref{fig:NumAnalysis}:
\begin{equation} \label{eq:laplaceeq}
	\nabla_{\perp}^{2}\mathbf{\Phi_{elec}(r)} = 0 \to \Phi_{elec}(x,y).
\end{equation}
As of the symmetries, the potentials of the other electrodes are simply found by coordinate rotation or mirroring, 
e.g. $\Phi_{B}(x,y)=\Phi_{A}(-x,y)$, etc. 
We now combine the electrode potentials $\Phi_{A}$ and $\Phi_{B}$, respectively $\Phi_{C}$ and $\Phi_{D}$ 
to compute the normalized H and V potential ratios, which are equivalent to the raw, normalized $\Delta / \Sigma$ 
beam position (see also Eq.~\ref{eq:SumDiff})
\begin{equation} \label{eq:potdiff}
	\hat{\Phi}_x (x,y) = \frac{\Phi_{A}-\Phi_{B}}{\Phi_{A}+\Phi_{B}} \;\;\;\;\text{ and }\;\;\;\; 
	\hat{\Phi}_y (x,y) = \frac{\Phi_{C}-\Phi_{D}}{\Phi_{C}+\Phi_{D}}.
\end{equation} 

\begin{figure}[t!]
\begin{subfigure}{0.3\textwidth}
  \centering
  \includegraphics[width=0.8\linewidth]{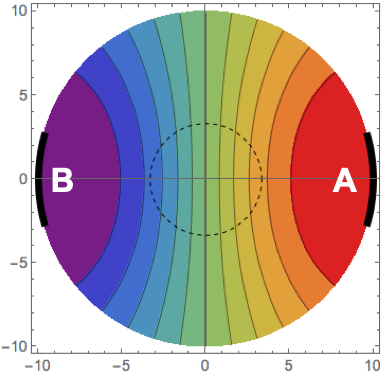}
  \caption{Analytical analysis}
  \label{fig:sfig14a}
\end{subfigure}
\hfill
\begin{subfigure}{0.3\textwidth}
  \centering
  \includegraphics[width=0.9\linewidth]{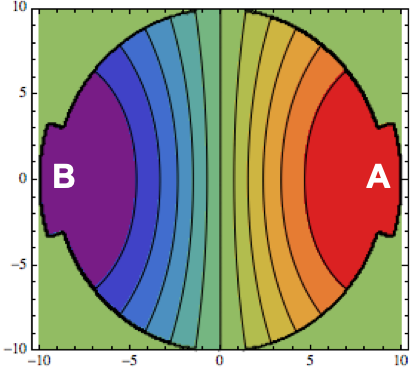}
  \caption{Numerical analysis.}
  \label{fig:sfig14b}
\end{subfigure}
\hfill
\begin{subfigure}{0.3\textwidth}
  \centering
  \includegraphics[width=0.97\linewidth]{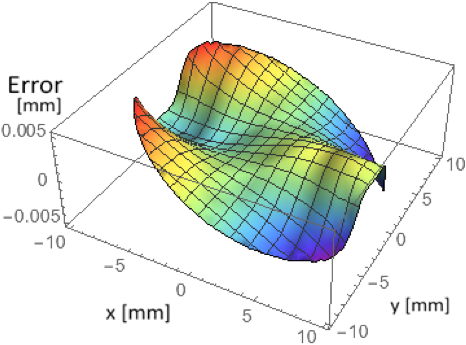}
  \caption{Difference: \\ analytical $-$ numerical.}
  \label{fig:sfig14c}
\end{subfigure}
\caption{Comparison of analytical and numerical analysis of a horizontal BPM pickup.}
\label{fig:compare}
\end{figure}

Figure~\ref{fig:compare} compares the 2D numerical result for an ``ideal'' horizontal BPM, having circular beam pipe of  $R = 10$~mm with electrodes spanning $\alpha = 30^\circ$ with the analytical result. The error reaches up to 5\% at some large beam displacements, but stays within $<1$\% for typical beam positions (see Fig.~\ref{fig:sfig14c}).

This 2D approach basically gives the position sensitivity $s(x,y)$ of an arbitrary BPM cross-section, 
thus cannot distinguish between, e.g.\ button and stripline configuration, and neglects fringe fields and any other 3D effects. 
The advantage lies in a very efficient computation of the BPM position characteristic, the beam-to-electrode coupling, 
and it also can be used to determine the characteristic impedance of a stripline electrode. 

\subsubsection{3D electromagnetic analysis}

Different options exists for a three-dimensional electromagnetic analysis of the BPM pickup,
among other simulation results, they all allow in some way to extract the position characteristic of a BPM pickup
with arbitrary shape of BPM body and electrodes:

\begin{description}
\item[S-parameter] The definition of a set of coaxial ports, one for each BPM electrode, plus two ports for the upstream
and downstream beam pipe enables the extraction of the scattering parameters based on a 3D electromagnetic analysis.
In this numerical simulation a thin metal conductor (``wire'') in $z$-direction, scanned in the $x$-$y$ plane similar to the
stretched-wire measurement method, replaces the beam.
As this coaxial stimulus arrangement excites TEM-fields, the result of this method is equivalent
to that of a relativistic beam ($\beta=1$) as stimulus signal.   
For the position characteristic the S-parameter between the upstream beam port and the pickup electrodes are relevant.
The analysis can be performed either in the frequency-domain, or in time-domain, the results may be valuable
to be compared with a practical, stretched-wire bench measurement setup.
\item[Wakefield] A wakefield simulation calculates the EM fields in the time-domain, 
driven by a line-current signal following a single bunch envelope function (often \textit{Gaussian}), 
moving at a given, constant velocity 
with a specified transverse \mbox{($x$-$y$)} displacement through the specified structure (here: a BPM pickup) 
along the $z$-direction.
While the primary goal of the wakefield simulation is the estimation of the wake-potential, 
which is proportional to the integrated force acting on the trailing
beam particles due to EM interaction with the surrounding structure by the leading particles,
the beam-like longitudinal excitation signal can also be used to study other characteristics of a BPM pickup,
e.g.\ the position characteristic (requires many runs with different beam offsets), the transfer impedance of the BPM electrodes in time- and frequency domain, unwanted beam-excited RF resonances, detailed field analysis, etc.
Therefore, waveguide ports, voltage signal and field monitors have to be added. 
\item[Particle-in-cell (PIC)] The PIC method offers the most flexible time-domain simulation of the response of
a BPM pickup to a beam which is defined in terms of macro-particles, defined by a particle emission model  
As the computational demands and times are substantially higher compared to the above 3D numerical methods,
PIC and other particle tracking methods are rarely used for the characterization of a BPM pick, 
except for e.g.\ for low-$\beta$ beams or other exotic beam conditions. 
\end{description} 

Usually the wakefield is the preferred numerical 3D method to analyze the BPM pickup characteristic, 
Fig.~\ref{fig:BPMpu} gives an example for a simplified button BPM geometry in \textit{CST Studio}.
The black and orange traces indicate the beam trajectory, respectively the wakefield integration path through the BPM
pickup structure, little arrows at the end of each coaxial port indicate the voltage signal monitors.
To analyze the position characteristic, the ``beam'' stimulus has to be scanned in reasonable steps,
typically $<1/10$ of the aperture,
in the $x$-$y$ plane
while recording the voltage signals at the four buttons. 
However, as of the transverse symmetry in this particular case it would be sufficient to scan $1/8$ of the area.
Beside the position characteristic  the wakefield simulation will give other relevant information,
like bunch response function of the BPM electrodes in time and frequency domain transfer impedance), 
wake-potential and beam coupling impedance, etc. 

\subsection{Correction of the position non-linearities of a broadband BPM pickup}

As analyzed in the previous chapter, all types of broadband (non-resonating), 2-dimensional BPM pickups 
with circular or similar beam-pipe cross-section are based on the wall current principle, 
and have a non-linear position behavior, 
regardless of their normalization principle 
(Figures~\ref{fig:PosCharDOS} and \ref{fig:PosCharLog}) or electrode orientation (Fig.~\ref{fig:PosChar45}).
This non-linear position behavior originates in Eq.~(\ref{eq:Ielec}), and is minimum for horizontal / vertical oriented
electrodes spanning an arc of $\alpha \approx 60^\circ$.
But also other two-dimensional cross-section geometries, e.g.\ elliptical or rectangular vacuum chambers with flat or curved
electrodes do have a non-linear position behavior, the only exception is the split-plane -- sometimes called
``shoe-box'' -- type BPM pickup, which consists out of a pair capacitive electrodes which is ``sliced'' in a 
three-dimensional way and discussed later.

However, all the popular button-style, or stripline BPMs have a non-linear position behavior that usually needs to be 
corrected, in particular if large beam displacements have to be monitored.
As the non-linearities of the BPM pickup are known and as the raw 
position data (uncorrected and uncalibrated) is made available in a digital format, 
a correction can be applied to linearize the pickup response.
Two linearization principles exists:
\begin{itemize}
 \item a look-up table
 \item an equation-based linearization based on a 1D or 2D polynomial fit of the normalized position characteristic
\end{itemize}
A look-up table based procedure directly corrects the digitized raw data, 
i.e.\ for each quantized value (raw ADC counts) the table responds with a
corrected value based on the equations or a numerical analysis as discussed in the previous chapter.
E.g., for a digitizer with an 8-bit analog-digital converter (ADC) the BPM pickup signals would be quantized into 256
values, thus requiring a 256-value look-up table.
If each of the four BPM electrodes is digitized individually, it requires four 256-entry look-up tables, each with a correction
following the analytical Eq.~(\ref{eq:Ielec}) or numerical derived position characteristic Eq.~(\ref{eq:laplaceeq}) of
the BPM electrode.

In practice however, things are a bit more complicated.
Any non-linearity, or change in the analog signal processing -- 
which also maps the dynamic signal range of the BPM electrode to the full-scale range of the ADC -- 
needs to be included in the correction for this procedure, thus the look-up table linearization
combines the BPM pickup response with the transfer characteristic of the analog signal conditioning section.
This ``mixing'' makes this concept more complicated in terms of maintenance and housekeeping aspects 
for large scale BPM systems,
and becomes even more complicated if the two opposite electrodes signals are combined before the digitalization.

A polynomial fit f the BPM pickup characteristic allows a clear separation between the pickup non-linearities 
and the following analog signal conditioning section.
The following example explains the procedure for the normalized $\Delta /\Sigma$ position response
on the simple example of a BPM pickup with circular cross-section: 

\subsubsection{Calibration with linear correction}

The linear calibration does not perform any correction of the non-linearities, it just scales the normalized
position characteristic of a pair of BPM electrodes, e.g. for the $\Delta /\Sigma$ normalization:
\begin{equation}
x_{\mathrm{raw}}  = \frac{A-B}{A+B}   \;\;\;\mbox{ or } \;\;\;  y_{\mathrm{raw}}  = \frac{C-D}{C+D}
\end{equation}
with $A$, $B$, $C$, and $D$ being the measured signal levels of the four BPM electrodes, usually in ADC counts.
The derived normalized ``raw'' horizontal and vertical positions range -1\ldots+1, and are unit-less. 
A simple, linear calibration is then:
\begin{equation}
x_{\mathrm{bpm}}  =  k_x x_{\mathrm{raw}} \approx x  \;\;\;\mbox{ or } \;\;\;  
y_{\mathrm{bpm}}  =  k_y y_{\mathrm{raw}} \approx y
\end{equation}
with $k_{x,y}=R/2$ for a BPM with circular cross-section, see also Eq.~\ref{eq:HorPosLin}.
As Figure~\ref{fig:sfig10b} shows, this linear ``fit'' can only be used for small beam displacements,
at large beam displacements it underestimates the true beam position.   

\subsubsection{Calibration with non-linear correction}
To better account for the non-linearities of the normalized position behavior a calibration with higher-order polynomials
can be applied.
Consider a pair of horizontal oriented electrodes as BPM pickup and beam displacements only in the horizontal
plane, $x\neq 0$, $y=0$. 
The function $x_{\mathrm{raw}}=f(x)$ between the true horizontal beam position $x$ 
and the measured value $x_{\mathrm{raw}}$
can be estimated analytical or numerically as described in the previous section.
By its nature we can assume $f$ to be smooth and invertible
\begin{equation}
x=f^{-1}(x_{\mathrm{raw}})
\end{equation}
The coefficients $c_i$ of a one-dimensional polynomial of power $p$ can be calculated to fit $f^{-1}$ and find an
approximate horizontal beam position:
\begin{equation} \label{eq:poly_1D}
x_{\mathrm{bpm}}^{1D} = \sum_{i=0}^{p}c_i x_{\mathrm{raw}}^i = U_p(x_{\mathrm{raw}}) \approx x .
\end{equation}
In case of a double symmetry of the BPM, e.g.\ circular cross-section, the polynomial for of the vertical axis
$y_{\mathrm{bpm}}^{1D}=U_p(y_{\mathrm{raw}})$ will be identical to $x_{\mathrm{bpm}}^{1D}$.
In other cases, e.g.\ elliptical or rectangular BPMs, a different set of polynomial coefficients $c_i$ for
$y_{\mathrm{bpm}}^{1D}\neq x_{\mathrm{bpm}}^{1D}$ has to be computed.
As of the symmetry, all even coefficients $c_{i=0,2,4,\ldots}=0$, and don't have to be considered. 
However, $c_0$ is an offset which comes to play in case of alignment errors or as ``electronics'' offset.
Typically, polynomials of order $p=3\ldots 7$ give a sufficient correction of the non-linearities.  
Figure~\ref{fig:1Dcorr} shows the remaining error of this 1D polynomial calibration procedure for the example of a
CERN LHC stripline BPM with circular cross-section.
The polynomial for was limited to $\mathbb{R}=68\%$ of the cross-section area, which seemed to be a good
compromise between remaining errors in the area of interest ($\thicksim$1/3 of the aperture) 
and reasonable order (here: $p=5$) of the polynomials.
While this calibration with 1D polynomials for the non-linear correction does a good job along the horizontal and
vertical axis, the position errors remain rather high for off-axis beam displacements ($x=r\sin\varphi$, $x=r\sin\varphi$)
for $\varphi\neq n\pi/2$ with $n\in(0,1,2,3)$.
\begin{figure}[t!]
\begin{center}
\includegraphics[width=0.5\textwidth]{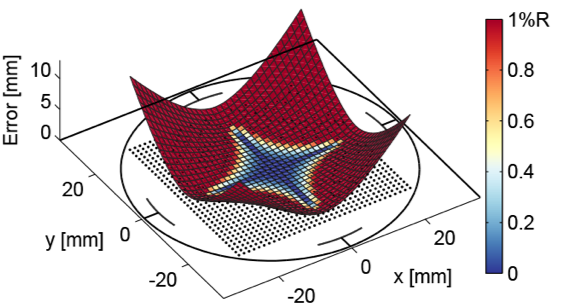}
\caption{Remaining position error after applying 1D correction polynomials $U_5(x)$, $U_5(y)$
for an area of $\mathbb{R}=68\%$ (courtesy \textit{A.\ Nosych}).}
\label{fig:1Dcorr}
\end{center}
\end{figure}

Consider an arbitrary beam offset $x,y \neq 0$ within the BPM aperture. 
The relation between raw and the true beam position can be linked by the functions: 
\begin{equation}
	\begin{cases}
		x = f(x_{\mathrm{raw}},y_{\mathrm{raw}})\\
		y = g(x_{\mathrm{raw}},y_{\textrm{raw}})
	\end{cases}
\end{equation}
where, in case of circular beam pipe and symmetries, $f=g$ and $y = f(y_{\mathrm{raw}},x_{\mathrm{raw}})$ 
(note the variable swap). 
By mapping and fitting $f(x,y)$ by a two-dimensional surface polynomial with $p$ and $q$ being the maximum powers for 
$x$ and $y$ respectively, 
a coupled relationship between the original, true beam position in each plane and its ``raw'' response 
is obtained for both planes:
\begin{equation} \label{eq:poly_2D}
	\begin{cases}
		x_{\mathrm{bpm}}^{\mathrm{2D}} = \sum_{i,j=0}^{p,q}(c_{ij} x_{\mathrm{raw}}^i y_{\mathrm{raw}}^j ) = 
		Q_{p,q}(x_{\mathrm{raw}},y_{\mathrm{raw}}) \approx x \\
		y_{\mathrm{bpm}}^{\mathrm{2D}} = \sum_{i,j=0}^{p,q}(c_{ij} y_{\mathrm{raw}}^i x_{\mathrm{raw}}^j ) = 
		Q_{p,q}(y_{\mathrm{raw}},x_{\mathrm{raw}}) \approx y
	\end{cases}
\end{equation}
\begin{figure}[t!]
\begin{subfigure}{0.3\textwidth}
  \centering
  \includegraphics[width=0.9\linewidth]{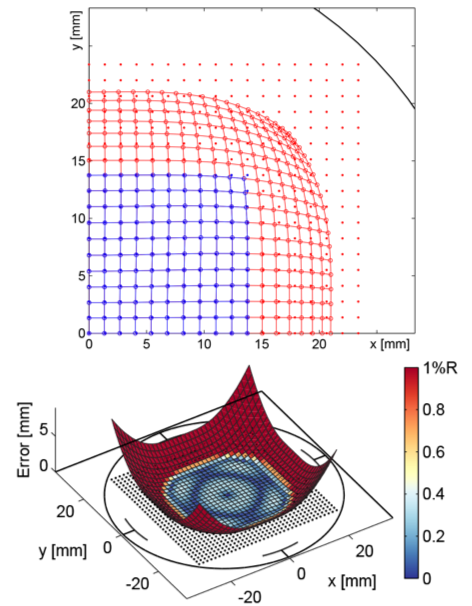}
  \caption{Map and errors for $Q_{3,2}$}
  \label{fig:sfig16a}
\end{subfigure}
\hfill
\begin{subfigure}{0.3\textwidth}
  \centering
  \includegraphics[width=0.9\linewidth]{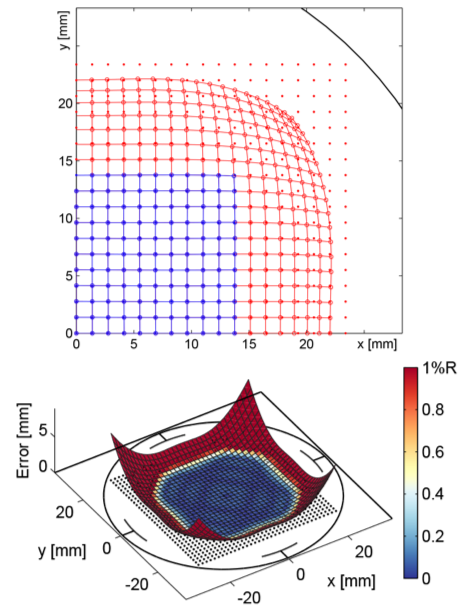}
  \caption{Map and errors for $Q_{5,4}$.}
  \label{fig:sfig16b}
\end{subfigure}
\hfill
\begin{subfigure}{0.3\textwidth}
  \centering
  \includegraphics[width=0.9\linewidth]{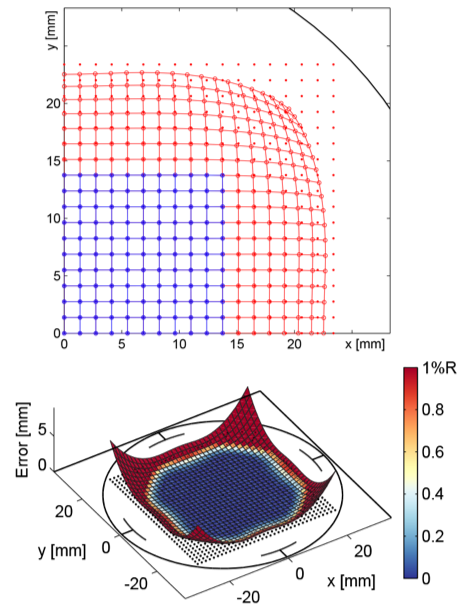}
  \caption{Map and errors for $Q_{7,6}$.}
  \label{fig:sfig16c}
\end{subfigure}
\caption{Pin-cushion effects and remaining errors for a calibration with 2D correction polynomials 
(courtesy \textit{A.\ Nosych}).}
\label{fig:2Dcorr}
\end{figure}
The 2D polynomial fit of the position characteristic returns a matrix of order $p,q$ with the polynomial coefficients
$c_{ij}$.  
In practice not only the even coefficients of the main rows are zero due to the symmetry conditions, 
$c_{i=0,2,4,\ldots ,j} = 0$, also many of the cross-coupling coefficients appear as very small value and can be
neglected, e.g.\ for the $5^{\mathrm{th}}$ order 2D polynomial fit of the LHC stripline BPM
\begin{equation} \label{eq:q54}
	Q_{5,4} \approx Q_{5,5} =
	 \begin{pmatrix}
	  \highlight{c_{10}} & c_{11} & \highlight{c_{12}} & c_{13} & \highlight{c_{14}} & c_{15} \\
	  \cdots & & & & & \\
	  \highlight{c_{30}} & c_{31} & \highlight{c_{32}} & c_{33} & c_{34} & c_{35} \\
	  \cdots & & & & & \\
	  \highlight{c_{50}} & c_{51} & c_{52} & c_{53} & c_{54} & c_{55}
	 \end{pmatrix}
\end{equation}
only the highlighted coefficients need to be considered.
Figure~\ref{fig:2Dcorr} visualizes the results of a 2D polynomial fit applied in the area $\mathbb{R}=40\%$
on the CERN LHC stripline BPM, using the reduced, minimum number of cross-coupling coefficients.

\subsection{Higher-order moments}
\begin{figure}[b!]
\begin{center}
\includegraphics[width=0.4\textwidth]{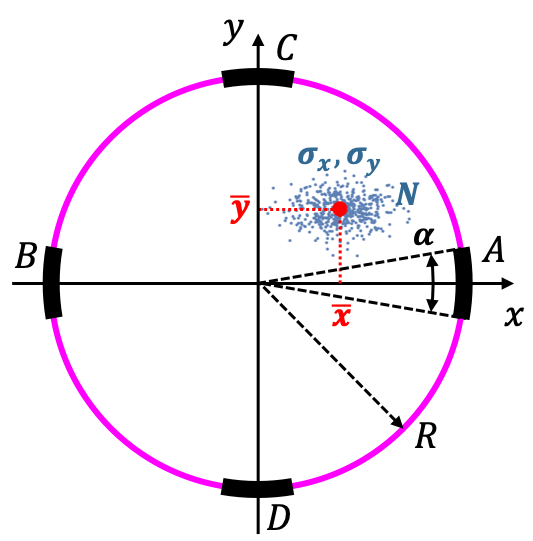}
\caption{A BPM pickup detects the center of charge of a beam of $N$ particles.}
\label{fig:HiOrdMom}
\end{center}
\end{figure}
Until now the charged particle beam was treated as a point charge traveling with constant velocity through the BPM pickup,
or as line charge.
In practice, the beam consists out of many particles, and as the EM field of each particle is linear and time-invariant, 
we can apply the superposition principle to evaluate the effect of many particles passing the BPM pickup. 
Figure~\ref{fig:HiOrdMom} illustrates the concept for our well known, ideal BPM with circular cross-section.
For most practical beam position monitoring applications we can ignore the transverse expansion of the beam,
and replace it by a single point charge of $Q=eN$, where $N$ is the number of particles in the beam and 
$e\approx 1.6\cdot 10^{-19}C$ the elementary charge.
The measured transverse position $\vec{r}=(\bar{x}$, $\bar{y})$ of $Q$ represents the 
center-of-charge of the beam:
\begin{equation}
\vec{r}=\frac{1}{eN}\sum_{i=1}^N q_i \vec{\rho}_i
\label{eq:cof}
\end{equation}
with $\vec{\rho}_i=(x_i, y_i)$ being the transverse position of the $i^{th}$-particle $q_i=e$. 

Strictly speaking Eq.~(\ref{eq:cof}) is only valid as long as the transverse dimensions of the beam are small compared
to the aperture of the BPM pickup, $\sigma_x, \sigma_y\ll R$.
The effect of the beam size on the signal of a BPM electrode can be studied by applying the superposition 
principle on Eq.~(\ref{eq:ElecAppr}), from which for the signal on electrode $A$ in Fig.~\ref{fig:HiOrdMom} follows:
\begin{equation}
I_{\mathrm{A}}=-\frac{I_{beam}}{2\pi} \left [
\alpha + 4 \sum_{i=1}^N \sum_{n=1}^{\infty} \frac{1}{n} \left ( \frac{r_i}{R} \right )^n 
\cos ( n\varphi_i ) \sin \left ( \frac{n\alpha}{2} \right ) \right ]
\label{eq:IelecAsumN} 
\end{equation}
with the position of the $i^{th}$-particle given in cylindrical coordinates ($x_i=r_i \cos\varphi_i$, $y_i=r_i \sin\varphi_i$).
Applying $Q\sum_{i=1}^N x_i^2/N=\sigma_x^2+\bar{x}^2$, respectively $Q\sum_{i=1}^N y_i^2/N=\sigma_y^2+\bar{y}^2$
Eq.~(\ref{eq:IelecAsumN}) can be rewritten in terms of \emph{moments}:
\begin{equation}
I_{\mathrm{A}}=-\frac{I_{beam}}{\pi} \bigg [ \underbrace{\frac{\alpha}{2}}_{\text{monopole}}+
\underbrace{\frac{2}{R}\sin \left (\frac{\alpha}{2}\right ) \bar{x}}_{\text{dipole}}  +
\underbrace{\frac{1}{R^2}\sin ( \alpha ) \left ( \sigma_x^2-\sigma_y^2+\bar{x}^2-\bar{y}^2 \right ) }_{\text{quadrupole}} 
+ \ldots  \bigg ]
\label{eq:mom}
\end{equation}

\begin{description}
\item[Monople moment] The first term in Eq.~(\ref{eq:mom}) is the monopole moment, 
which is only proportional to the beam intensity. 
It often is called the \emph{common} mode, as it appears in all BPM electrodes.
\item[Dipole moment] The second term is of interest for the beam position measurement, it is proportional
to the mean value of the beam displacement (here: $\bar{x}$) times the beam intensity.
\item[Quadrupolar moment] and other higher-order moments can be defined as of the transverse expansion of the beam.
The quadrupolar moment includes information of the transverse beam dimensions $\sigma_x$, $\sigma_y$, and is 
$\propto (\Delta\mathrm{size}+\Delta\mathrm{pos})\times \mathrm{int}$.
\end{description}
The higher-order moments exist because of the non-linear position behavior of a point charge in a BPM with circular cross-section, a BPM pickup with linear position characteristic would contain only monopole and dipolar moments.
The constants between the moments scale with $\sim 1/R$, which in practice is more than an order of magnitude,
thus, the dipolar moment is typically $10\ldots 100$ smaller than the common mode, while the quadrupolar moment
is smaller by a similar amount compared to the dipolar moment, which makes the detection of the quadrupolar moment 
very challenging. 
To extract information about the transverse beam size, two or more BPM pickups need to be located along the beam-line, 
preferable with no magnetic elements in between, at a location with
$\sigma_x\neq\sigma_y$ and where $\bar{x}-\bar{y}=0$, aligned with a very high accuracy $\ll 1/R^2$.
Imperfections of the symmetry of the BPM electrodes have to be in the same oder of magnitude, a BPM with a small
aperture $R$ wrt.\ the beam size dimension is preferable.
To insure $\bar{x}-\bar{y}=0$ the BPM pickups can be mounted on remote controlled translation stages, which allows
to center the beam in the BPM pickups without moving the beam.

\subsection{Bunched beam signals}

For the analysis of the position characteristic of a BPM pickup based the sensitivity function $s(x,y)$ the point charge
approach is sufficient.
However, to estimate the waveform and signal power out of a BPM pickup electrode in Eq.~(\ref{eq:transfer}),
the bunch current signal $i_{\mathrm{bunch}}(t)$ or $I_{\mathrm{bunch}}(\omega)$ needs to be known -- 
as discussed in this section -- 
along with the transfer impedance $Z(\omega)$ of the pickup electrode -- discussed in the next section.

A point-like charge $q=zeN$, with $z$ being the charge state (for ions), $e\approx 1.6\cdot 10^{-19}C$ the elementary
charge and $N$ the number of particles, 
 traveling with relativistic velocity $\beta=v/c_0=1$ in a 
perfectly conducting vacuum chamber (Fig.~\ref{fig:WallCurrent})
has a bunch current:
\begin{equation}
i_{\mathrm{bunch}}(t)=q\cdot\delta (t)= -i_w(t)
\end{equation}
which is cancelled by the wall current $i_w(t)$, originated  by the image charges $q_w=-q$ distributed around
the azimuth of the beam pipe wall.
In some cases, e.g.\ very short bunches, this $\delta$-signal approach for the bunch current is an acceptable
approximation, with the \textit{Dirac} ($\delta$) function defined as $\int_{-\infty}^{+\infty}\delta (t)=1$.
In most other cases the longitudinal distribution of the particles in the RF bucket is approximated by an
analytic function to describe the longitudinal bunch current distribution vs.\ time, e.g.\
\begin{equation}
i_{\mathrm{bunch}}(t)=\frac{zeN}{\sqrt{2\pi}\sigma_t}e^{-\frac{t^2}{2\sigma_t^2}}
\label{eq:gaussTD}
\end{equation}
for a \textit{Gaussian} particle distribution of length $\sigma_t$ (in time), see also Fig.~\ref{fig:sfig18a}.
\begin{figure}[t]
\begin{subfigure}{0.45\textwidth}
  \centering
  \includegraphics[width=0.9\linewidth]{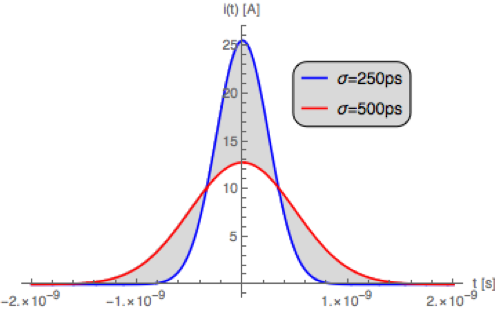}
  \caption{\textit{Gaussian} bunch current.}
  \label{fig:sfig18a}
\end{subfigure}
\hfill
\begin{subfigure}{0.45\textwidth}
  \centering
  \includegraphics[width=0.95\linewidth]{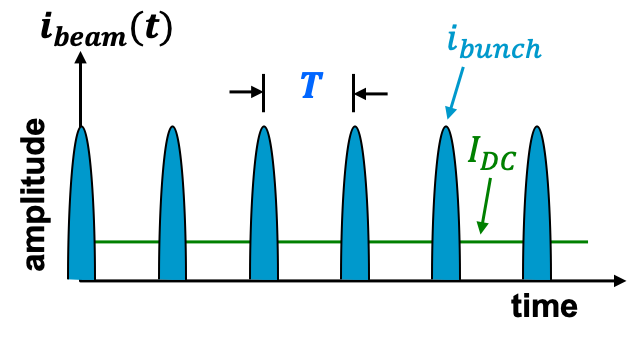}
  \caption{Beam current made out of bunches of same intensity and shape.}
  \label{fig:sfig18b}
\end{subfigure}
\caption{Bunch and beam current signals in the time-domain.}
\label{fig:BeamBunchCurrent}
\end{figure}
In a ring accelerator of circumference $l_c=vT_{\mathrm{rev}}=v/f_{\mathrm{rev}}$, 
RF buckets can be filled with charged particles at equidistance times intervals
\begin{table}[b!]
\begin{center}
\caption{Harmonic amplitude factors for various bunch shapes.}
\label{tab:Am}
\begin{tabular}{ccc}
\hline\hline
\textbf{bunch shape}             & \textbf{harmonic amplitude factor} \boldmath$A_m$
                                                & \textbf{comments}\\
\hline
$\delta$-function   & 1             & for all harmonics\\
\textit{\textbf{Gaussian}}     & \boldmath$\exp \left [\frac{(m\omega\sigma)^2}{2} \right ] $           
 & \boldmath$\sigma=$ \textbf{RMS bunch length} \\
 parabolic & $3 \left ( \frac{\sin\alpha}{\alpha^3}-\frac{\cos\alpha}{\alpha^2}\right )$ 
 & $\alpha=m\pi t_b/T$ \\
(cos)$^2$ & $\frac{\sin(\alpha -2)\frac{\pi}{2}}{(\alpha -2)\pi}+\frac{\sin\frac{\alpha\pi}{2}}{\frac{\alpha\pi}{2}}+
\frac{\cos(\alpha +2)\frac{\pi}{2}}{(\alpha +2)\pi}$ 
 & $\alpha=2m t_b/T$ \\
 triangular & $\frac{2(1-\cos\alpha )}{\alpha^2}$  & $\alpha=m\pi t_b/T$ \\
 square & $\frac{\sin\alpha}{\alpha}$  & $\alpha=m\pi t_b/T$ \\
\hline\hline
\end{tabular}
\end{center}
\end{table}
$$
T=\frac{2\pi}{\omega}=\frac{h}{f_{RF}}
$$
where $h=f_{RF}/f_{\mathrm{rev}}$ is the harmonic number.
The resulting beam current is composed out of charged particle bunches,
in the ideal case of equal intensity and shape,
spaced by an equidistant time $T$, and can be written in terms of a \textit{Fourier} series, see also Fig.~\ref{fig:sfig18b}:
\begin{equation}
i_{\mathrm{beam}}(t)=\langle I_{DC}\rangle + 2 \langle I_{DC}\rangle \sum_{m=1}^{\infty} A_m \cos (m\omega t)
\label{eq:iTDfseries}
\end{equation}
with the average beam current $\langle I_{DC}\rangle=zeN/T$ and frequency harmonics spaced by $\omega = 2\pi f$.
The bunch shape in Eq.~(\ref{eq:iTDfseries}) is defined by a harmonic amplitude factor. 
Table~\ref{tab:Am} lists the harmonic amplitude factor $A_m$ for some typical bunch shape functions,
with $t_b$ being the bunch length (in time) at the base, and a normalization of $A_m\rightarrow 1$ for
$\omega\rightarrow 0$.
\begin{figure}[h!]
\begin{subfigure}{0.32\textwidth}
  \centering
  \includegraphics[width=0.95\linewidth]{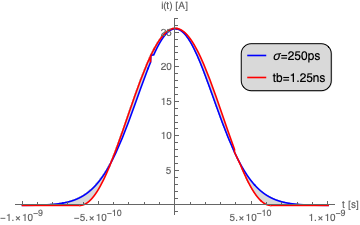}
  \caption{Time-domain.}
  \label{fig:sfig19a}
\end{subfigure}
\hfill
\begin{subfigure}{0.32\textwidth}
  \centering
  \includegraphics[width=0.95\linewidth]{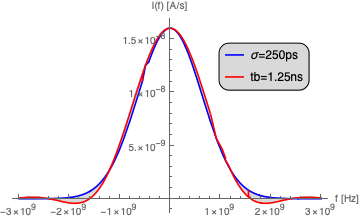}
  \caption{Frequency-domain.}
  \label{fig:sfig19b}
\end{subfigure}
\hfill
\begin{subfigure}{0.32\textwidth}
  \centering
  \includegraphics[width=0.95\linewidth]{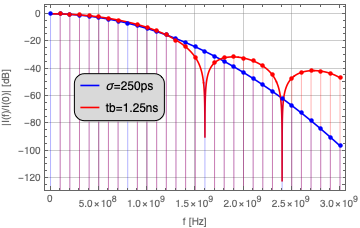}
  \caption{$20 [\log_{10}|I(f)|/\log_{10}|I(0)|]  $.}
  \label{fig:sfig19c}
\end{subfigure}
\caption{Bunch signals with \textit{Gaussian} and cos$^2$ shape in time and frequency domain.}
\label{fig:FourierTransform}
\end{figure}
\begin{equation}
\begin{aligned}
&\text{time-domain}    &\qquad   \text{freq}&\text{uency-domain} \\ 
i_{Gauss}(t)&=\frac{zeN}{\sqrt{2\pi}\sigma_t}e^{-\frac{t^2}{2\sigma_t^2}} &\qquad
I_{Gauss}(f )&=eNe^{-2(\pi f\sigma_t)^2} \\[1em]
i_{cos^2}(t)&=
\smash{\left\{\begin{array}{c@{}c@{}}
\frac{N}{t_b} \left ( 1+\cos\frac{2\pi t}{t_b} \right ), \;\; -t_b/2<t<t_b/2 \\ [\jot] 
0,  \qquad\qquad\text{elsewhere} 
\end{array}\right.}
&\qquad
I_{cos^2}(f)&=\frac{eN\sin\pi f t_b}{\pi f t_b \left [1-(f t_b)^2\right ]} 
\end{aligned}
\label{eq:iftrans}
\end{equation}

For many practical cases however, it is more convenient to use the \textit{Fourier} transformation instead of a
\textit{Fourier} series expansion with infinite sums. 
Figure~\ref{fig:FourierTransform} illustrates the two examples given in Eq.~(\ref{eq:iftrans}) of a bunch shape
following a \textit{Gaussian} and a raised-cosine (cos$^2$) function, having similar parameters.
The differences between the two bunch shapes become more obvious compering them in the frequency-domain,
at higher frequencies on a logarithmic scale (Fig.~\ref{fig:sfig19c}).
For illustration, bunch harmonics spaced by $f=100$~MHz have been indicated in the logarithmic plot, demonstrating
how this concept of the \textit{Fourier} transformation also covers the line spectrum of bunches, e.g.\ in a ring accelerator.

\begin{figure}[h]
\hfill
\begin{subfigure}{0.41\textwidth}
  \centering
  \includegraphics[width=0.99\linewidth]{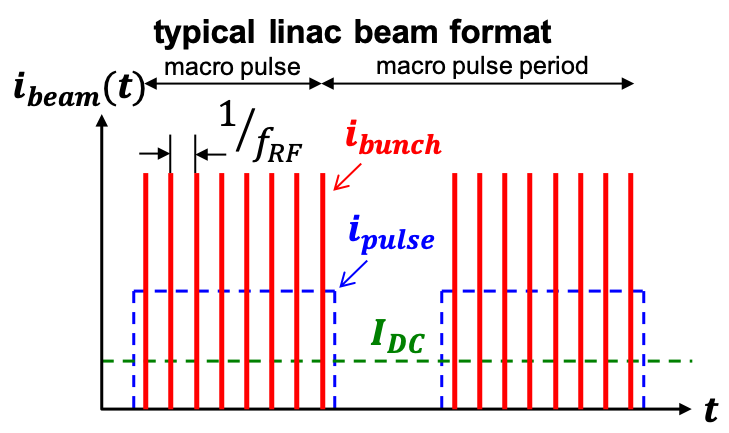}
  \caption{Ideal beam format in a linac.}
  \label{fig:sfig20a}
\end{subfigure}
\hfill
\begin{subfigure}{0.45\textwidth}
  \centering
  \includegraphics[width=0.99\linewidth]{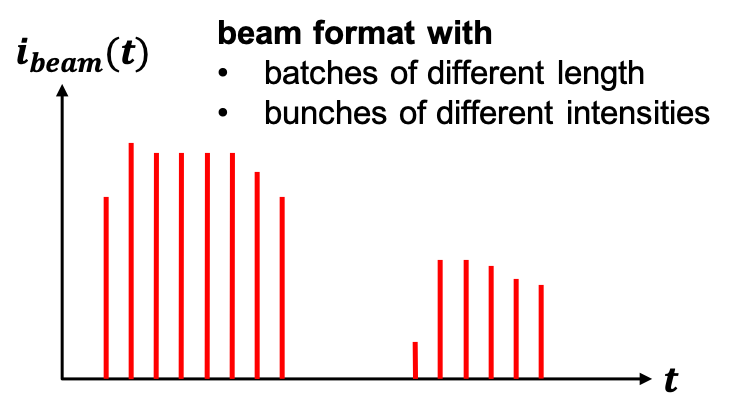}
  \caption{Beam batches with different bunch intensities.}
  \label{fig:sfig20b}
\end{subfigure}
\hspace*{5mm}
\caption{Examples of bunched beam formats.}
\label{fig:beamFormatting}
\end{figure}

However, in practice a beam of equidistant bunches of same intensity rarely exists, instead
beam-free gaps for the injection / extraction or other purposes have to be provided, the beam in a linac is usually pulsed,
and furthermore it is challenging to ensure all bunches in the accelerator to have the same intensity.
Examples for different beam formats are shown in Fig.~\ref{fig:beamFormatting}, which will cause the related
beam spectrum to be more complicated.
However, for the evaluation of the response of a BPM pickup, characterized by the transfer impedance $Z(\omega)$ or $z(t)$,
a single bunch excitation signal $I_{\mathrm{bunch}}(\omega)$ or  $i_{\mathrm{bunch}}(t)$ with the
bunch-length of relevance is sufficient.
As the BPM pickup is a linear, time-invariant system the
response to other beam formats can be simply evaluated by superposition of the response signal $v_{\mathrm{elec}}(t)$
to a single bunch with the appropriate time-delays and intensity factors.

\subsection{Examples of BPM pickups}

\subsubsection{The ``button'' BPM}
\label{sec:button}
\begin{figure}[ht]
\begin{subfigure}{0.28\textwidth}
  \centering
  \includegraphics[width=0.95\linewidth]{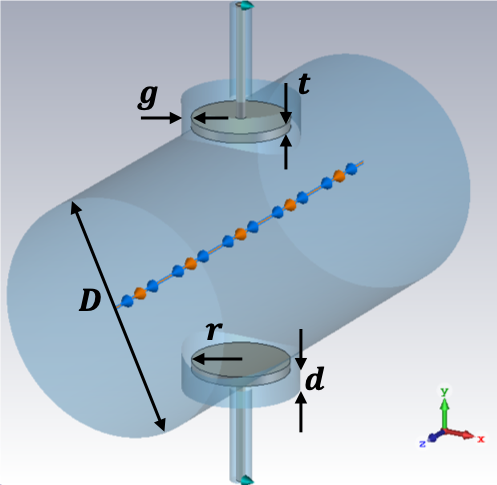}
  \caption{Button BPM \\ (for the vertical plane).}
  \label{fig:sfig21a}
\end{subfigure}
\hfill
\begin{subfigure}{0.4\textwidth}
  \centering
  \includegraphics[width=0.91\linewidth]{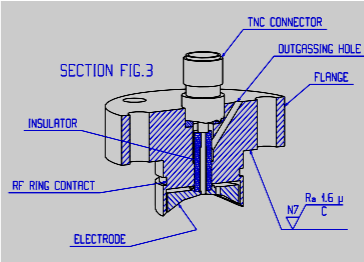}
  \caption{Cross-section view of a CERN button feedthrough.}
  \label{fig:sfig21b}
\end{subfigure}
\hfill
\begin{subfigure}{0.28\textwidth}
  \centering
  \includegraphics[width=0.88\linewidth]{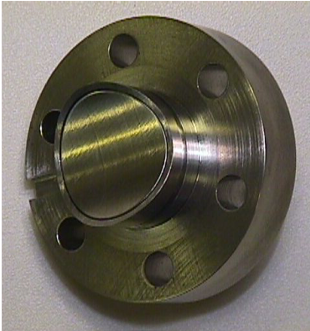}
  \caption{Flange mount LHC button BPM vacuum feedthrough.}
  \label{fig:sfig21c}
\end{subfigure}
\caption{The button BPM pickup.}
\label{fig:buttonPU}
\end{figure}
The button-style BPM is the most popular type of BPM pickup because of its simplicity, robustness and compact design,
and has well defined, reproducible characteristics at relatively moderate costs. 
In most cases it consists out of four symmetrically arranged circular electrodes, the ``buttons'', each
located at the end of a coaxial transmission-line connector of usually $Z_0=50\;\Omega$ characteristic impedance,
similar as illustrated in Fig.~\ref{fig:BPMpu}.
In practice, button electrode, coaxial connector and vacuum feedthrough barrier are integrated in a single element,
the button BPM feedthrough, as shown in Fig.~\ref{fig:sfig21b} and Fig.~\ref{fig:sfig21c}.
Different styles are available, often custom designs, for flange mounting or welding, with shape and dimensions
optimized for the application.
Figure~\ref{fig:sfig21a} shows a pair of vertical oriented button electrodes with the variables 
for the most relevant mechanical dimensions defined.
To avoid confusion with the electrical symbols, we define $D$ as the diameter of the 
beam-pipe (assuming a circular cross-section), $r$ as the radius of the metallic button electrode, $t$ being the thickness
of the button, $g$ as the width of the gap between the rim of the button electrode
 and the beam-pipe, and $d$ as the distance between button electrode and beam-pipe.}
 
\begin{figure}[ht]
\begin{center}
\includegraphics[width=0.7\textwidth]{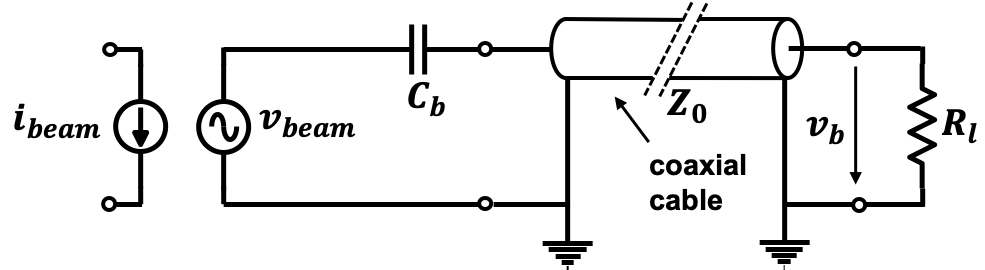}
\caption{Equivalent circuit of a button electrode.}
\label{fig:buttonEquiCircuit}
\end{center}
\end{figure}
\begin{figure}[t!]
\begin{center}
\includegraphics[width=0.75\textwidth]{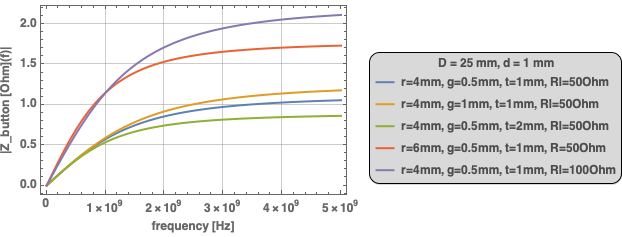}
\caption{Transfer impedance of a button BPM electrode.}
\label{fig:buttonZ}
\end{center}
\end{figure}

Electrically, the button electrode couples capacitively to the beam, it is dominated by its capacitance $C_b$ 
to ground (the beam-pipe), and forms a high-pass filter with the load impedance $R_l$.
Figure~\ref{fig:buttonEquiCircuit} shows the equivalent circuit of a button electrode, where
\begin{equation}
v_{\mathrm{beam}}(t)\simeq
-\frac{1}{C_b} \overbrace{i_{\mathrm{beam}}(t)\underbrace{\frac{r}{2D}}_{\phi}}^{\Delta i_{\mathrm{beam}}(t)}
\underbrace{\frac{2r}{c_0}}_{\Delta t}
=-i_{\mathrm{beam}}(t) \frac{r^2}{Dc_o C_b}
\label{eq:vbeam}
\end{equation}
is the beam related generator voltage at the button, no load impedance assumed. 
This induced voltage is proportional to the interception area of the button, 
given by the coverage factor $\phi\approx r/(2D)$ (see also Eq.~(\ref{eq:cover})),
the transit time $\Delta t= 2r/c_0$, and inverse proportional to the button capacitance $C_b$.
Eq.~(\ref{eq:vbeam}) holds for relativistic beams $v\simeq c_0$ and a 
beam current signal $i_{\mathrm{beam}}(t)$ that resembles a bunch length longer
then the button diameter $\sigma_s \gg 2r$.
For a centered beam ($x=y=0$) the transfer impedance of a button electrode follows as:
\begin{equation}
Z_b(\omega )=\frac{V_b(\omega )}{I_{\mathrm{beam}}(\omega )}=\phi R_l \frac{\omega_1}{\omega_2}
\frac{j\omega /\omega_1}{1+\omega /\omega_1}
\label{eq:Zbutton}
\end{equation}
with:
$$
\text{1/time constant: } \omega_1=\frac{1}{R_l C_b}\; , \qquad
\text{1/transit time: } \omega_2=\frac{c_0}{2r}\; , \qquad
\text{coverage factor: } \phi\approx \frac{r}{2D}
$$
The button-to-ground capacitance can be estimated as
$$
C_b \approx \frac{\pi\epsilon t_b}{\ln \left ( \frac{r+g}{r} \right) }  + \frac{\pi\epsilon r^2 }{d} + C_{\mathrm{fringe}}
$$%
but in practice a measurement or a numerical computation of $C_b$ will give better, more accurate results,
the same is true for the coverage factor $\phi$.
Figure~\ref{fig:buttonZ} show some examples of $Z_b(\omega)$ for different dimensions of the button ($r$, $t$) 
and the gap ($g$), as well as for a variation of the load impedance ($R_l$). 
Typical values of the transfer impedance for high frequencies are around $Z\approx 1\;\Omega\;@\; f\gtrsim 1$~GHz,
the 3~dB cutoff-frequency is rather high, typically $f_{3dB} \approx 500\ldots 2000$~MHz.
A lower cutoff frequency and increase of $Z_b$ could be achieved by increasing the load impedance $R_l$, however,
usually $R_l$ is fixed to 50~$\Omega$ to stay compatible with the impedance standard in RF technology.
Increasing the diameter $2r$ of the button will increase $Z_b$ and result in a higher coupling (sensitivity) to the beam,
and -- as of the increase of $C_b$ -- will also lower $f_{3dB}$,
but simultaneously will lower the position sensitivity.
Moreover, a larger button size will also result in lowering the frequencies of trapped eigenmodes, which
can lead to an increase of unwanted wake field effects.
An increase of the button thickness can reduce the beam coupling (wake) impedance, but increases the
weight of the button, which has to be handled by the thin pin / vacuum feedthrough construction.
Lowering the gap is even more beneficial to reduce the beam coupling impedance,  
the limit is set by manufacturing tolerances.

\subsubsection{The stripline BPM}

\begin{figure}[ht]
\begin{subfigure}{0.54\textwidth}
  \centering
  \includegraphics[width=0.99\linewidth]{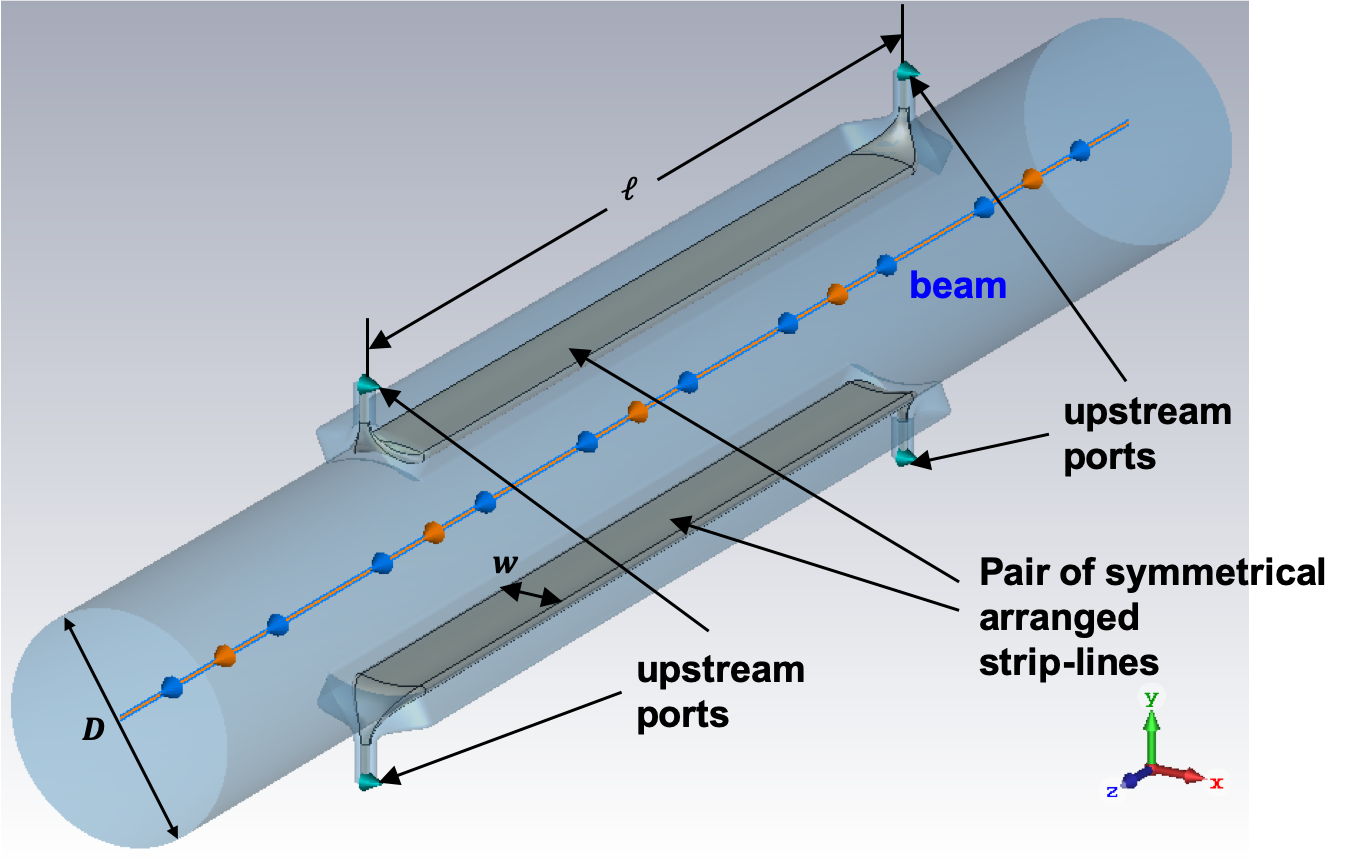}
  \caption{Stripline BPM (for the vertical plane).}
  \label{fig:sfig24a}
\end{subfigure}
\hfill
\begin{subfigure}{0.45\textwidth}
  \centering
  \includegraphics[width=0.99\linewidth]{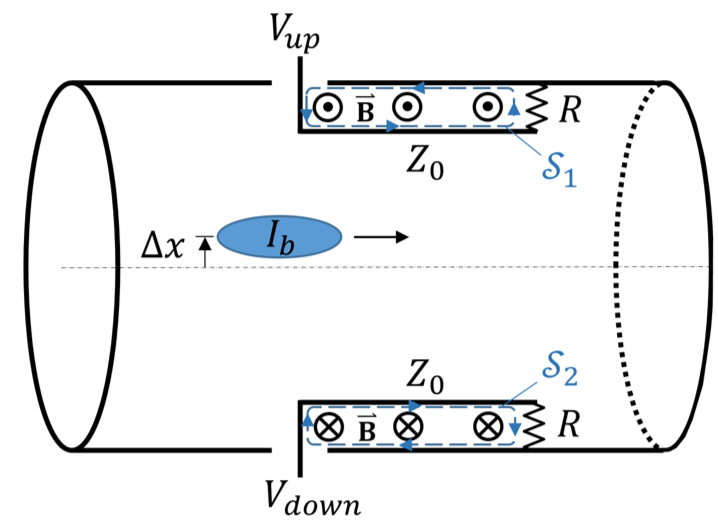}
  \caption{Coupling to the magnetic field components of the beam for a pair of stripline electrodes.}
  \label{fig:sfig24b}
\end{subfigure}
\caption{The stripline BPM pickup.}
\label{fig:striplinePU}
\end{figure}
Similar to the button BPM, the stripline BPM is a broadband coupler.
Two or four stripline-like electrodes of length $\ell$ and width $w$ are arranged symmetrically 
in the beam-pipe and form the stripline BPM pickup, see Fig.~\ref{fig:sfig24a}.
Each stripline is a TEM transmission-line of characteristic impedance $Z_0$, usually matched to the
50~$\Omega$ impedance of the coaxial feedthrough ports, located at both ends.
Unlike a button, which couples capacitively to the beam, the stripline electrode is an electro-magnetic
coupler, Fig.~\ref{fig:sfig24b} illustrates how stripline and beam-pipe form a ``loop'' that also couples to the
magnetic field components of the beam.

\begin{figure}[t!]
\begin{center}
\includegraphics[width=0.95\textwidth]{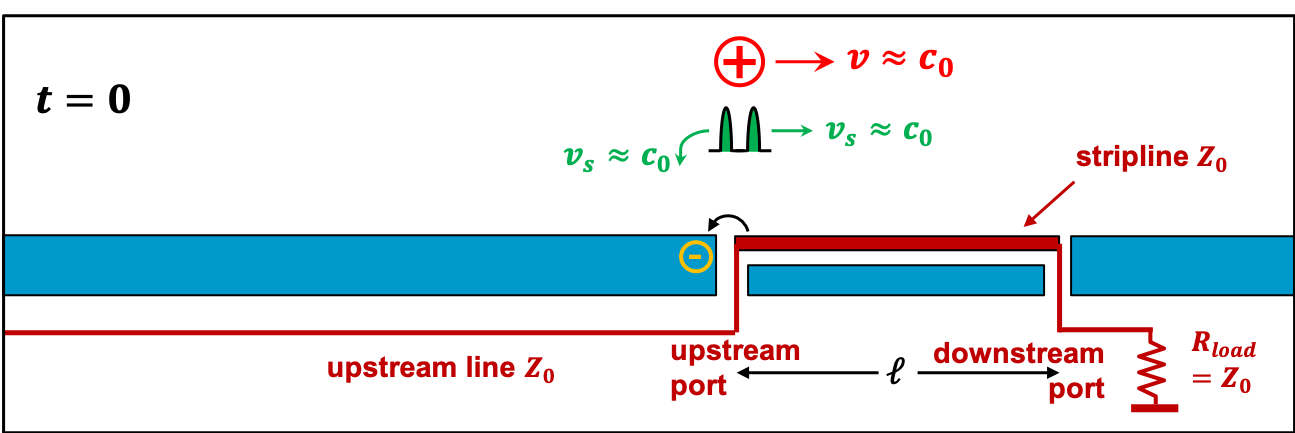} \\
\vspace*{3mm}
\includegraphics[width=0.95\textwidth]{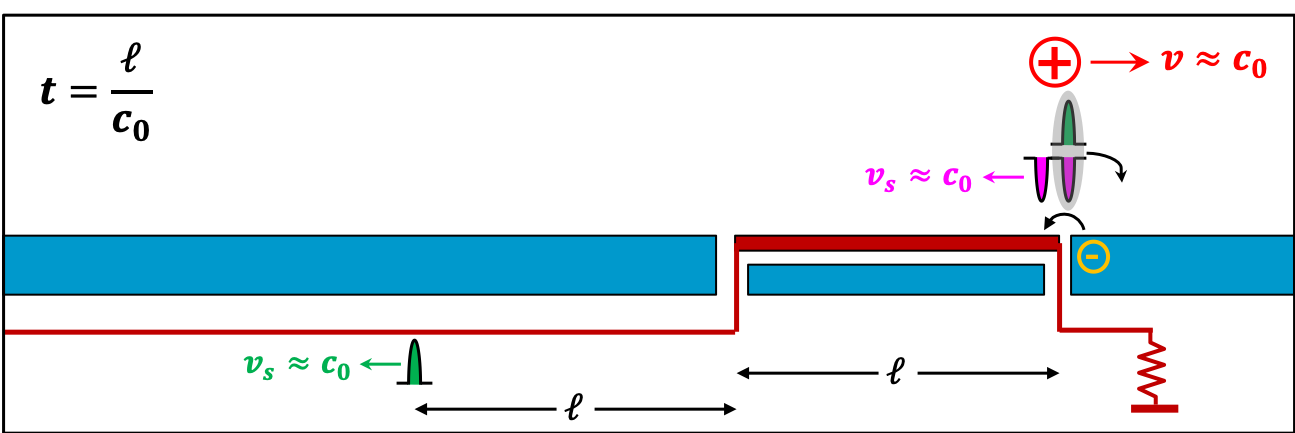} \\
\vspace*{3mm}
\includegraphics[width=0.95\textwidth]{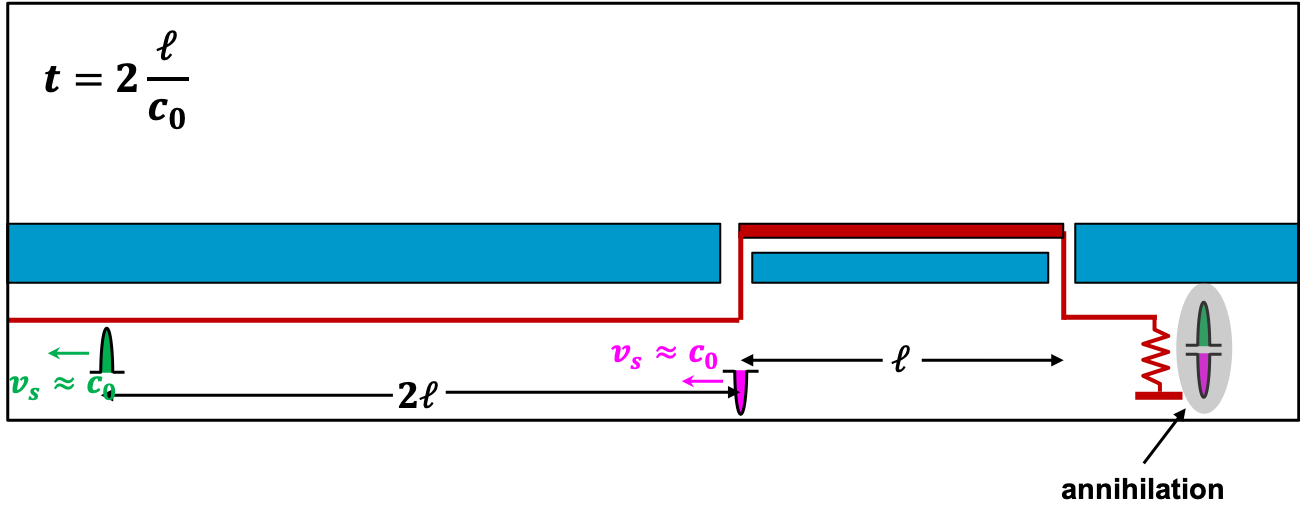}
\includegraphics[width=0.45\textwidth]{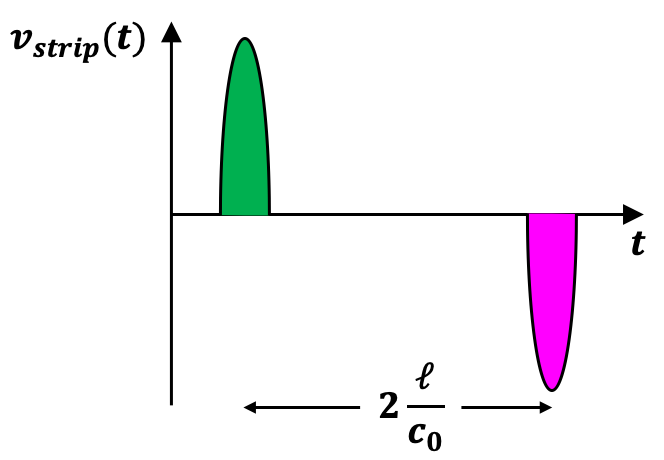}
\caption{Operational principle of a stripline BPM electrode.}
\label{fig:striplinePrinciple}
\end{center}
\end{figure}

Electrically, stripline electrode and beam form a 4-port network, similar to that of a directional coupler,
but in case of a stripline BPM, two of the ports are beam signal related waveguide ports.
An intuitive  explanation of the principle of operation is illustrated in the time-domain
in Fig.~\ref{fig:striplinePrinciple}, assuming the signal velocity along
the stripline and the velocity of a point charge being the same, close to the speed-of-light, $v_s=v\simeq c_0$.

\begin{description}
\item[$\mathbf{t=0}$]
 A positive charged beam particle enters the upstream portion of the stripline electrode.
 The corresponding image charges (wall current) induces a point charge on the upstream end of the stripline, 
 causing a positive, pulse-like 
 signal with an intensity that depends on
 the distance (beam position) between charged particle and electrode, the width $w$ of
 the electrode, and the intensity $q$ of the beam particle.
 As the induced image charge ``sees'' a transmission-line of same characteristic impedance $Z_0$ in both directions --
 upstream through the vacuum feedthrough as coaxial transmission-line and downstream as stripline electrode --
 it splits in two parts of equal amplitude, traveling with $v_s\simeq c_0$ in opposite directions. 
\item[$\mathbf{t=\ell / c_0}$]
The beam particle exits the downstream portion of the stripline electrode.
Likewise to $t=0$, an image charge is induced on the downstream end of the stripline, now with opposite sign,
causing a negative, pulse-like signal which also splits off into two equal parts traveling in opposite directions.
At the same time the positive image charge signal that is linked to the beam charge 
arrives at the downstream end of the stripline,
its signal has the same intensity but opposite sign as the freshly induced signal portion.
Both signal portions compensate each other and no signal appears at the termination resistor outside 
at the downstream port. 
\item[$\mathbf{t=2\ell /c_0}$]
The negative signal from the downstream induction exits the upstream port, following the positive signal pulse
at a distance of $2\ell$, which will generate a ``doublet''-pulse  voltage signal with a time
delay between positive and negative parts of $\Delta t=2\ell / c_0$, which is then
observable in a load resistor $R_l=Z_0$ (not shown).
\end{description}
As of the symmetry, a beam charge entering from the opposite end will cause the same signal pattern in 
the other port:\newline
The stripline BPM has \emph{directivity}, in the ideal case the beam signal appears only at the upstream port.
The time-domain transfer impedance of an ideal, lossless stripline electrode is given as a $\delta$-doublet signal
(see also Fig.~\ref{fig:sfig26a}):
\begin{figure}[ht]
\begin{subfigure}{0.45\textwidth}
  \centering
  \includegraphics[width=0.95\linewidth]{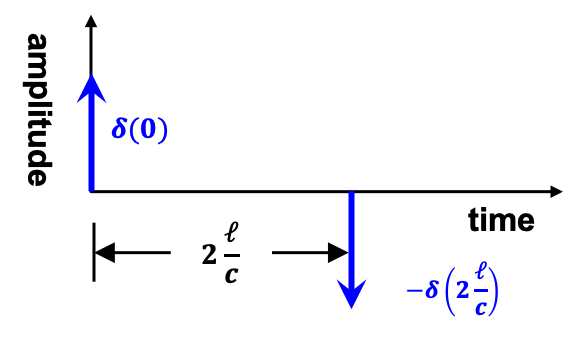}
  \caption{Time-domain.}
  \label{fig:sfig26a}
\end{subfigure}
\hfill
\begin{subfigure}{0.45\textwidth}
  \centering
  \includegraphics[width=0.95\linewidth]{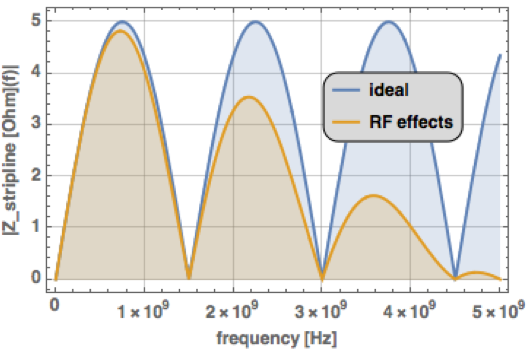}
  \caption{Frequency-domain.}
  \label{fig:sfig26b}
\end{subfigure}
\caption{Transfer characteristic of a stripline BPM electrode.}
\label{fig:striplineTDFD}
\end{figure}
\begin{equation}
z(t)=\phi\frac{Z_0}{2} \left [ \delta(t)-\delta \left ( t-2\frac{\ell}{c_0} \right) \right]
\label{eq:striplineTD}
\end{equation}
with $\phi=\frac{1}{\pi}\arcsin (w/D)$ being the geometric coverage factor and $Z_0$ the characteristic impedance of the stripline.
The frequency-domain transfer impedance of an ideal stripline electrode is then the
\textit{Fourier} transformation of Eq.~(\ref{eq:striplineTD}):
\begin{equation}
Z(\omega)
=\phi\frac{Z_0}{2} \left ( 1-e^{-j2\omega\frac{\ell}{c_0}} \right )
=j\phi Z_0 e^{-j\omega\frac{\ell}{c_0}}\sin \left ( \omega\frac{\ell}{c_0} \right )
\label{eq:striplineFD}
\end{equation}
The modulus of Eq.~(\ref{eq:striplineFD}), $|Z(f)|$ is visualized in Fig.~\ref{fig:sfig26b} for $\ell=100$~mm, 
$\phi=1$, and $Z_0=50\;\Omega$ as parameters of the stripline electrode.
 The lobes of $|Z(\omega)|$ peak at
$f_c=(2n-1)c_0/(4\ell)$,
each has a bandwidth $f_{BW}=f_{hi}-f_{lo}=f_{c1}$ with $f_{lo}=f_{c1}/2$, $f_{hi}=3f_{lo}$, and $f_{c1}=c_0/(4\ell)$.
In most applications the length $\ell$ of the stripline is chosen to operate within the first lobe.
\begin{figure}[b!]
\begin{center}
\includegraphics[width=0.75\textwidth]{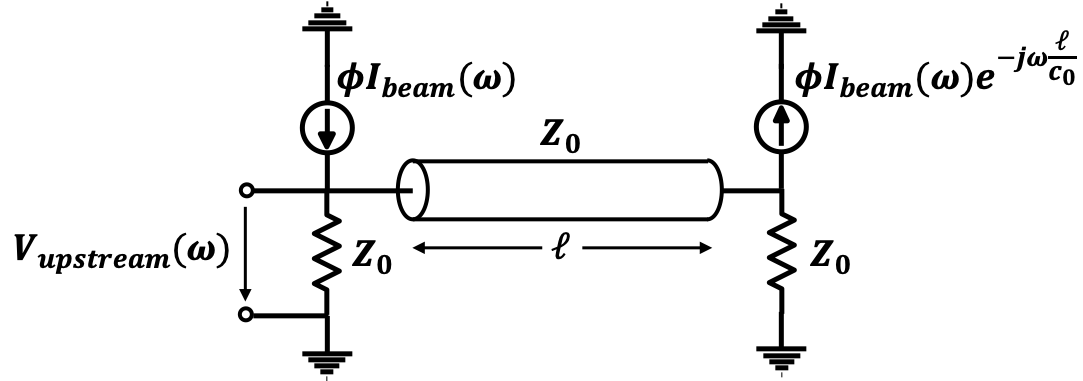}
\caption{Equivalent circuit of a stripline electrode.}
\label{fig:equivStrip}
\end{center}
\end{figure}
Figure~\ref{fig:equivStrip} shows the equivalent circuit for a stripline BPM electrode.

If the stripline electrodes are rather wide, their coupling between each other cannot be neglected.
In that case, and depending if the BPM has two (single-plane) or four (dual-plane) electrodes the characteristic
line impedance $Z_0$ needs to be replace by $Z_0=\sqrt{Z_{0e} Z_{0o}}$ for two electrodes, or by
$Z_0=\sqrt{Z_{0s} Z_{0q}}$ for four electrodes.
The concept of odd and even mode impedances $Z_{0e}$, $Z_{0o}$ for two coupled TEM transmission-lines is well
known, it just requires a simple electrostatic analysis of the cross-section geometry, similar to that of a single electrode.
In case of four electrodes, $Z_{0s}$ represents the sum mode, equivalent to the even mode of two electrodes but
now for all four electrodes,
and $Z_{0q}$ is a quadrupolar mode with always two opposite electrodes in even mode but one pair in odd mode
wrt.\ the other pair.  

To ensure the annihilation of the signals at the downstream port, the signal velocity on the stripline $v=\beta_s c_0$ and
the velocity of the beam $v=\beta c_0$ must be identical, $v_s=v$.
If the strip-electrode includes dielectric material $e_r>1$, e.g.\ ceramic supports, $v_s\neq v$,
the directivity is broken, and some parts of the upstream beam signal will also appear at the downstream port,
therefore:
\begin{align}
z_{\mathrm{downstream}}(t)&=
\phi\frac{Z_0}{2} \left [ \delta\left(t-\frac{\ell}{\beta_s c_0}\right)-\delta \left ( t-\frac{\ell}{\beta c_0} \right) \right]  \\
\intertext{while the upstream port couples:}
z_{\mathrm{upstream}}(t)&=
\phi\frac{Z_0}{2} \left [ \delta(t)-\delta \left ( t-\frac{\ell}{\beta_s c_0}-\frac{\ell}{\beta c_0} \right) \right] 
\label{eq:stripTD}
\end{align}
The frequency-domain equivalent of the above equations is then:
\begin{align*}
Z_{\mathrm{downstream}}(\omega) &
=\phi\frac{Z_0}{2} \left [ e^{-j\omega\frac{\ell}{\beta_s c_0}}-e^{-j\omega\frac{\ell}{\beta c_0}} \right ]
=-j\phi Z_0 e^{-j\frac{\omega}{2}\frac{\ell}{\beta_s c_0}} e^{-j\frac{\omega}{2}\frac{\ell}{\beta c_0})}
\sin \left [ \frac{\omega}{2}\left(\frac{\ell}{\beta_s c_0}-\frac{\ell}{\beta c_0}\right) \right ] \\
Z_{\mathrm{upstream}}(\omega) &
=\phi\frac{Z_0}{2} \left [ 1-e^{-j\omega\left(\frac{\ell}{\beta_s c_0}+\frac{\ell}{\beta c_0}\right)} \right ]
=j\phi Z_0 e^{-j\frac{\omega}{2}\left(\frac{\ell}{\beta_s c_0}+\frac{\ell}{\beta c_0}\right)}
\sin \left [ \frac{\omega}{2}\left(\frac{\ell}{\beta_s c_0}+\frac{\ell}{\beta c_0}\right) \right ]
\end{align*}

The stripline BPM is a TEM coupler, in the ideal case electric and magnetic fields have only transverse components,
and the coupler operates for a beam of relativistic velocity.
In practice, due of the transition from a coaxial-line to the stripline, in and near the vacuum feedthroughs at both ports,
 those discontinuities
``disturb'' the TEM field propagation, causing a perturbation of the characteristic impedance and therefore signal reflections,
which degrade the performance of the BPM, in particular limits its directivity at higher frequencies.
In a very simplistic approach the width $w$ of the stripline causes a reflection near both ports with a 
time delay $\Delta t=w / c_0$, which will alter the frequency-domain transfer characteristic: 
\begin{equation}
Z_{RF}(\omega)
=j\phi Z_0 e^{-j\omega\left(\frac{\ell}{c_0}+\Delta t\right)} \frac{\sin(\omega \Delta t)}{\omega \Delta t} 
\sin \left ( \omega\frac{\ell}{c_0} \right )
\label{eq:striplineFDRF}
\end{equation}
This RF effect of Eq.~(\ref{eq:striplineFDRF}) is visualized in Fig.~\ref{fig:sfig26b} for $\Delta t=100$~ps, 
i.e.\ stripline width $w=30$~mm, 
in comparison
with the ideal response (\ref{eq:striplineFD}).
As the stripline basically is a lossless coupler, the difference between $Z(\omega)$ and $Z_{RF}(\omega)$
will appear as unwanted signal at the downstream port and limit its directivity.
The TEM operation of the stripline BPM, as basis of the theory presented here, can only be ensured for $\ell\gg w$.
If $\ell\approx w$ the stripline converges into an electrostatic or button electrode, and the TEM-based theory becomes
invalid.

Compared to a button BPM, the stripline BPM is a more complicated, fragile and costly pickup, it also requires more
real estate.
Still, the stripline offers several advantages, e.g.:
\begin{itemize}
\item The matched line-impedance terminated at the downstream port acts as almost perfect 50~$\Omega$ signal source,
and minimizes reflections between the BPM pickup and the read-out electronics.
\item In case of hadron beams, the length of the stripline can be tuned to match the bunch length, thus improving
the signal levels, therefore the signal-to-noise ratio (S/N) and the resolution.
\item The downstream port can be utilized to feed test or calibration signals.
\item A stripline BPM with good directivity can be used in particle colliders near the interaction point (IP) to
provide separate beam position signals for the counterrotating beams. 
\end{itemize}

Stripline BPM are also popular in hadron linacs, often with downstream ports short circuited. 
This safes 2 or 4 vacuum feedthroughs and simplifies the mechanical construction, while stiffening the stripline
mechanics.

\subsubsection{The split-plane BPM}
\begin{figure}[ht]
\begin{center}
\includegraphics[width=0.3\textwidth]{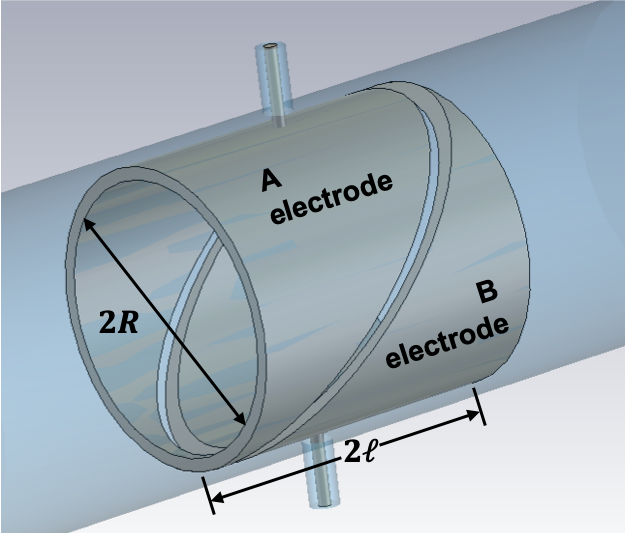}
\caption{The split-plane BPM.}
\label{fig:splitPlane}
\end{center}
\end{figure}
Consider a hollow metallic cylinder of length $2\ell$ and diameter $2R$ floating in the beam-pipe as shown in 
Fig.~\ref{fig:splitPlane}.
The cylinder has a diagonal cut, which forms two electrodes $A$ and $B$, 
each connected to a coaxial vacuum feedthrough to deliver the corresponding beam signal.
Based of Eq.~(\ref{eq:WCclosed}), a beam current $I_{\mathrm{beam}}(r,\varphi)$ induces 
a wall current $I_{\mathrm{elec}}$ on the split-plane electrode:
\begin{equation}
I_{\mathrm{elec}} =
-\frac{I_{\mathrm{beam}}}{2\pi} \frac{\ell}{R} \int_0^{2\pi} 
\frac{(1+\cos\Phi)\left(R^2-r^2\right)}{R^2+r^2-2Rr\cos(\Phi-\varphi)} d\Phi
\label{eq:splitInt}  
\end{equation}
for a beam position $x=r\cos\varphi$, $y=r\sin\varphi$, where $\ell(1+\cos\Phi)$ describes the
variation of the electrode length wrt.\ the integration path $d\Phi$.
The solution of Eq.~(\ref{eq:splitInt}) results in:
\begin{equation}
I_{\mathrm{elec}} =
-\frac{I_{\mathrm{beam}}}{2\pi} \frac{\ell}{R} \left(1+\frac{r\cos\varphi}{R}\right)
=-\frac{I_{\mathrm{beam}}}{2\pi} \frac{\ell}{R} \left(1+\frac{x}{R}\right)
\label{eq:splitPlane}
\end{equation} 
offering a perfectly linear position characteristic.
The linear position behavior is reflected in the normalization of the two horizontal electrodes $A$ and $B$:
\begin{equation}
\frac{\Delta}{\Sigma}=\frac{A-B}{A+B}=\frac{1}{R}x=\mathrm{hor.\; position}
\label{eq:splitPlaneDOS}
\end{equation}
The position sensitivity (\ref{eq:splitPlaneDOS}) of the split-plane BPM is half of the sensitivity of
two-dimensional BPM electrodes, e.g.\ buttons or striplines, approximated by Eq.~\ref{eq:HorPosLin}.

\begin{figure}[b!]
\hfill
\begin{subfigure}{0.4\textwidth}
  \centering
  \includegraphics[width=0.9\linewidth]{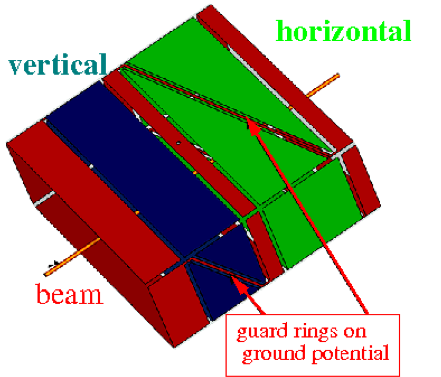}
  \caption{BPM layout.}
  \label{fig:sfig29a}
\end{subfigure}
\hfill
\begin{subfigure}{0.4\textwidth}
  \centering
  \includegraphics[width=0.9\linewidth]{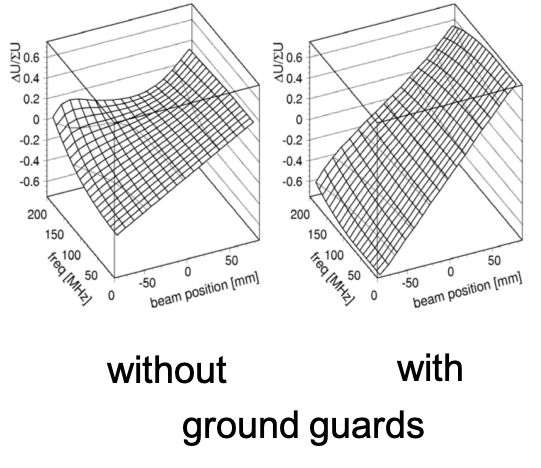}
  \caption{Position characteristic.}
  \label{fig:sfig29b}
\end{subfigure}
\hspace*{5mm}
\caption{The ``shoe-box'' BPM without, and with ground-guards (courtesy \textit{P.~Kowina}).}
\label{fig:shoebox}
\end{figure}
Figure~\ref{fig:shoebox} shows a variant of the split-plane BPM, the arrangement of split electrodes in
a rectangular cross-section of the beam-pipe, the so-called ``shoe-box'`BPM.
This configuration, as any other linear cut electrode configuration, offers the same linear position characteristic.
However, numerical studies show the coupling along the cut between the electrodes to be an error source,
limiting the linearity in particular at higher frequencies, see Fig.~\ref{fig:sfig29b} (left).
A narrow \emph{ground-guard} between the electrodes reduces the cross-coupling and 
improves the linearity, also at higher frequencies, see Fig.~\ref{fig:sfig29b} (right).

Split-plane BPMs were popular in the early days -- particular in hardon accelerators -- when the BPM read-out electronics
was entirely based on analog signal-processing, and a correction of a non-linear position behavior was difficult
to implement.
The electrode dimensions are large, in the order of the beam-pipe diameter, which results in a high 
electrode-to-ground capacitance and limits the usable frequency range.
At higher frequencies the electrodes develop eigenmodes which results in an unwanted, high beam coupling
impedance, therefore this type of BPM should not be used in accelerators with short bunches. 
 
\subsubsection{The cavity BPM}  
  
The types of BPM pickups discussed so far were based on beam image currents, induced into the pickup electrodes, and
have a broadband characteristic:\newline 
Their transfer impedance $Z(\omega)$
covers a wide frequency range with a position behavior independent of the operation frequency.
For a pair of stripline electrodes, the maximum achievable value of the transfer impedance is limited to half of the
characteristic impedance of the electrodes ($Z_0/2$), but in practice the transfer impedance of a broadband
BPM rarely exceeds a few-$\Omega$ (see Fig.~\ref{fig:buttonZ} and Fig.~\ref{fig:sfig26b}).
The BPM transfer impedance, together with the intensity and spectrum of the bunched beam, defines the
achievable output signal level at the ports (vacuum feedthroughs) of the BPM, 
therefore the signal-to-noise ratio, and finally the resolution of the beam position measurement.

\begin{figure}[b!]
\begin{center}
\includegraphics[width=0.9\textwidth]{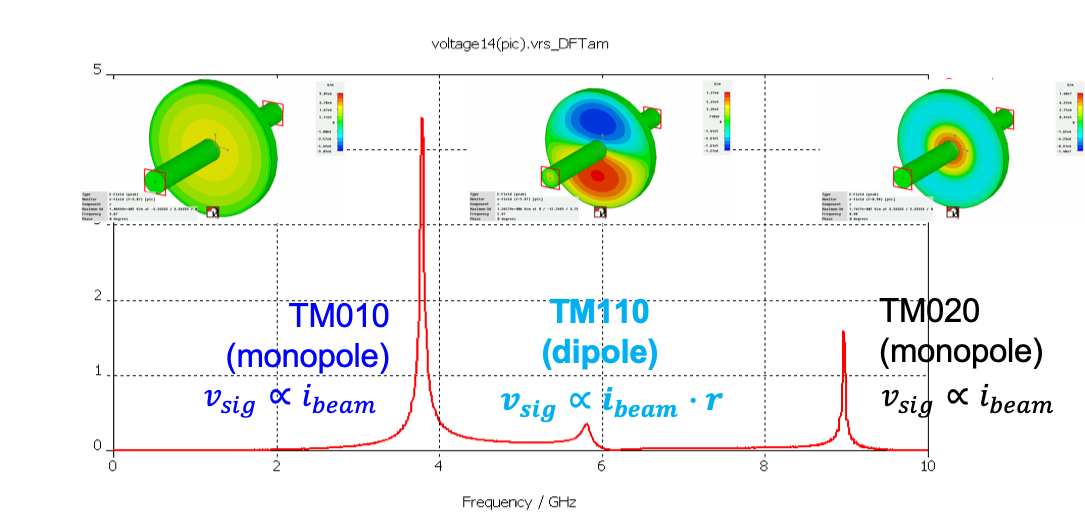}
\caption{Beam excited eigenmodes in a cylindrical cavity (courtesy \textit{D.~Lipka}).}
\label{fig:cavityModes}
\end{center}
\end{figure}
A beam-excited, passive resonant cavity can be used a beam detector, and offers, similar to an accelerating
structure, a high shunt impedance $R_s$ (typically $k\Omega/mm$) at the resonant frequency.
Figure~\ref{fig:cavityModes} shows the first three beam excited eigenmodes of a cylindrical resonator,
the so-called ``pillbox'' cavity.
For beam position measurements the TM110 dipole mode is of relevance,
\begin{equation}
v_{\mathrm{TM110}}\propto i_{\mathrm{beam}}\cdot r
\label{eq:TM110}
\end{equation}
it returns a signal which is proportional to the beam intensity and the beam displacement $|\vec{r}|=\sqrt{x^2+y^2}$
wrt.\ the center of the resonator.
In contrast to the signals from broadband BPM electrodes, a dipole-like mode signal from a cavity resonator 
has no common-mode signal contribution, for a centered beam $r=0$ the TM110-mode signal vanishes, 
see Eq.~(\ref{eq:TM110}).
For broadband BPM pickups the difference signal of Eq.~(\ref{eq:delta}) has to be arranged externally, with help
of the read-out electronics, in case of the cylindrical cavity the TM110 mode already is a $\Delta$-signal.

\begin{figure}[t!]
\hfill
\begin{subfigure}{0.4\textwidth}
  \centering
  \includegraphics[width=0.9\linewidth]{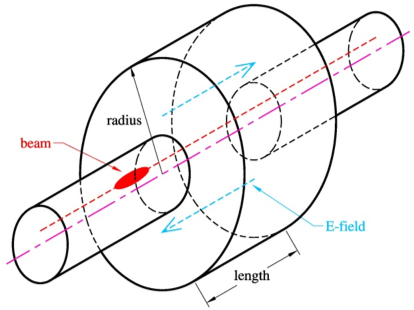}
  \caption{Beam excited, passive cylindrical \\ resonator.}
  \label{fig:sfig31a}
\end{subfigure}
\hfill
\begin{subfigure}{0.37\textwidth}
  \centering
  \includegraphics[width=0.9\linewidth]{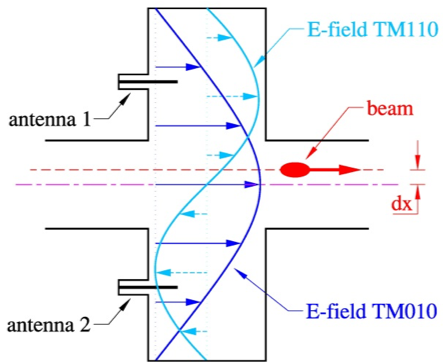}
  \caption{Electric fields of the monopole \\ and the dipole mode.}
  \label{fig:sfig31b}
\end{subfigure}
\hspace*{10mm}
\caption{Operation principle of a ``pillbox'' cavity BPM.}
\label{fig:cavityBPMprinciple}
\end{figure}
A single bunch or a train of bunches with a sufficient wide spectrum is be able to excite the TM110 eigenmode, 
but it also may excite unwanted resonances like
the TM010 and TM020 monopole modes and other higher-order modes, as illustrated in Fig.~\ref{fig:cavityModes}.
The principle of operation of a cylindrical ``pillbox''cavity BPM of radius $R$ and length (height) $h$ is sketched in
Fig.~\ref{fig:cavityBPMprinciple}, with a beam passing the resonator with some displacement $dx$ away from the
$z$ symmetry axis. 
The velocity vector of the beam points dominantly to the $z$-direction, therefore it couples to the eigenmodes
that have longitudinal E-field components, the so-called E or TM modes.
As the E-field always has to be zero at the conductive boundary wall of the lateral area, i.e.\ the cavity rim,
the diameter $2R$ defines many (but not all) of the possible E-field mode patterns, the E-field of the 
fundamental TM010 monopole mode and the TM110 dipole mode are indicated in the 
longitudinal section sketch of Fig.~\ref{fig:sfig31b}.
The E-field of the dipole mode (TM110) is \emph{zero} at the center, along the $z$ symmetry axis, where the E-field
of the monopole mode (TM010) has its maximum.
Therefore, a centered beam will only excite the TM010, plus some other higher-order modes, a displaced beam $dx$
will also excite the TM110 dipole mode, which changes its sign as the beam crosses the center.
The beam excited fields are coupled to a coaxial port, usually loaded with $R_l=50\;\Omega$, by means of a pin
(capacitive coupling to the E-field), or loop antenna (inductive coupling to the H-field).
For symmetry reasons a symmetric arrangement of coupling antennas is preferred, Fig.~\ref{fig:sfig31b} shows two
pin-antennas which will detect the same mode signal, except the TM110 dipole mode signals will have opposite sign.

The parameters of the cavity resonances (frequencies and Q-values) are defined by its geometric shape, 
the dimensions and the conductivity of the metallic material.
For simple geometries, e.g.\ a cuboid (brick-shape) or a cylinder (``pillbox''), the \textit{Laplace} equation
of the vector potential
\begin{equation}
\Delta\Psi+k_0^2\epsilon_r\mu_r\Psi=0
\label{eq:laplace}
\end{equation}
can be solved analytically in form an orthonormal series expansion. 
$k_0=2\pi/\lambda_0$ is the free space wave number in (\ref{eq:laplace}), and is $k_0^2=\omega^2\epsilon_0\mu_0$, while
$\lambda_0=2\pi c_0/\omega$ is the free space wave length.
Without going into the details, lets summarize the theory first 
for a hollow cuboid in \textit{Cartesian} coordinates of dimensions $0<x<a$, $0<y<b$, $0<z<c$ as boundary conditions. 
A product ``Ansatz'' for the vector potential:
$$
\Psi= X(x)Y(y)Z(z)
$$
with the separation condition $k_x^2+k_y^2+k_z^2=k_0^2\epsilon_r\mu_r$
finds a general solution for the vector potential as orthonormal series expansion:
\begin{equation}
\resizebox{0.93\hsize}{!}{%
$\psi=\left\{\begin{array}{cc} A\cos\left(k_x x\right)+B\sin\left(k_x x\right) \\
\grave{A}e^{-jk_x x}+\grave{B}e^{-jk_x x} 
\end{array}\right\}
\left\{\begin{array}{cc} C\cos\left(k_y y\right)+D\sin\left(k_y y\right) \\
\grave{C}e^{-jk_y y}+\grave{D}e^{-jk_y y} 
\end{array}\right\}
\left\{\begin{array}{cc} E\cos\left(k_z z\right)+F\sin\left(z_z z\right) \\
\grave{E}e^{-jk_z z}+\grave{F}e^{-jk_z z} 
\end{array}\right\}
\begin{array}{ll}
\rightarrow\text{standing waves} \\
\rightarrow\text{traveling waves}
\end{array}
$}
\label{eq:seriesBrick}
\end{equation} 
with $k_x=m\pi/a$,  $k_y=n\pi/b$, and $k_z=p\pi/c$.
In our case the standing wave solution of Eq.~(\ref{eq:seriesBrick}) will give the results for the fields 
and for the frequencies of the eigenmodes:
\begin{equation}
f_{mnp}=\frac{c_0}{2\pi\epsilon_r\mu_r}\sqrt{\left(\frac{m\pi}{a}\right)^2+\left(\frac{n\pi}{b}\right)^2+\left(\frac{p\pi}{c}\right)^2}
\label{eq:fbrick}
\end{equation}

While there are a few examples using a cuboid as high resolution cavity BPM, often a cylinder is preferred,
e.g.\ for manufacturing advantages on a precision lathe.
The orthonormal series expansion procedure of (\ref{eq:laplace}) for a cylinder is the same as for the cuboid, 
but as of the cylindrical coordinates, now cylindrical functions come into play.  
For a hollow cylinder its dimensions $0<\rho<R$ and $0<z<h$, as well as the angle $0<\varphi<2\pi$ will
set the boundary conditions for the product approach:
$$
\Psi=R(\rho)F(\varphi)Z(z)
$$
This results in a general solution for the vector potential, again as orthonormal series expansion, 
but now in cylindrical coordinates:
\begin{equation}
\resizebox{0.93\hsize}{!}{%
$\psi=\left\{\begin{array}{cc} AJ_m\left(k_{\rho}\rho\right)+BY_m\left(k_{\rho}\rho\right) \\
\grave{A}H_m^{(2)}\left(k_{\rho}\rho\right)+\grave{B}H_m^{(1)}\left(k_{\rho}\rho\right)
\end{array}\right\}
\left\{\begin{array}{cc} C\cos(m\varphi)+D\sin(m\varphi) \\
\grave{C}e^{-jm\varphi}+\grave{D}e^{-jm\varphi} 
\end{array}\right\}
\left\{\begin{array}{cc} E\cos\left(k_z z\right)+F\sin\left(k_z z\right) \\
\grave{E}e^{-jk_z z}+\grave{F}e^{-jk_z z} 
\end{array}\right\}
\begin{array}{ll}
\rightarrow\text{standing waves} \\
\rightarrow\text{traveling waves}
\end{array}
$}
\label{eq:seriesPillbox}
\end{equation}
Here the cylindrical function are:
\begin{description}
\item[$J_m$] \textit{Bessel} function of first kind of order $m$.
\item[$Y_m$] \textit{Bessel} function of second kind of order $m$ (also called \textit{Weber} function or \textit{Neumann}
function).
\item[$H_m^{(1)}$] \textit{Hankel} function of first kind of order $m$ (for outward traveling waves, also called
\textit{Bessel} function of third kind).
\item[$H_m^{(2)}$] \textit{Hankel} function of second kind of order $m$ (for inward traveling waves, also called
\textit{Bessel} function of third kind).
\end{description}
Here, two separation conditions are used:
$$
\left(\frac{j_{mn}}{R}\right)^2+k_z^2=k_0^2 \qquad\qquad
\left(\frac{j'_{mn}}{R}\right)^2+k_z^2=k_0^2
$$ 
with $k_z=p\pi/h$, and $j_{mn}$ being the $n^{th}$ root of $J_m(x)$ and $j'_{mn}$ being the $n^{th}$ root of $J'_m(x)$, with $J'_m$ being the derivative of the Bessel function $J_m$. 
Eq.~(\ref{eq:seriesPillbox}) can be expressed in form of TM$_{mnp}$ and TE$_{mnp}$ eigenmodes, 
with the frequencies found by the separation conditions:
\begin{align}
\begin{split}
f_{\mathrm{TM}mnp}&=\frac{c_0}{2\pi\epsilon_r\mu_r}\sqrt{\left(\frac{j_{mn}}{R}\right)^2+\left(\frac{p\pi}{h}\right)^2} \\
f_{\mathrm{TE}mnp}&=\frac{c_0}{2\pi\epsilon_r\mu_r}\sqrt{\left(\frac{j'_{mn}}{R}\right)^2+\left(\frac{p\pi}{h}\right)^2}
\end{split}
\label{eq:fpillbox}
\end{align} 

The analytical approach gives a rough estimation of the cavity dimensions and the related eigen-frequencies for a given
shape.
The ports of the beam-pipe act as waveguides, therefore most of the cavity eigemodes with 
$f_{\mathrm{cavity}}>f_{\mathrm{TE}11}$ 
will not be trapped, assuming a beam-pipe with circular cross-section. 
As the waveguide ports alter the boundary conditions, a so-called ``mode matching''
approach is required to correctly solve the cavity/beam-pipe geometry in an analytical way.
In practice a numerical analysis often is more convenient, which also allows to include the coaxial port antennas and
other details, and provides additional information, e.g.\ due to the finite conductive of the cavity walls.

\begin{figure}[ht]
\begin{center}
\includegraphics[width=0.9\textwidth]{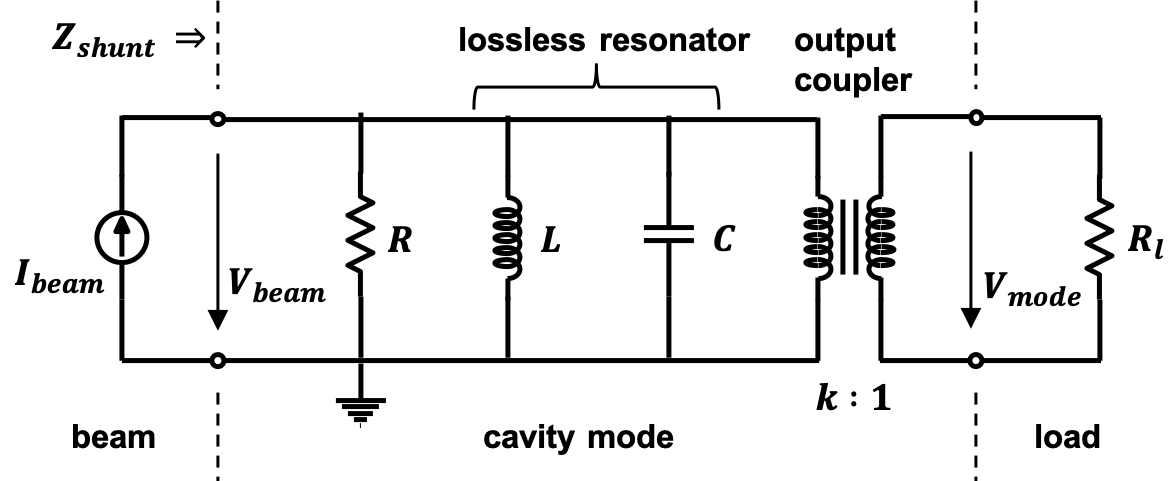}
\caption{Equivalent circuit of a resonant mode of a cavity BPM.}
\label{fig:cavityEqui}
\end{center}
\end{figure}
Electrically, each mode can be described as $RLC$ equivalent circuit (Fig.~\ref{fig:cavityEqui}),
the circuit parameters can be derived from analytical formulas, from a numerical analysis, or from
RF measurements in the laboratory. 
The resonant frequency is given by $\omega_{\mathrm{mode}}=1/\sqrt{LC}$, the losses are represented by $R$, and
the geometry (or shape) of the resonator is reflected by 
$R/Q=\omega_{\mathrm{mode}}L=1/(\omega_{\mathrm{mode}}C)=\sqrt{L/C}$, which is an equivalent to the
characteristic impedance of a transmission-line.
The output power delivered at a load impedance $R_l$ is maximum if $R$ is matched to $R_l$, i.e.\
at critical coupling (no reflections).
This can be achieved by tuning the coupling to a value $k=\sqrt{R/R_l}$.

\begin{figure}[b!]
\hfill
\begin{subfigure}{0.4\textwidth}
  \centering
  \includegraphics[width=0.9\linewidth]{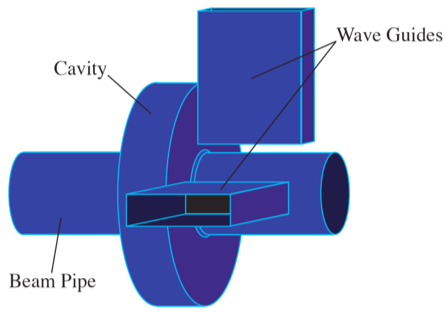}
  \caption{A cavity BPM with waveguides attached.}
  \label{fig:sfig33a}
\end{subfigure}
\hfill
\begin{subfigure}{0.37\textwidth}
  \centering
  \includegraphics[width=0.9\linewidth]{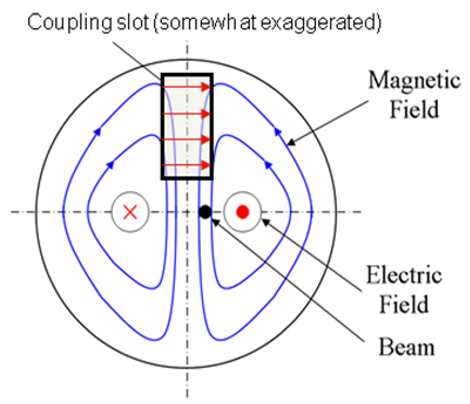}
  \caption{Waveguide coupling to the TM110 dipole mode.}
  \label{fig:sfig33b}
\end{subfigure}
\hspace*{10mm}
\caption{A common-mode free cavity BPM.}
\label{fig:cavityCMfree}
\end{figure}
As Fig.~\ref{fig:cavityModes} suggests, the TM010 monopole mode is very strong and always present,
and causes a major limitation to detect a weak TM110 dipole mode signal, 
despite the fact that it is located at a higher frequency,
The reason lies in the finite Q-value of the resonant modes, with the ``tail'' of $f_{\mathrm{TM}010}$ leaking into
$f_{\mathrm{TM}110}$.
The TM010 mode is equivalent the common-mode (intensity) signal contribution in a broadband BPM, except
in case of the cavity BPM there is some benefit due to the frequency separation.
Figure~\ref{fig:sfig33a} shows a pillbox cavity BPM with two slot-coupled rectangular waveguides attached,
oriented in the horizontal and vertical plane.
The waveguides of width $a$ act as very efficient high-pass filter, and are dimensioned such that the cutoff frequency of the
TE10 waveguide mode lies between the monopole and dipole-mode frequency of the cavity resonator,
$f_{\mathrm{TM}010}<f_{\mathrm{TE}10\mathrm{wg}}=1/(2a\sqrt{\epsilon\mu})<f_{\mathrm{TM}010}$.
To minimize the monopole-mode leakage of this so-called ``common-mode free'' cavity BPM, the rectangular
coupling slot between cavity resonator and waveguide has to be made reasonable narrow, the coupling
mechanism is shown qualitatively in Fig.~\ref{fig:sfig33b}.

A cavity BPM based on a cylindrical pillbox resonator requires the two TM110 mode \emph{polarizations} to be
perfectly orthogonal, aligned to the horizontal and vertical plane.
Usually this is achieved by precise manufacturing of the coupling slots or antennas, which cause an
``imperfection'' for the cylinder and gives the TM110 fields a boundary to align the polarization axis.
In practice, a residual cross-coupling between the planes of $<-40$~dB can be achieved, while the frequencies 
of the two TM110 polarization are almost identical.
Alternatively two cuboid (brick-style) resonators with different resonant frequencies
can be arranged to detect horizontal and vertical beam displacements separately with minimum cross-talk.

While the operational principle seems to be simple, the realization of a BPM system based on resonant cavities is non-trivial.
Along with the high transfer impedance goes a high beam coupling impedance that can have negative
effects on the beam quality, e.g.\ beam instabilities, beam breakup, etc., 
therefore cavity BPMs are not used in storage rings.
Beside the precision manufacturing and calibration of the cavity BPM pickup, the RF front-end 
and signal processing is more demanding compared to read-out electronics for broadband BPMs,
moreover, a separate TM010 monopole mode resonator is required
as reference for the beam intensity normalization and as beam phase reference to
detect the phase of the dipole mode. 

\subsubsection{Other types of BPM pickups} 

The presented BPM pickup types, button-style, stripline, split-plane and cavity BPM are the most common used
BPM pickups.
However, there exists a larger variety of other BPM pickups, e.g.
\begin{description}
\item[Exponentially tapered stripline BPM] This modified stripline BPM alters the transfer function such that the lobes
of $|Z(\omega)|$ have a wider bandwidth.
Similar techniques are used in the RF technology, known as multi-element direction coupler.
\item[Re-entrant BPM] is a ``hybrid'' between broadband, stripline-style and resonant, cavity-style BPM pickup.
\item[Resonant button or stripline BPM] An external reactive circuit element is used to change the broadband behavior
of a traditional capacitive pickup, e.g.\ button BPM, or a stripline BPM into a resonant circuit. 
While the beam position characteristic stays unchanged, the transfer function is altered towards a band-pass characteristic,
with a higher $|Z(\omega_{\mathrm{res}})|$ at the resonance frequency.
\item[Inductive BPM] The wall current is detected by a symmetric arrangement of inductive couplers.
%\item[...]
\end{description}
The list is incomplete, but in common of all type of BPM pickups is the symmetric arrangement of beam coupling elements,
and the better the symmetry is realized in practice, the better is the performance potential in terms of accuracy and
achievable resolution.

Other, more ambitious electromagnetic BPM detectors proposed will operate in domain of optical wavelengths,
but require a short bunch length.
The beam field could be observed by a symmetric arrangement of electro-optical crystals, similar 
to an electro-optical modulator, or could generate diffraction or \textit{Cherenkov} radiation with radiators
and detectors in a symmetrical setup to extract the beam position information.

\subsection{Broadband BPM pickup response to a \textit{Gaussian} bunch}
\label{sec:BPMgaussResponse}

The response of a broadband BPM pickup electrode, button or stripline, to a single bunch in terms of 
time-domain output signal
waveform and level is of fundamental relevance, as this is the signal to be detected and processed 
by the read-out electronics.
From the single bunch response it is straightforward -- applying the superposition principle -- to analyze more complex beam
formats, like multi-bunch trains with or without equidistant bunch spacing, also with varying bunch intensities. 

\begin{figure}[t]
\begin{subfigure}{0.45\textwidth}
  \centering
  \includegraphics[width=0.99\linewidth]{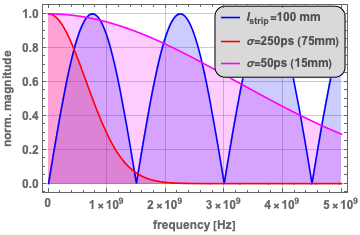}
  \caption{Relative (normalized) levels of the stripline transfer impedance and the \textit{Gausian} bunch signal.}
  \label{fig:sfig34a}
\end{subfigure}
\hfill
\begin{subfigure}{0.49\textwidth}
  \centering
  \includegraphics[width=0.99\linewidth]{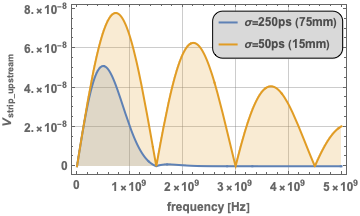}
  \caption{Output voltage (magnitude) of a 100~mm long stripline electrode ($\phi=0.1$)
  to a \textit{Gaussian} bunch.}
  \label{fig:sfig34b}
\end{subfigure}
\caption{Stripline BPM response to a single \textit{Gaussian} bunch 
in the frequency domain (magnitude).}
\label{fig:stripGaussFD}
\end{figure}
The frequency-domain output signal for the upstream (coupled) port of an ideal, air-dielectric stripline electrode, 
i.e.\ $\beta_s=\beta$, neglecting
high frequency and fringe field effects, to a relativistic ($\beta=1$) \textit{Gaussian} beam bunch is given by the 
multiplication of
the frequency-domain \textit{Gaussian} bunch stimulus in Eq.~(\ref{eq:iftrans}), and the transfer impedance of the
stripline electrode Eq.~(\ref{eq:striplineFD}): 
\begin{equation}
V_{\mathrm{strip}_{\mathrm{upstream}}}(\omega) = j\phi Z_0 e N 
e^{-\frac{(\omega \sigma_t)^2}{2}}
 e^{-j\omega\frac{\ell}{c_0}} 
\sin \left ( \omega\frac{\ell}{c_0} \right )
\label{eq:stripGaussFD}
\end{equation}
and is illustrated in Fig.~\ref{fig:stripGaussFD}.

\begin{figure}[b]
\begin{subfigure}{0.32\textwidth}
  \centering
  \includegraphics[width=0.99\linewidth]{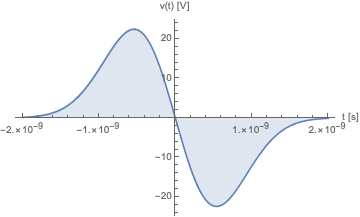}
  \caption{Bunch length $\sigma=150$~mm.}
  \label{fig:sfig35a}
\end{subfigure}
\hfill
\begin{subfigure}{0.32\textwidth}
  \centering
  \includegraphics[width=0.99\linewidth]{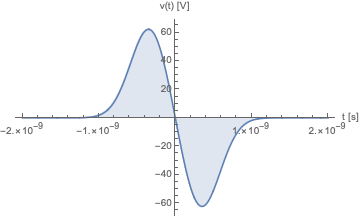}
  \caption{Bunch length $\sigma=75$~mm.}
  \label{fig:sfig35b}
\end{subfigure}
\hfill
\begin{subfigure}{0.32\textwidth}
  \centering
  \includegraphics[width=0.99\linewidth]{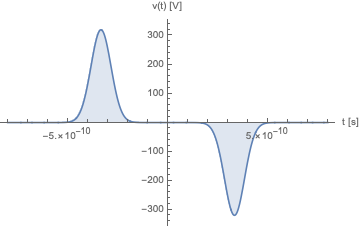}
  \caption{Bunch length $\sigma=15$~mm.}
  \label{fig:sfig35c}
\end{subfigure}
\caption{Stripline BPM response to a single \textit{Gaussian} bunch 
in the time domain.}
\label{fig:stripGaussTD}
\end{figure}
The time-domain output signal at the upstream port of a stripline electrode -- with the impulse response according to
Eq.~(\ref{eq:stripTD}) -- to a \textit{Gaussian} bunch following Eq.~(\ref{eq:gaussTD}) is simply:
\begin{align}
\begin{split}
v_{\mathrm{strip}_{\mathrm{upstream}}}&=\phi\frac{Z_0}{2}\frac{eN}{\sqrt{2\pi}\sigma_t}
\left[e^{-\frac{\left(t+\tau\right)^2}{2\sigma_t^2}}-e^{-\frac{\left(t-\tau\right)^2}{2\sigma_t^2}}\right] \\
\text{with: } \tau&=\frac{\ell}{2c_0}\left(\frac{1}{\beta}+\frac{1}{\beta_s}\right)
\end{split}
\label{eq:stripGauss} 
\end{align}
For relativistic beams ($\beta=1$) and air-stripline electrodes ($\epsilon_r=1\rightarrow\beta_s=1$)  $\tau$ in
Eq.~(\ref{eq:stripGauss}) simplifies $\beta=\beta_s=1\rightarrow\tau=\ell/c_0$.
Figure~\ref{fig:sfig35a} and \ref{fig:sfig35b} show the output signal waveforms of a stripline electrode of length $\ell=100$~mm
for a \textit{Gaussian} bunch ($\beta=1$) of $\sigma=150$~mm and 75~mm length, respectively, 
positive and negative parts of the
doublet impulse response waveform are overlapping.
A shorter bunch of $\sigma=15$~mm length for the same stripline length separates the those parts, 
see Fig.~\ref{fig:sfig35c}, and in the
general case $t_{\mathrm{bunch}}<\ell/2$, with $t_{\mathrm{bunch}}$ being the \emph{total} bunch length, 
the stripline BPM can also be used as detector to monitor the longitudinal bunch profile.
Please note, the instantaneous signal voltages out of a stripline BPM electrode for short bunches can be
high, e.g.\ several 100 volts, in our example computed for $N=10^{11}$ charges per bunch. 
 
\begin{figure}[t]
\begin{subfigure}{0.45\textwidth}
  \centering
  \includegraphics[width=0.99\linewidth]{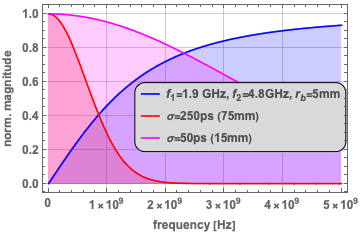}
  \caption{Relative (normalized) levels of the button BPM transfer impedance and the \textit{Gausian} bunch signal.}
  \label{fig:sfig36a}
\end{subfigure}
\hfill
\begin{subfigure}{0.49\textwidth}
  \centering
  \includegraphics[width=0.99\linewidth]{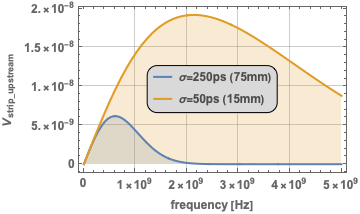}
  \caption{Output voltage (magnitude) of a 10~mm diameter button BPM electrode
  to a \textit{Gaussian} bunch.}
  \label{fig:sfig36b}
\end{subfigure}
\caption{Button BPM response to a single \textit{Gaussian} bunch 
in the frequency domain (magnitude).}
\label{fig:buttonGaussFD}
\end{figure}

\begin{figure}[b!]
\begin{subfigure}{0.32\textwidth}
  \centering
  \includegraphics[width=0.95\linewidth]{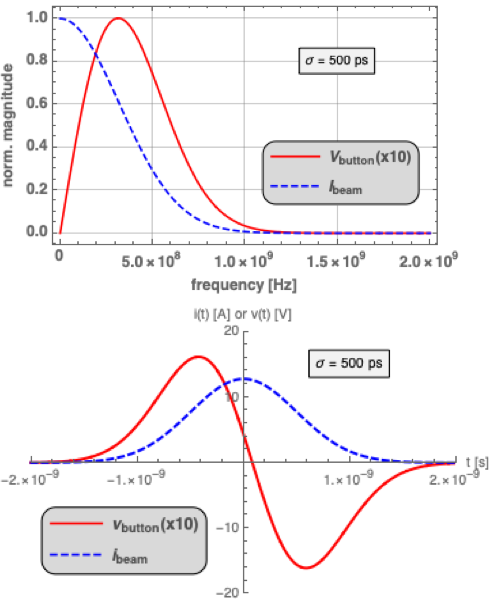}
  \caption{$\sigma_t=500$~ps.}
  \label{fig:sfig37a}
\end{subfigure}
\hfill
\begin{subfigure}{0.32\textwidth}
  \centering
  \includegraphics[width=0.95\linewidth]{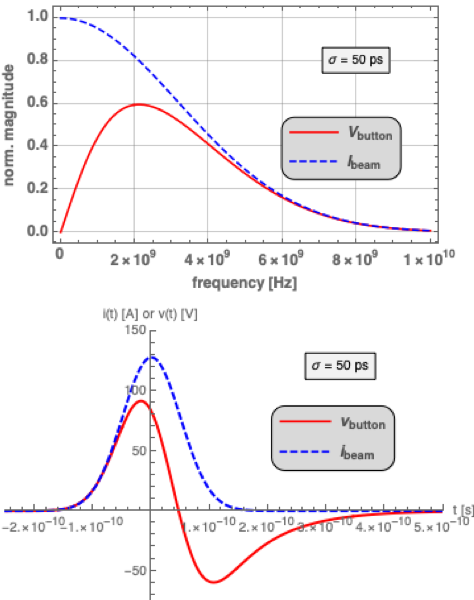}
  \caption{$\sigma_t=50$~ps.}
  \label{fig:sfig37b}
\end{subfigure}
\hfill
\begin{subfigure}{0.32\textwidth}
  \centering
  \includegraphics[width=0.95\linewidth]{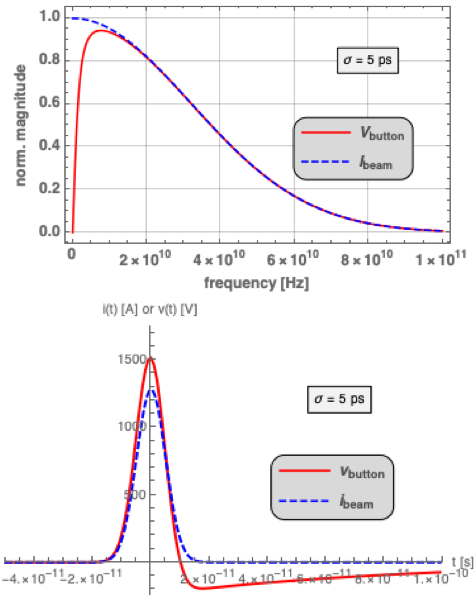}
  \caption{$\sigma_t=5$~ps.}
  \label{fig:sfig37c}
\end{subfigure}
\caption{Button BPM signal in the frequency-domain (upper row), \\ and the time-domain (lower row)
to a single \textit{Gaussian} bunch.}
\label{fig:buttonGaussiFFT}
\end{figure}
Likewise for the stripline BPM, the frequency domain response of a button BPM electrode is computed by:
\begin{equation}
V_{\mathrm{button}}(f) = Z_{\mathrm{button}}(f) I_{\mathrm{bunch}}(f)
\label{eq:vButtonFD}
\end{equation}
with $I_{\mathrm{bunch}}(f)$ as e.g.\ \textit {Gaussian} bunch according to Eq.~\ref{eq:iftrans}, 
and $Z_{\mathrm{button}}(f)$ given by Eq.~\ref{eq:Zbutton}, examples see Fig.~\ref{fig:buttonGaussFD}.

For the time-domain output signal of a button BPM electrode to a \textit{Gaussian} bunch there
unfortunately is no closed-form analytical expression.
However, if the frequency spectrum of the bunch stays below the high-pass cut-off frequency of the button electrode,
$f_1=\omega_1/(2\pi)$ in Eq.~(\ref{eq:Zbutton}), the button is purely differentiating the bunch signal,
so that:
\begin{align}
\begin{split}
v_{\mathrm{button}}&\approx\frac{A}{\pi D} \frac{R_l}{\beta c_0}\frac{di_{\mathrm{bunch}(t)}}{dt} \\
&=\frac{r^2 R_l}{D\beta c_0}\frac{eN}{\sqrt{2\pi}\sigma_t^3}te^{-\frac{t^2}{2\sigma_t^2}} \quad\text{for: }
f_{\text{3dB}} \ll f_1
\end{split}
\label{eq:vButton}
\end{align} 
with $A=r^2\pi$ being the surface are of the button electrode, and all other parameters as defined in
section \ref{sec:button}.
The 3~dB cut-off frequency for a \textit{Gaussian} bunch spectrum, used in Eq.~(\ref{eq:vButton}),
is found from Eq.~(\ref{eq:iftrans}):
$$
f_{\text{3dB, Gauss}}=\frac{\sqrt{\frac{\ln\sqrt{2}}{2}}}{\pi} \frac{1}{\sigma_t} \approx\frac{0.1325}{\sigma_t}
$$
Figure~\ref{fig:sfig37b} shows a button output signal with same bunch parameters as for the
response of a $\ell=100$~mm long stripline electrode in Fig.~\ref{fig:sfig35c}. 
As Figure~\ref{fig:sfig37a}
indicates, the transfer function of the $r=4$~mm button is not well matched to the spectrum of the
\textit{Gaussian} bunch of length $\sigma=150$~mm, therefore the output voltage is substantially lower
compered to that of the stripline electrode, see Fig.~\ref{fig:sfig35a}%
\footnote{To better visualize bunch current and button voltage in Fig.~\ref{fig:sfig37a}, 
the button output signal voltage
was magnified by $\times$10, i.e.\ the voltage scale has to be divided by 10!}.

Short bunches in a button BPM are better matched to the higher frequency range of the button transfer impedance, 
including and beyond $f_1$, as Fig.~\ref{fig:sfig37c} indicates.
To include that regime we have to evaluate Eq.~(\ref{eq:vButtonFD})
and numerically compute the time-domain button output signal by applying 
the inverse discrete \textit{Fourier} transformation
(iDFT or iFFT).
For a \textit{Gaussian} bunch the button output signal is of form:
\begin{equation}
v_{\mathrm{button}}(t)\propto\int_{-\infty}^{+\infty}\frac{jfe^{-2\left(\pi f\sigma_t\right)^2}}{f_1+jf}e^{-j2\pi f t}df 
\label{eq:vButtonTD}
\end{equation}
Fig.~\ref{fig:buttonGaussiFFT} shows the response in time and frequency-domain for
three different cases of the bunch length, based on Eq.~(\ref{eq:vButtonFD}) and (\ref{eq:vButtonTD}).

\begin{figure}[b!]
\begin{subfigure}{0.32\textwidth}
  \centering
  \includegraphics[width=0.95\linewidth]{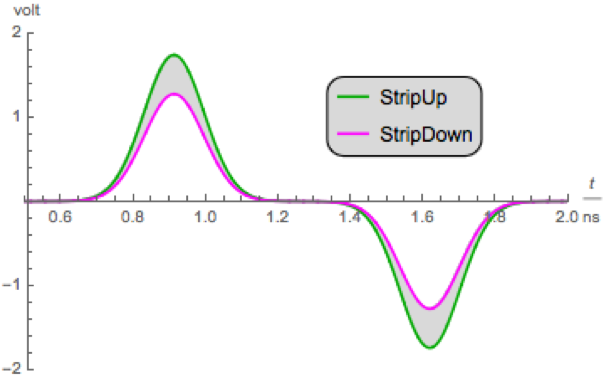}
  \caption{Time-domain, stripline.}
  \label{fig:sfig38a}
\end{subfigure}
\hfill
\begin{subfigure}{0.32\textwidth}
  \centering
  \includegraphics[width=0.95\linewidth]{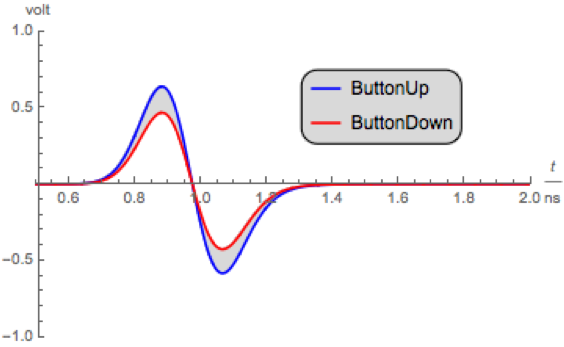}
  \caption{Time-domain, button.}
  \label{fig:sfig38b}
\end{subfigure}
\hfill
\begin{subfigure}{0.32\textwidth}
  \centering
  \includegraphics[width=0.95\linewidth]{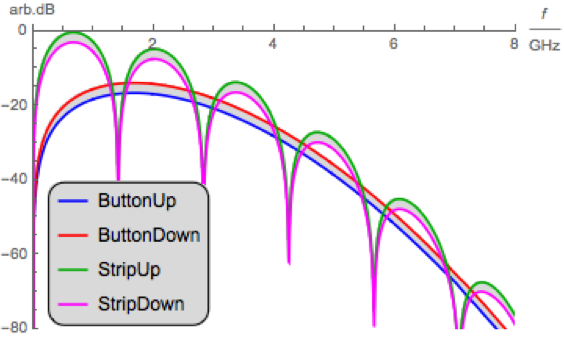}
  \caption{Frequency-domain.}
  \label{fig:sfig38c}
\end{subfigure}
\caption{Broadband BPM signals as response to a
to a single \textit{Gaussian} bunch ($eN=100$~pC, $\sigma=25$~mm, $\beta=1$, beam displacement: 1~mm) 
as result of a 3D numerical analysis.}
\label{fig:numAnalysis}
\end{figure}
Evidently, the computation of the output signal of the BPM electrodes based on analytical expressions has
some limitations, it also misses many details like high frequency effects at the vacuum feedthrough transition,
geometrical details of the BPM pickup, effects of low-$\beta$ beams, losses, etc.
A numerical analysis of the electromagnetic problem, using a good approximation of the exact BPM pickup shape, 
and a beam field as
stimulus signal allows a more realistic estimation of the electrode signals.
Figure~\ref{fig:numAnalysis} shows the results of a so-called \emph{wakefield} simulation for the button and
stripline BPMs as illustrated in Fig.~\ref{fig:sfig21a} and Fig.~\ref{fig:sfig24a}, for both cases the beam-pipe diameter
is $D=25$~mm and the coverage factor is $\phi\approx 0.1$, the length of the stripline electrode is $\ell=100$~mm. 
As a result, the time-domain output signal waveforms for both pickup electrodes are computed
(Fig.~\ref{fig:sfig38a} and Fig.~\ref{fig:sfig38b}), as well as the frequency domain response (Fig.~\ref{fig:sfig38c}), here
normalized to the peak value of the upper stripline electrode.
Fig.~\ref{fig:sfig38c} also demonstrates the broadband behavior of the position characteristic, in this example
both BPM pickups have a sensitivity of 2.7~dB/mm.

\subsection{Beam position monitoring of low-$\beta$ beams}

\begin{figure}[h]
\begin{subfigure}{0.32\textwidth}
  \centering
  \includegraphics[width=0.95\linewidth]{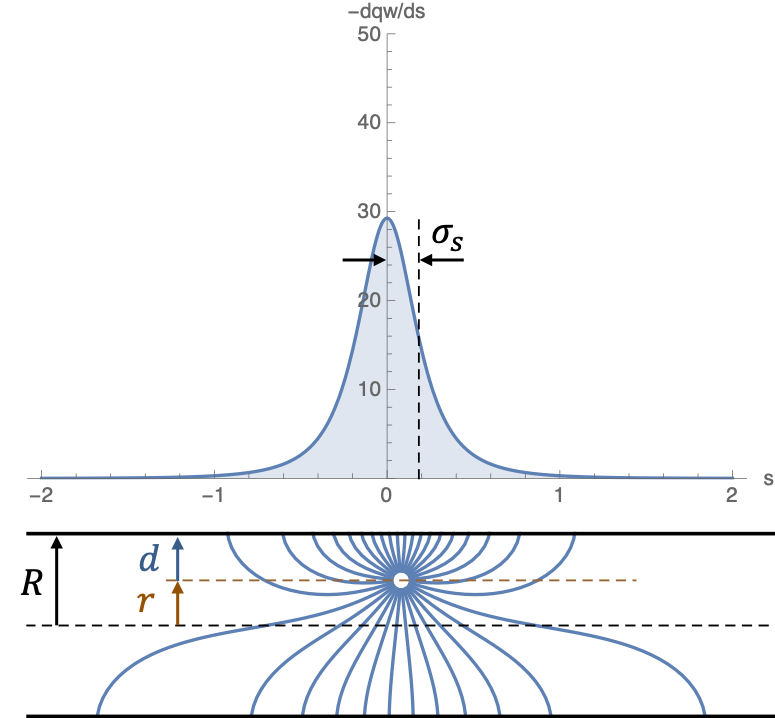}
  \caption{$\beta=0$}
  \label{fig:sfig39a}
\end{subfigure}
\hfill
\begin{subfigure}{0.32\textwidth}
  \centering
  \includegraphics[width=0.95\linewidth]{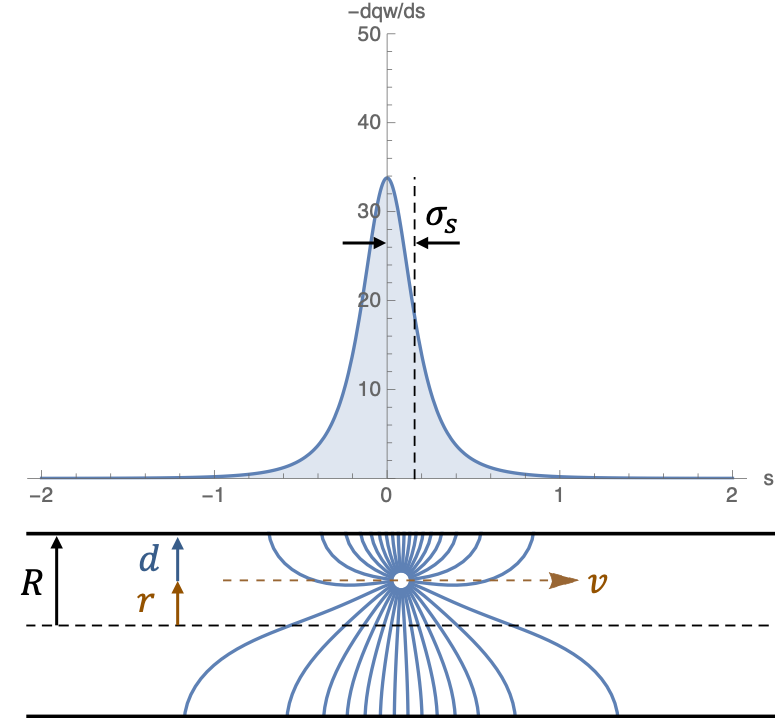}
  \caption{$\beta=0.5$}
  \label{fig:sfig39b}
\end{subfigure}
\hfill
\begin{subfigure}{0.32\textwidth}
  \centering
  \includegraphics[width=0.95\linewidth]{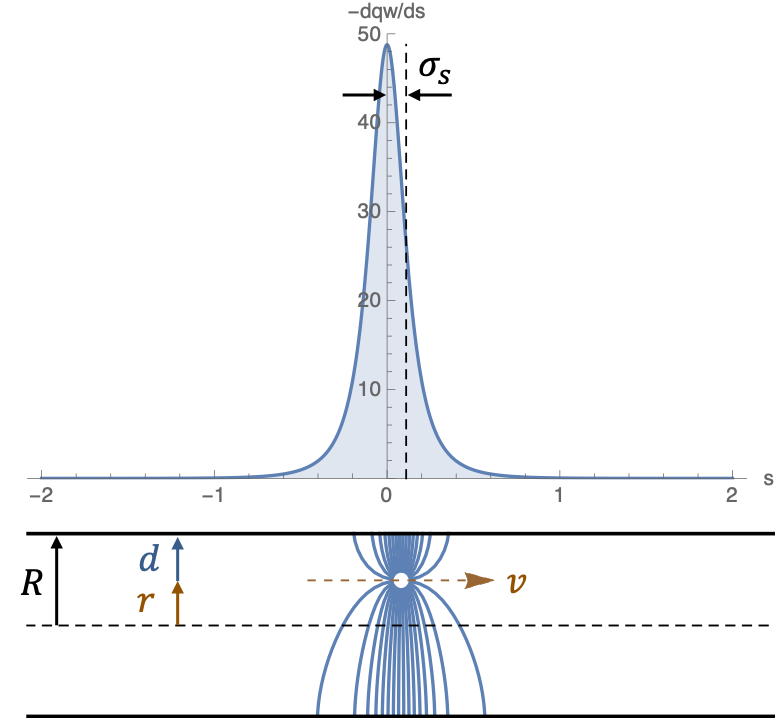}
  \caption{$\beta=0.8$}
  \label{fig:sfig39c}
\end{subfigure}
\caption{Electric field lines and wall current density for an off-center point charge moving at a low-$\beta$ velocity.}
\label{fig:lowBeta}
\end{figure}
Until now we considered the beam moving at a relativistic velocity $v=\beta c_0$, with $\beta\simeq 1$.
In that situation all electromagnetic field components of the charged particles are purely transverse, the so-called
TEM field configuration, see Fig.~\ref{fig:WallCurrent}.
As a consequence, the longitudinal wall current density $J_w(s)=dq_w/ds$ is always a replica of the
longitudinal particle distribution, the line-current $i_{\mathrm{beam}}(t)$, with $t=s/(\beta c_0)$. 
At lower beam velocities $\beta\ll1$, the EM-field of a charged particle also develops longitudinal field components, 
Fig.~\ref{fig:PointCharge} shows the effect for a moving point charge in free space. 
In Figure~\ref{fig:lowBeta} the electric field lines and the related -- for the beam monitoring more interesting --
longitudinal wall current density $J_w(s)$ 
are analyzed for a point charge,
moving with constant velocity $v=\beta c_0$ in a circular beam-pipe of radius $R$ at a displacement
$|\vec{r}|=\sqrt{x^2+y^2}$. 
While there is no closed-form analytical expression for $J_w(s, r)$ the RMS-value of the current distribution
can be expressed simply by:
\begin{equation}
\sigma_s=\frac{d}{\sqrt{2 \gamma}}
\end{equation} 
where $d$ is the distance between the point charge and the beam-pipe wall, which is the difference
$\vec{r}-\vec{R}$ between the beam at the transverse position ($r$, $\varphi$) 
and a location on beam-pipe surface ($R$, $\Phi$), see also Fig.~\ref{fig:BPMcircular}. 
As of the additional longitudinal component of the wall current distribution, the electrostatic problem of low-$\beta$ beams
now requires to solve the \textit{Laplace} equation in three-dimension. 
As a result, for the configuration of Fig.~\ref{fig:BPMcircular}, the wall current density around the azimuth is:
\begin{equation}
J_w(\omega, R, \Phi, r, \varphi)=
-\frac{I_{\mathrm{beam}}(\omega)}{2\pi R}
\left\{ \frac{I_0(gr)}{I_0(gR)}+2\sum_{n=1}^{\infty} \frac{I_n(gr)}{I_n(gR)}  \cos \left[ n (\Phi-\varphi) \right ] \right\}
\label{eq:WCapprlb}
\end{equation}
where $I_{\mathrm{beam}}(\omega)=I_{\mathrm{beam}}(m\omega_0)=\sqrt{2}\langle I_{DC}\rangle A_m$
is expressed as the RMS amplitude at a specific frequency harmonic $m$.
$I_n$ is a modified \textit{Bessel} function of order $n$, and 
\begin{equation}
g(\omega)=\frac{2\pi}{\gamma\lambda}=\frac{m\omega_0}{\beta\gamma c_0}=\frac{\omega}{\beta\gamma c_0}
\label{eq:g}
\end{equation}
is a frequency depending function.
Eq.~\ref{eq:WCapprlb} is similar to Eq.~\ref{eq:WCappr} for relativistic beams. 
Again, by integration over the
electrode surface defined by the angle $\alpha$, see Fig.~\ref{fig:BPMcircular}, we find the beam position characteristic
for a pair of BPM electrodes $A$ and $B$ in a beam-pipe with circular cross-section of radius $R$, 
now for non-relativistic beams:
\begin{equation}
I_{\mathrm{elec}}=R\int_{-\alpha/2}^{+\alpha/2} J_w(\omega, R, \Phi, r, \varphi) \mathrm{d}\Phi
\label{eq:Ieleclb}
\end{equation}
For the two horizontal arranged electrodes $A$ and $B$ follows:
\begin{equation}
I_{\mathrm{elec}}=-\frac{I_{beam}}{2\pi} s_{\mathrm{elec}} \left ( \omega, r/R,\varphi, \alpha \right )
\label{eq:ElecApprlb}
\end{equation}
with the sensitivity functions for the $A$ and $B$ electrodes based on (\ref{eq:WCapprlb}):
\begin{align}
s_A \left( \omega, r/R,\varphi, \alpha \right ) &= 
\alpha \frac{I_0(gr)}{I_0(gR)}+ 4 \sum_{n=1}^{\infty} \frac{1}{n} \frac{I_m(gr)}{I_m(gR)} 
\sin \left [n \left ( \frac{\alpha}{2}-\varphi \right ) \right ]
\label{eq:Aeleclb} \\
s_B \left( \omega, r/R,\varphi, \alpha \right ) &= 
\alpha \frac{I_0(gr)}{I_0(gR)}+ 4 \sum_{n=1}^{\infty} \frac{1}{n} \frac{I_m(gr)}{I_m(gR)} 
\sin \left [n \left ( \pi+\frac{\alpha}{2} -\varphi\right ) \right ]
\label{eq:Beleclb}
\end{align}

\begin{figure}[t!]
\begin{subfigure}{0.32\textwidth}
  \centering
  \includegraphics[width=0.99\linewidth]{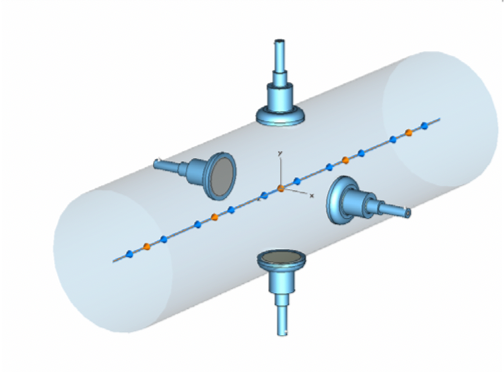}
  \caption{$R=30$~mm button BPM \\ for numerical analysis}
  \label{fig:sfig40a}
\end{subfigure}
\hfill
\begin{subfigure}{0.32\textwidth}
  \centering
  \includegraphics[width=0.95\linewidth]{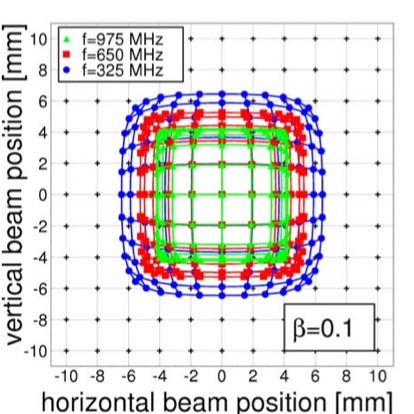}
  \caption{Position characteristic at $\beta=0.1$}
  \label{fig:sfig40b}
\end{subfigure}
\hfill
\begin{subfigure}{0.32\textwidth}
  \centering
  \includegraphics[width=0.95\linewidth]{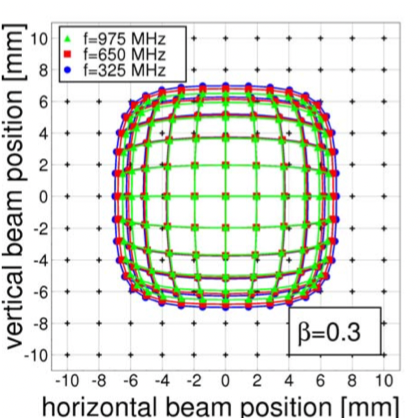}
  \caption{Position characteristic at $\beta=0.3$}
  \label{fig:sfig40c}
\end{subfigure}
\caption{Button BPM position characteristic for low-$\beta$ beams, $\beta\in$ (0.1, 0.3), 
observed at frequencies $f\in$ (325, 650, 975 MHz) (courtesy \textit{P.~Kowina}).}
\label{fig:GSIlb}
\end{figure}

Similar to section~\ref{sec:PosChar} for a relativistic beam, we can combine (\ref{eq:Aeleclb}) and (\ref{eq:Beleclb})
in terms of a $\Delta/\Sigma$ or log-ratio normalization and find the normalized position characteristic. 
However, unlike for the relativistic case Eq.~(\ref{eq:WCclosed}), a closed form expression has never been
worked out for the wall current distribution in the non-relativistic case. 
The main difference for the BPM position characteristic between relativistic and non-relativistic beams is the dependence
on the frequency in the latter, given by the term Eq.~(\ref{eq:g}).

Figure~\ref{fig:GSIlb} shows a numerical study of the effects for low-$\beta$ beams on the position characteristic 
wrt.\ operation frequency of the read-out electronics, which qualitatively verifies the analytical analysis
discussed above.
Still, the situation may become more complicated if the read-out system operates over a broad range of frequencies,
e.g.\ for single bunch signal processing, as the impulse response of the position signal, i.e.\ its shape, is a function
of the frequency content modulated by the beam velocity.

\section{BPM Signal Processing} 

The signals from the BPM electrodes need to be conditioned and processed to extract the beam intensity independent
beam position information.
This is realized by a set of electronics hardware (analog, RF and digital), digital gateware and real-time software
that follows the discussed normalization procedure, typically the $\Delta/\Sigma$-normalization.
The exact concept, design, layout, and implementation of the BPM signal processing system depends on a variety of
factors, e.g.\ type of accelerator, beam parameters and formatting, performance requirements of the beam
position measurement, environmental
conditions, etc., but also on less
technical factors, like budget and manpower resources, infrastructure, as well as on 
laboratory standards, rules and regulations.

\begin{figure}[t!]
\begin{center}
\includegraphics[width=0.8\textwidth]{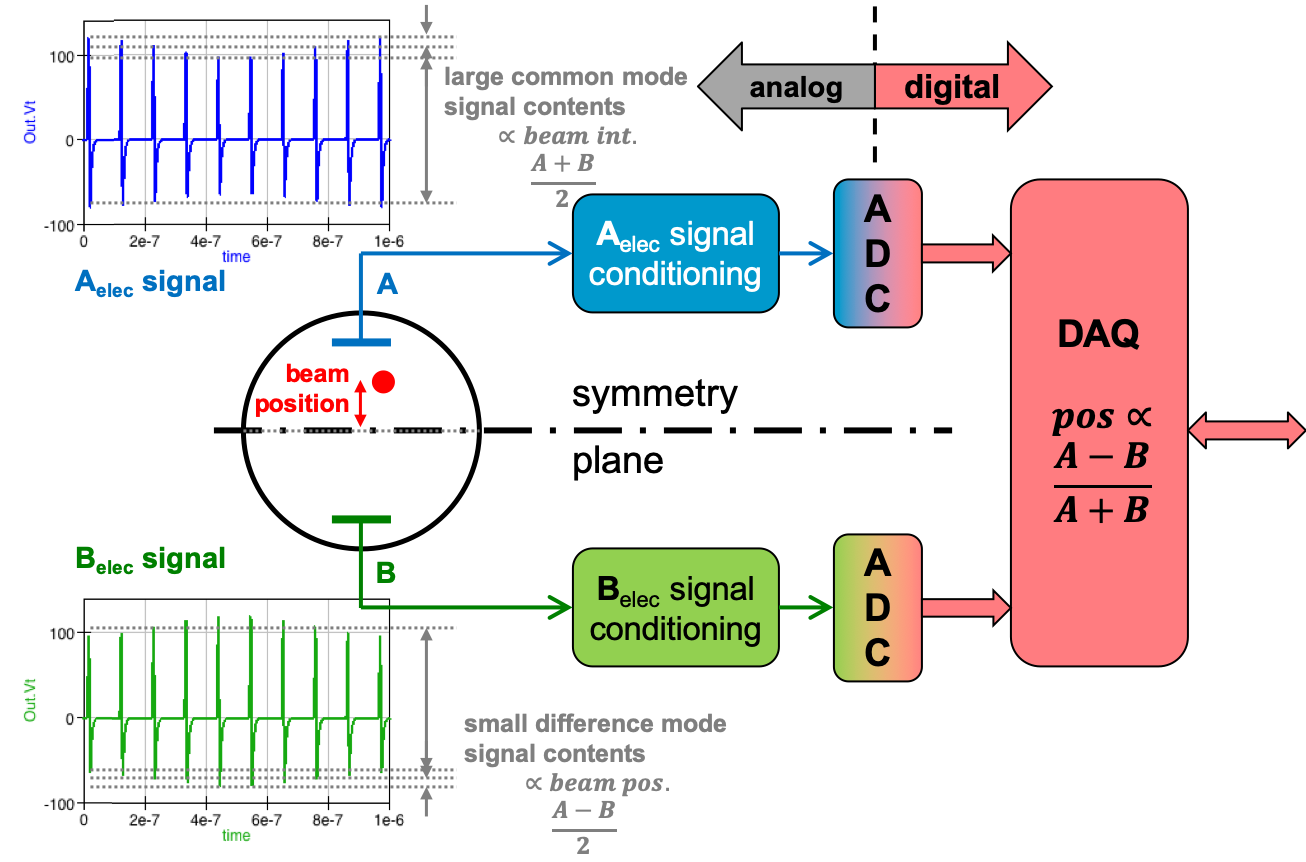}
\caption{Symmetry concept of a BPM system.}
\label{fig:BPMsymmetry}
\end{center}
\end{figure}
As Fig.~\ref{fig:BPMsymmetry} illustrates, the BPM signal processing is a continuation of the symmetry concept 
of the BPM pickup, used to detect a small asymmetry, 
i.e.\ the beam displacement with a perfect symmetric system in presence of a high common-mode signal.
The more perfect the symmetry of the entire BPM system -- pickup and signal processing --
 the higher is its performance potential, BPM resolution and accuracy.

\subsection{Read-out electronics}

\begin{figure}[t!]
\begin{center}
\includegraphics[width=0.8\textwidth]{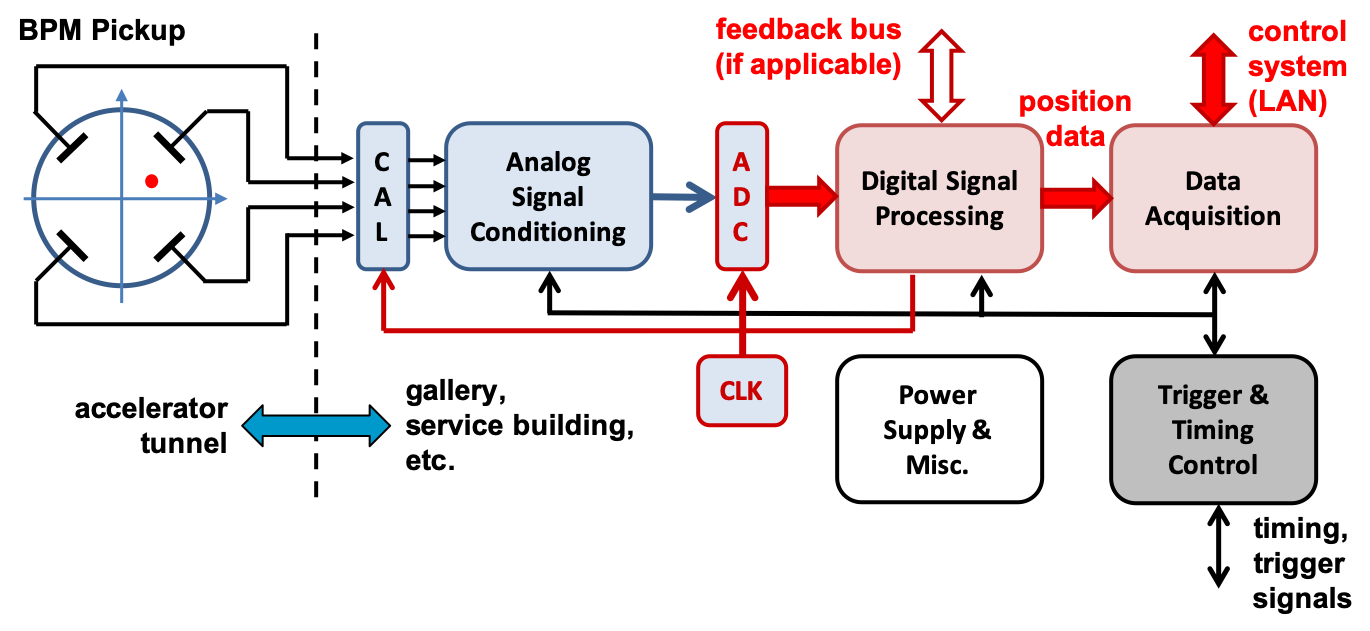}
\caption{Read-out (front-end) electronics for a BPM pickup.}
\label{fig:BPMelectronics}
\end{center}
\end{figure}

Figure~\ref{fig:BPMelectronics} shows the building blocks of a typical BPM read-out electronics. 
Each BPM pickup
is equipped with this so-called \textit{front-end} electronics, processing one or both (horizontal and vertical) planes. 
In most cases it is preferred to locate the BPM electronics outside the accelerator tunnel to avoid damage of the
semiconductors and other elements from the ionizing radiation during machine operation.
This requires rather long coaxial cables, TEM transmission-lines, between the BPM pickup in the accelerator tunnel
and the read-out electronics located in a gallery or service building, often installed in temperature stabilized racks.
Sometimes, e.g.\ to detect the position of beam with very low intensity, some signal conditioning elements, e.g.\
pre-amplifiers, RF filters, hybrids, etc.\ have to be installed in the tunnel in close proximity to the BPM pickup, 
or a special impedance-matching
network requires to be located directly on the BPM feedthrough. 

\begin{figure}[b!]
\begin{center}
\includegraphics[width=0.9\textwidth]{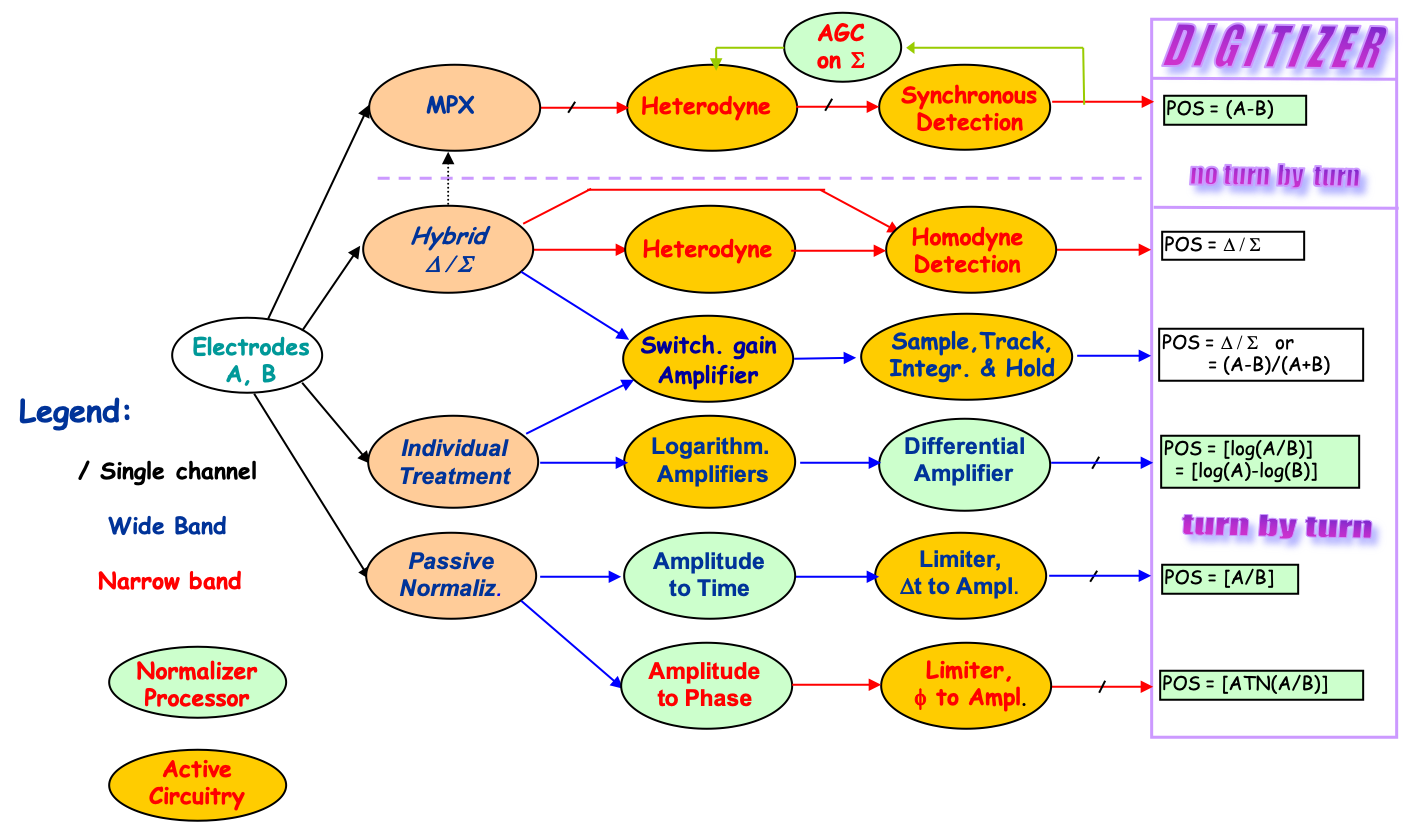}
\caption{BPM analog signal processing concepts (courtesy \textit{G.~Vismara}).}
\label{fig:BPManalog}
\end{center}
\end{figure}
\begin{figure}[t]
\begin{center}
\includegraphics[width=0.95\textwidth]{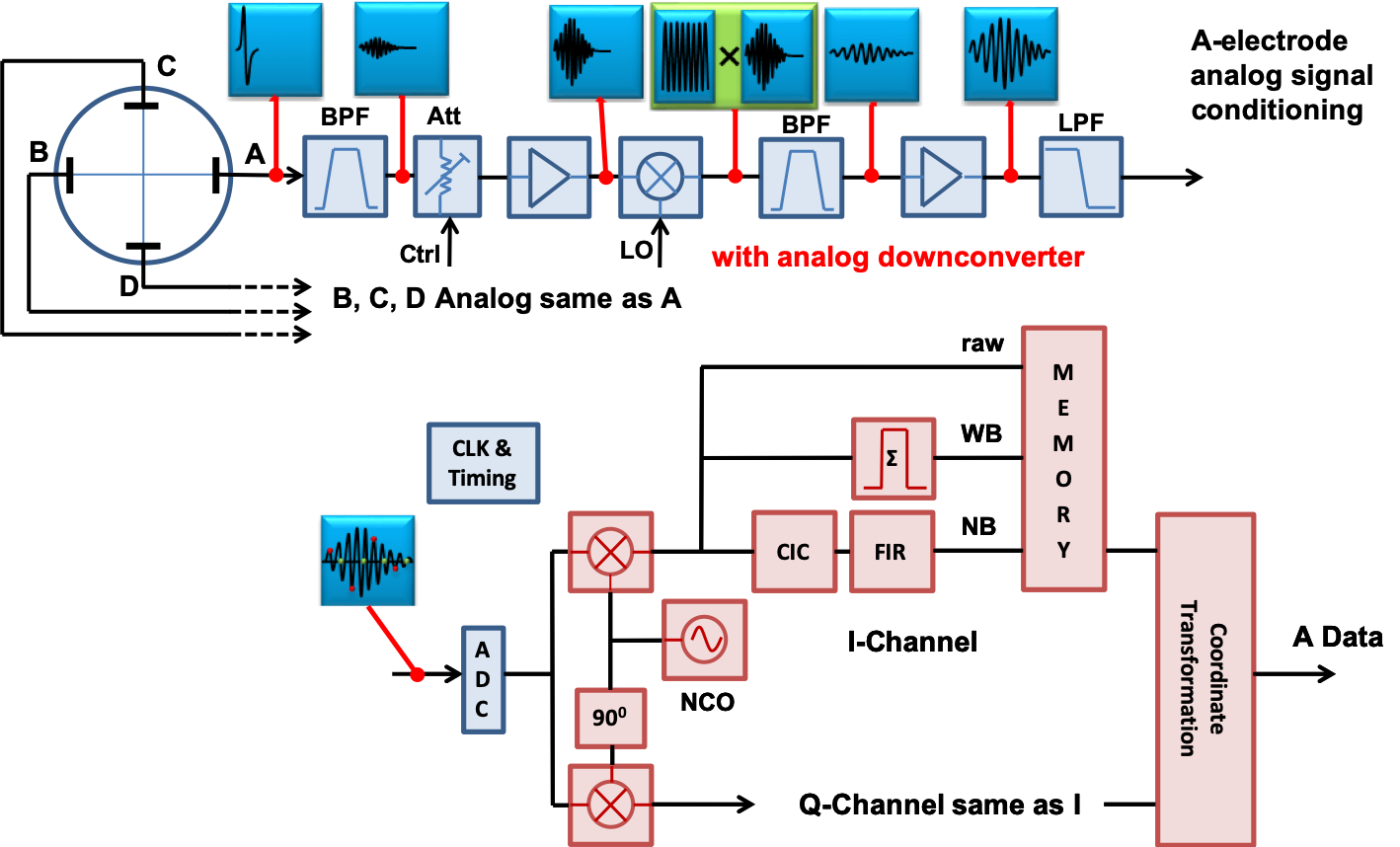}
\caption{BPM signal processing with analog and digital electronics.}
\label{fig:BPMdigital}
\end{center}
\end{figure}

There is a very large variety of BPM signal processing concepts and post-processing schemas, impossible to cover even
a few of them in this tutorial.
Moreover, many of the different BPM signal processing concepts reflect the present state-of-the-art in electronics
and telecommunication technologies, but will be quickly outdated as these techniques advance.
Figure~\ref{fig:BPManalog} shows an overview of the various BPM analog signal processing concepts which
were popular until the last millennium; Fig.~\ref{fig:BPMdigital} shows a more ``modern'' concept, utilizing 
a digital based signal processor which was developed
in the beginning of the millinium, and is based on I/Q demodulation techniques known from the telecom industry. 
The latter technique requires four separate
processing channels, one for each BPM electrode. 

For the BPM read-out concepts, the technology advances of analog-to-digital (ADC) converters have the greatest
influence on selecting a particular signal processing schema.
Clearly, the trend goes to digital signal processing techniques, which prevents imperfections and drift effects 
in analog and RF circuit sections due to the aging of
electronics components, their tolerances, and the variation of their nominal values with the ambient temperature. 
The digital hardware also enables more complex arithmetics on the BPM signals, typically implemented as
gateware in a field-progamable-gate-array (FPGA).
Still, some minimalistic conditioning of the BPM electrode signals with classical analog and/or RF techniques remains,
to adapt the signal levels and modify the waveforms to the optimal ADC operational conditions.
Also, the generation of test or calibration signals, necessary to ensure long term reliability and performance of the
BPM processor, requires some analog and RF electronics.

\subsection{BPM resolution}

The resolution of a beam position monitor, 
i.e.\ the smallest detectable change of the beam position is -- beside accuracy and long term
stability (reproducibility) -- the most important performance parameter of a BPM system.
Both, BPM pickup and the following signal processor define the achievable BPM resolution.

Consider a broadband BPM, e.g.\ a pair of button electrodes $A$ and $B$.
For small beam displacements we consider only the linear term of the normalization function, 
the beam position is detected as:
\begin{align}
\text{pos}&=k_{\mathrm{PU}}\frac{\Delta_{\mathrm{meas}}}{\Sigma_{\mathrm{meas}}}
=k_{\mathrm{PU}}\frac{\Delta_{\mathrm{meas}}+\Delta_{\mathrm{noise}}}{\Sigma_{\mathrm{meas}}+\Sigma_{\mathrm{meas}}}
\simeq k_{\mathrm{PU}}\frac{\Delta_{\mathrm{PU}}}{\Sigma_{\mathrm{PU}}}+
\underbrace{k_{\mathrm{PU}}\frac{\Delta_{\mathrm{noise}}}{\Sigma_{\mathrm{PU}}}}_{\text{resolution}} \\
k_{\mathrm{PU}}\frac{\Delta_{\mathrm{noise}}}{\Sigma_{\mathrm{PU}}}&=\text{resolution}
=k_{\mathrm{PU}}\frac{A_{\mathrm{noise}}-B_{\mathrm{noise}}}{A_{\mathrm{PU}}+B_{\mathrm{PU}}}
\end{align}
with $k_{\mathrm{PU}}\approx 2/R$, see Eq.~(\ref{eq:HorPosLin}).
Assuming small beam displacements:
\begin{align*}
A_{\mathrm{PU}}\simeq B_{\mathrm{PU}} &= S \\
A_{\mathrm{noise}}\simeq B_{\mathrm{noise}} &= N 
\end{align*}
and $A_{\mathrm{noise}}$, $B_{\mathrm{noise}}$ being uncorrelated, we find:
\begin{equation}
\text{resolution}=k_{\mathrm{PU}}\frac{\sqrt{2}N}{2S} = \frac{k_{\mathrm{PU}}}{\sqrt{2}}\left(\frac{S}{N}\right)^{-1}
\end{equation}
The \emph{noise} $N$ is contributed by the signal processor, and has a minimum level of:
\begin{equation}
v_{\mathrm{noise}}=\sqrt{4kTR\Delta f}
\label{eq:noise}
\end{equation}
with $k=1.38\cdot 10^{-23}$~J/K (\textit{Bolzmann} constant), $T=300$~K (typical operating temperature), and
$R=50\;\Omega$ (typical load impedance of the signal source, here: BPM electrode).
In practice the thermal noise level of Eq.~\ref{eq:noise} is the lower limit of  $v_{\mathrm{noise}}$,
passive components including cables (counting as insertion loss), and the so-called \emph{noise figure} of
gain stages (amplifiers) will always results in a higher value of  $v_{\mathrm{noise}}$.\newline
$\Delta f$ is the 3~dB \emph{overall} bandwidth of the BPM processing electronics, this includes \emph{all} 
filter, averaging, and other bandwidth limiting elements on the BPM signal or data (analog, digital, and software).\newline
The \emph{signal} level $S$ can be estimated from the discussion and equations in section \ref{sec:BPMgaussResponse},
selecting a frequency band as defined by the input filters of the front-end electronics.

\section{Summary}

This summary of the tutorial on BPM systems, presented at the CAS2018 on beam instrumentation,
covered the more elementary topics and aspects on beam position monitoring.
The presented material is of common knowledge, to be found in textbooks, papers and conference contributions,
see the bibliography below.
Typos and errors found in the CAS presentation have been corrected, including some ``clean-up'' of the 
used symbols, also some topics required clarifications and enhancements.

As this introduction on BPM systems was set in frame of a school tutorial, many technology aspects have been
omitted, and the focus was set on the physics and engineering basics of the beam position detectors,
the beam signals, and a few, basic aspects of the BPM signal processing.
The equations and mathematical expression have not been derived from the fundamental field theory and principles,
instead known basic solutions have been introduced as is, on which the formalism for the beam
position monitoring was build upon.
With the presented material the reader should be able to design a BPM system, in particular to select a
BPM pickup for given beam conditions and parameters, to evaluate its characteristics, 
and to estimate the performance of the BPM in terms of resolution potential.

For more details on this subject I recommend the literature below and the contributions to the 
workshops and conference series
of the BIW, DIPAC and IBIC.

\section*{Bibliography}

J.~Cuperus, \emph{Monitoring of Particle Beams at High Frequencies},
Nuclear Instruments and Methods, 145 (1977) 219, CERN/PS/LIN-76-7

\noindent
D.~A.~Goldberg and G.~R.~Lambertson, \emph{Dynamic Devices -- A Primer on Pickups and Kickers},
AIP Conference Proceedings, 
249,  (1992), pp. 537-600

\noindent
R.~E.~Shafer, \emph{Beam Position Monitoring},
AIP Conference Proceedings, 
249,  (1992), pp. 601-636

\noindent
R.~E.~Shafer, \emph{Beam Position Monitor Sensitivity for Low-$\beta$ Beams},
AIP Conference Proceedings,
319,  (1994), pp. 303-308

\noindent
D.~P.~McGinnis, \emph{The Design of Beam Pickup and Kickers},
AIP Conference Proceedings,
333, (1995), pp. 64-85

\noindent
F.~Marcellini, M.~Serio, M.~Zobov, \emph{DA$\mathit{\Phi}$NE Broad-Band Button Electrodes},
INFN-LNF, Accelerator Division, DA$\Phi$NE Technical Note CD-6, Frascati, Italy, Jan.~1996

\end{document}